\documentclass[aps,prb,amsmath,superscriptaddress,amssymb,longbibliography,twocolumn]{revtex4-2}

\usepackage{amsfonts}
\usepackage{amsbsy}
\usepackage{amsmath}
\usepackage{bm}
\usepackage{braket}
\usepackage{color}
\usepackage{epsfig}
\usepackage{subfigure}
\usepackage{eqnarray}
\usepackage{float}
\usepackage{graphicx}
\usepackage[colorlinks,urlcolor=blue,linkcolor=red,anchorcolor=blue,citecolor=red]{hyperref}
\usepackage{enumitem}
\usepackage{placeins}

\newcommand{\ttau}{\tilde\tau}

\begin{document}
\title{Phases and Quantum Phase Transitions in Anisotropic Antiferromagnetic Kitaev-Heisenberg-\texorpdfstring{$\Gamma$}{} magnet}
\author{Animesh Nanda}
\email{animesh.nanda@icts.res.in}
\affiliation{International Centre for Theoretical Sciences, Tata Institute of Fundamental Research, Bangalore 560089, India}
\author{Adhip Agarwala}
\email{adhip@pks.mpg.de}
\affiliation{Max Planck Institute for the Physics of Complex Systems, Dresden, Germany}
\affiliation{International Centre for Theoretical Sciences, Tata Institute of Fundamental Research, Bangalore 560089, India}
\author{Subhro Bhattacharjee}
\email{subhro@icts.res.in}
\affiliation{International Centre for Theoretical Sciences, Tata Institute of Fundamental Research, Bangalore 560089, India}

\begin{abstract}
We study the Kitaev-Heisenberg-$\Gamma$ model with antiferromagnetic Kitaev exchanges in the strong anisotropic (toric code) limit to understand the phases and the intervening phase transitions between the gapped $Z_2$ quantum spin liquid and the spin-ordered (in the Heisenberg limit) as well as paramagnetic phases (in the pseudo-dipolar, $\Gamma$, limit).  We find that the paramagnetic phase obtained in the large $\Gamma$ limit has no topological entanglement entropy and is proximate to a gapless critical point of a system described by equal superposition of differently oriented stacked one-dimensional $Z_2\times Z_2$ symmetry protected topological phases. Using a combination of exact diagonalization calculations and field theoretic analysis we map out the phases and phase transitions to reveal the complete phase diagram as a function of the Heisenberg, the Kitaev and the pseudo-dipolar interactions. Our work shows a rich plethora of unconventional phases and phase transitions and provides a comprehensive understanding of the physics of anisotropic Kitaev-Heisenberg-$\Gamma$ systems along with our recent paper [Phys.  Rev.  B 102, 235124 (2020)~[\onlinecite{nanda2020phases}]] where the ferromagnetic Kitaev exchange was studied.
\end{abstract}
\maketitle


\section{Introduction}

Systems of interacting spins on a lattice serve as a rich playground for exploring novel quantum phases as well as associated phase transitions that are brought about by the interplay of symmetries and competing interactions \cite{diep2013frustrated}. In addition to the broken symmetry phases, we now know of a plethora of quantum spin-liquids (QSLs) ~\cite{anderson1973resonating,wen2017colloquium,anderson1987resonating,wen2002quantum,savary2016quantum,balents2010spin,lee2008end,lee2006doping} and symmetry protected topological (SPT)~\cite{wen2017colloquium,Chen_PRB_2013,Chen_PRB_2013, Ryu_PS_2015, Senthil_ARCMP_2015} phases that can be realised in lattice spin-systems-- often of direct relevance to  candidate materials. In this regard, magnetic systems with strong spin-orbit coupling are leading to spin Hamiltonians without full $SU(2)$ spin-rotation symmetry have been particularly interesting on both theoretical \cite{witczak2014correlated,hermanns2018physics} and experimental front~\cite{takagi2019concept,broholm2020quantum} by providing, respectively, explicit solutions of novel magnetic phases~\cite{PhysRevLett.105.027204,PhysRevLett.112.207203,PhysRevB.89.014414,trebst2017kitaev,nussinov2013compass,PhysRevB.79.024426} and their possible materials realisations~\cite{kitaev2003fault,kitaev2006anyons,PhysRevLett.90.016803,PhysRevLett.98.247201,PhysRevLett.102.017205,PhysRevB.82.064412,PhysRevLett.108.127203,PhysRevB.85.180403,PhysRevLett.108.127204,PhysRevLett.113.197201,Banerjee2016,Banerjee2017,banerjee2018excitations,kasahara2018majorana,lee2014heisenberg,PhysRevLett.119.057203,PhysRevB.93.064406,PhysRevX.1.021002,takagi2019concept}. 

In parallel with the novel phases, these lattice systems allows us to pose concrete questions about the nature of the quantum phase transitions associated with QSL and SPT phase. These transitions generically are not captured by the conventional order parameter based theories of phase transitions~\cite{chaikin1995principles} as they fail to capture the non-trivial structure of the entanglement pattern in the QSLs and the SPTs~\cite{wen2017colloquium,senthil2006quantum,Senthil_ARCMP_2015}. Intense research over the last two decades have fleshed out several paradigmatic features of the theory of such unconventional quantum phase transitions-- in particular continuous transitions or quantum critical points~\cite{senthil2006quantum,senthil2004deconfined,senthil2004quantum}. Central to these ideas is the construction of the critical theory of such critical points-- dubbed as deconfined quantum critical points (DQCP)~\cite{senthil2006quantum,senthil2004deconfined,senthil2004quantum}-- in terms of the fractionalised fields (instead of the order parameter), transforming under the projective representation of the microscopic symmetries~\cite{wen2002quantum}, which interact with each other with emergent fluctuating gauge fields. The construction and controlled understanding of such critical theories of possible DQCPs, particularly in context of experimentally relevant situation is therefore crucial for novel quantum ordered phases of matter.

In order to obtain a controlled understanding of transitions out of an exactly solvable $Z_2$ QSL, in a recent paper \cite{nanda2020phases}  we presented our results for the phases and phase transitions for the anisotropic or Toric code limit~\cite{kitaev2003fault} of Kitaev-Heisenberg-$\Gamma$ (pseudo-dipolar) magnet where the Kitaev interactions are {\it ferromagnetic}. By systematic analysis of the symmetries of the low energy excitations of the $Z_2$ QSL-- the Ising magnetic and the electric charges-- we obtained the critical theory for transitions out of the QSL to both a magnetically ordered phase (driven by the Heisenberg interactions) and a trivial paramagnet phase (driven by the pseudo-dipolar interactions). Central to our analysis were the non-trivial implementation of the time-reversal symmetry and the transition symmetries on the gauge charges. In particular, the magnetic and electric charges transformed into each other under primitive lattice translations enforcing an electromagnetic self-dual structure on the description of the resultant {\it anyon permutation} protected deconfined critical point`\cite{nanda2020phases}. 

\begin{figure}
\includegraphics[]{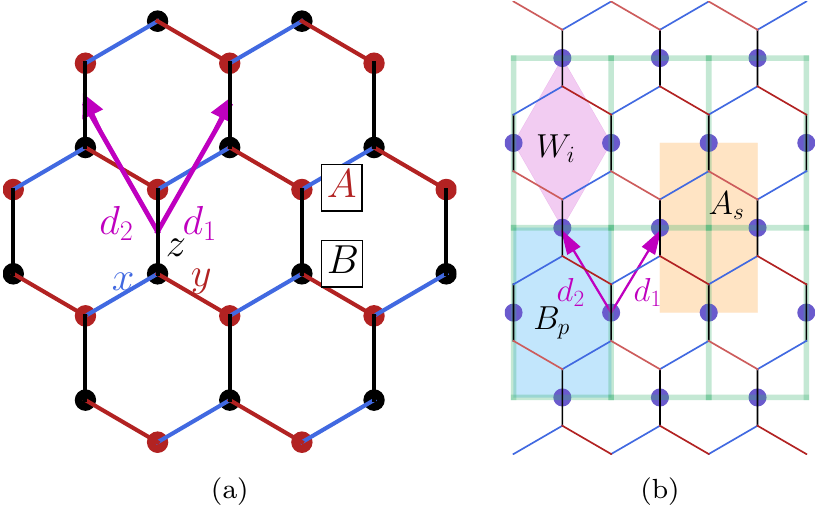}
\label{fig_kitaevtoric}
\caption{(a) Kitaev's honeycomb model is described on a bipartite hexagonal lattice with two sublattices A (red) and B (black). The spin exchanges are defined on three bonds labelled by $x,y,z$ (shown in blue, red and black lines respectively). (b) The anisotropic limit in the $z$ direction leads to an square lattice where new degrees of freedom $\tau$ (shown in grayish blue) sits on the bonds. Lattice vectors $\bf d_1$ and $\bf d_2$ are $\left(\frac{1}{2},\frac{1}{2}\right)$ and $\left(-\frac{1}{2},\frac{1}{2}\right)$ defined in the units of square lattice bond length which is assumed to be same in horizontal and vertical direction.}
\end{figure}

In this paper we present our results of the same class of systems, but with the Kitaev interactions being {\it antiferromagnetic} to reveal a richer physics (compared to the ferromagnetic case of Ref. [\onlinecite{nanda2020phases}]). Exploiting the energy hierarchy in the anisotropic Kitaev interactions, we distill low energy degrees of freedom to show that the difference in the physics arises due to an inherent feature of the interplay of symmetries and correlations alluded above-- the microscopic antiferromagnetic interactions lead to low energy degrees of freedom that have very different symmetry properties from the ferromagnetic case. Our starting point remains the Heisenberg-Kitaev-Pseudo-dipolar ($KJ\Gamma$) Hamiltonians on the honeycomb lattice of the form~\cite{PhysRevLett.112.077204,kitaev2006anyons,PhysRevLett.102.017205,PhysRevLett.105.027204}
\begin{align}
\mathcal{H}=&J\sum_{\langle p,q\rangle}\boldsymbol{\sigma_p}\cdot\boldsymbol{\sigma_{q}}+\sum_{\langle p,q\rangle \alpha}\left[\Gamma\left(\sigma_{p}^{\beta}\sigma_{q}^{\gamma}+\sigma_{q}^{\beta}\sigma_{p}^{\gamma}\right)-K_{\alpha}\sigma_{p}^{\alpha}\sigma_{q}^{\alpha}\right]
\label{eq_hamiltonian}
\end{align}
where $\alpha=x,y,z$ are the three bonds of the honeycomb lattice (Fig.~\ref{fig_kitaevtoric}) and $\sigma^\alpha_p$ denotes the Pauli matrices denoting the spin-1/2s at the sites of the honeycomb lattice for e.g., at $p,q$.   We are now interested in the antiferromagnetic Kitaev limit, {\it i.e.} $K_{\alpha}<0$ such that the anisotropic limit is obtained by taking, as in Ref.~[\onlinecite{nanda2020phases}], $|K_z|\gg \{ |J|, |K_x|=|K_y|\equiv|K|, |\Gamma| \}$. 

The low energy non-Kramers doublet stabilised in the anisotropic limit of the antiferromagnetic Kitaev exchange is different from the Ferromagnetic case and leads to a different realisation of the microscopic symmetries resulting in a different low energy Hamiltonian for these doublets. While this reflects in a distinct symmetry enrichment for the QSL or distinct spin-orders in the Heisenberg limit, the most startling effect occurs in the large pseudo-dipolar ($\Gamma$) limit whence the Hamiltonian, to the lading order in perturbation theory, leads to a superposition of stacked $Z_2\times Z_2$ spin SPT phases. The resultant phase is, according to our numerical calculations, a gapless critical point which is trivially gapped out by higher order (hence weaker) perturbations. Remarkably this gapless critical point supports edge modes that do not hybridise with the bulk modes due to subsystem symmetries. Interestingly in recent studies investigating the role of pseudo-dipolar interactions in both isotropic and anisotropic Kitaev Hamiltonians \cite{, Wachtel_PRB_2019,yamada2020ground,  Gohlke_PRB_2018, Buessen_PRB_2021} have found gapless phases \cite{luo2021gapless, Wang_PRL_2019} (often for ferromagnetic Kitaev exchanges). The relevance of these other gapless phases to our work is not immediately clear and needs to be further explored.

Our numerical studies on small spin clusters reveal the general structure of the phase diagram indicating that the $Z_2$ QSL is destroyed via proliferation and condensation of its gauge charges-- both electric and magnetic. While the transition to the paramagnetic phase in the large $\Gamma$ limit turns out to be discontinuous, a the continuous transition to the spin-ordered state (from the QSL) is driven Heisenberg coupling via a deconfined critical point. We construct a critical continuum field theory in terms of the soft modes of the electric and magnetic charges via a mutual $Z_2$ Chern-Simons (CS) theory and show that the direct transition between the QSL and the spin-ordered phase is described by a self-dual modified Abelian Higgs field theory-- in agreement with the critical theory obtained by us in the ferromagnetic case using a mutual $U(1)$ CS theory in Ref. [\onlinecite{nanda2020phases}]. The overall summary of our phase diagram is then illustrated in Fig.~\ref{fig_full_pd_schematic}.

The rest of this paper work is organised follows. We start with a discussion of the anisotropic limit of Eq.~\ref{eq_hamiltonian}, its low energy degrees of freedom and effective interactions in section~\ref{sec_hamil} and derive the action of symmetries on them as well as the low energy effective Hamiltonian that captures the low energy physics. We show that the nature of the low energy degree of freedom-- an effective non-Kramers spin-$1/2$-- is different from the ferromagnetic case leading to a  different symmetry transformation and low energy Hamiltonian. We start our analysis of the effective low energy Hamiltonian in Section \ref{sec_phase_and_PD}. In particular we examine the three different limits dominated by the Kitaev, the Heisenberg and the pseudo-dipolar interactions. While in the first two case a $Z_2$ QSL and various spin-ordered phases are stabilised respectively, similar to the FM case~\cite{nanda2020phases}-- albeit with important difference in the symmetry implementation, the limit where the pseudo-dipolar interactions dominate turns out to be startlingly different. In this limit, the leading order interactions lead to a superposition of stacked $Z_2\times Z_2$ SPTs with edge modes and special sub-system symmetries which are weakly lifted by higher order interactions. The equal superposition of SPT lead to a gapless critical point according to our finite size exact diagonalisation calculations. The gapless point, accordingly to our analysis, is fragile and immediately opens up a small gap due to higher order perturbations. In Section \ref{sec_numerics_afm} we present the results of our exact diagonalisation calculations on the leading order low energy Hamiltonian to obtain an estimate of the phases and phase boundaries. This analysis shows that the transition out of the QSL is brought about by the proliferation and condensation of the its excitations-- the Ising electric and magnetic charges. With these ingredients we consider the physics of the phase transitions in Section \ref{sec_phasetrans}. We find that contrary to the FM case~\cite{nanda2020phases}, the transition between the QSL and the paramagnetic phase in the large $\Gamma$ limit is a first order transition. For the continuous transition between the QSL and the spin-ordered phase in the large Heisenberg limit, we develop the critical theory in terms of the soft Ising electric and magnetic charge modes of the QSL. Using a mutual $Z_2$ CS theory to implement the mutual semionic statistics between the electric and magnetic charges of the QSL, we construct the continuum critical theory in addition to the mutual $U(1)$ CS theory implemented in Ref. [\onlinecite{nanda2020phases}]. Both these approaches consistently lead to a self-dual modified Abelian Higgs's theory that describes the deconfined critical point for the direct continuous transition between the QSL and the spin-ordered phases. Finally we summarise our results regarding the phase diagram obtained for the anisotropic limit of $JK\Gamma$ model with antiferromagnetic Kitaev exchange  in Section. \ref{sec_summary}. Various details of the calculations are summarised in different appendices.  Throughout this paper, we shall continue to use several notations elaborately introduced in Ref.~[\onlinecite{nanda2020phases}] and here we briefly summarised the relevant portions. 

\section{The low energy spin model in the anisotropic limit}
\label{sec_hamil}
Similar to Ref. [\onlinecite{nanda2020phases}], the effective low energy Hamiltonian is obtained by re-writing Eq.~\ref{eq_hamiltonian} as $\mathcal{H}=\mathcal{H}_{0}+\mathcal{V}$ where $\ \mathcal{H}_0 $ is given by
\begin{equation}
\mathcal{H}_{0}=(|K_z|+J)\sum_{\langle p,q\rangle,z}\sigma_p^z\sigma_q^z
\label{eq_exact_part}
\end{equation}
where the sum is over only the $z$-bonds (Fig.~\ref{fig_kitaevtoric}(a)). $\mathcal{V}$ stands for the rest of the terms in Eq.~\ref{eq_hamiltonian} which can be treated as perturbation in the anisotropic limit. 

Similar to the FM case, for $\mathcal{V}=0$ the system breaks up into isolated bonds and each bond has two ground states. However,  contrary to the FM~\cite{nanda2020phases}, in the present AFM case of $\mathcal{H}_0$, the two spins on each $z$-bond are anti-aligned with respect to each other in the ground state manifold. So the ground states and the excited states are:

\begin{align}
\text{Ground States:}~|\uparrow\downarrow\rangle,~~|\downarrow\uparrow\rangle \\
\text{Excited States:}~|\uparrow\uparrow\rangle,~~|\downarrow\downarrow\rangle
\label{eq_tau}
\end{align}
which is exactly opposite to the FM case~\cite{nanda2020phases}. We define a new degree of freedom for the two fold ground state manifold of $\mathcal{H}_0$ in Eq.~\ref{eq_exact_part} as
\begin{equation}
|+\rangle \equiv \ket{\uparrow\downarrow}~;~|-\rangle \equiv \ket{\downarrow\uparrow}
\label{eq_tau_def_afm}
\end{equation}
The $\tau^z$ operator defined on each $z$-bond acts on this ground state space as: $\ \tau^z\ket{\pm}=\pm\ket{\pm} $ which in terms of the underlying $\sigma$ spins is, 
\begin{align}\label{eq_tau_def}
\tau^z=\
(\sigma_A^z-\sigma^z_B)/2
\end{align}
where the subscripts $A$ and $B$ labels the two spins belonging to the two different sublattices participating in a particular $z$-bond. The $\tau$-spins therefore reside on the links of a square lattice with lattice vectors ${\bf d_1}\ \&\ {\bf d_2}$ as shown in Fig.~\ref{fig_kitaevtoric}(b). To this end we define the lattice points where the $\tau$-spins reside:
\begin{align}
i\equiv (i_1,i_2)=i_1{\bf d_1}+i_2{\bf d_2},
\end{align}

\subsection{Symmetry transformation of $\tau$ spins}\label{subsec_symm}

\begin{figure}
\centering
\includegraphics[]{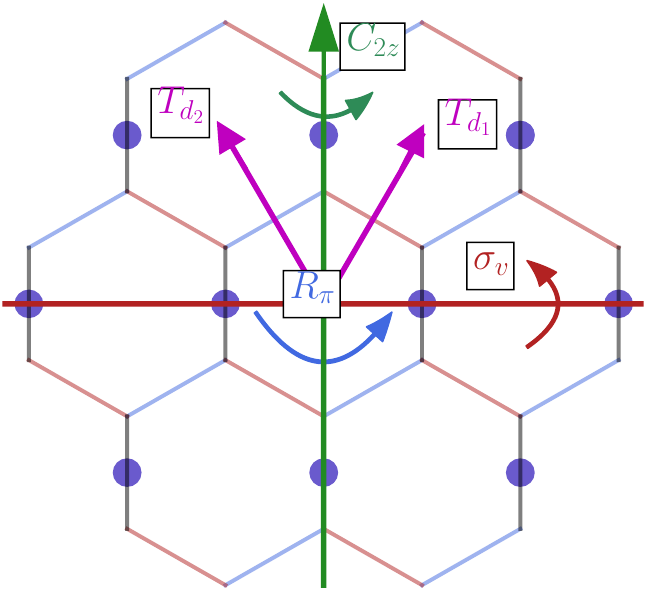}
\caption{The symmetries in the toric code limit of Kitaev's honeycomb model, {\it e.g.} $T_{d_1}$, $T_{d_2}$, $R_\pi$, $\sigma_v$ and $C_{2z}$ are shown. See the corresponding text in section~\ref{subsec_symm}.}
\label{fig_symm_details}
\end{figure}

Starting with the symmetries of the honeycomb lattice and focusing on the anisotropic limit we derive the symmetry transformation of the $\tau$-spins under the generators of symmetry group given by (also see Fig.~\ref{fig_symm_details}):

\begin{itemize}
    \item Time reversal, $\mathcal{T}$.
    \item Translations in the honeycomb plane, $T_{d_1}$ and $T_{d_2}$. These acts on the $(i_1,i_2)$ as $T_{d_1}:(i_1,i_2)\rightarrow(i_1+1,i_2)$ and $T_{d_2}:(i_1,i_2)\rightarrow(i_1,i_2+1)$.
    \item Reflection about the z-bond, $\sigma_v$. On the lattice this acts as: $\sigma_v:(i_1,i_2)\rightarrow(-i_2,-i_1)$.
    \item $\pi$-rotation about the z-bond, $C_{2z}$, which gives  $C_{2z}:(i_1,i_2)\rightarrow(i_2,i_1)$.
\end{itemize}

Additionally, we will consider the square lattice translation symmetries $T_x\equiv T_{d_1}-T_{d_2}$ and $T_y\equiv T_{d_1}+T_{d_2}$ and the rotational symmetry about the center of the hexagonal plaquette $R_{\pi}\equiv C_{2z}\sigma_v$. A detailed discussion of these symmetries and the way they act in both real and spin space was given in \cite{nanda2020phases}. The symmetry transformations for the $\tau$ spins are summarised in Table \ref{table_tau_symm} where interestingly time-reversal takes $\tau^{x(y)}_i \rightarrow \tau^{x(y)}_i$. 

Notably, while the symmetries are same as that of the ferromagnetic case~\cite{nanda2020phases}, the symmetry table, particularly the action of the point group symmetries are rather different in the present case due to the difference in the make-up of the ground state doublet. This, as we shall see, has profound influence on the nature of the effective Hamiltonian and ultimately in the phases obtained. In particular this directly affects the physics in the large $\Gamma$ limit.

\begin{table}
\begin{center}
 \begin{tabular}{|c| c| c| c|} 
 \hline
 Symmetry & $ \tau^x_{i}$ & $\tau^y_{i}$ & $\tau^z_{i}$  \\ [0.5ex] \hline \hline
 $\mathcal{T}$ & $ \tau^x_{i}$ & $\tau^y_{i}$ & $-\tau^z_{i}$  \\  [0.5ex] \hline
 $\sigma_v$ & $ \tau^x_{(\bar{i}_2,\bar{i}_1)}$ & $\tau^y_{(\bar{i}_2,\bar{i}_1)}$ & $\tau^z_{(\bar{i}_2,\bar{i}_1)}$ \\ [0.5ex] \hline
 $C_{2z}$ & $\tau^x_{(i_2,i_1)}$ & $-\tau^y_{(i_2,i_1)}$ & $-\tau^z_{(i_2,i_1)}$ \\ [0.5ex] \hline
 $R_{\pi}$ & $\tau^x_{(\bar{i}_1,\bar{i}_2)}$ & $-\tau^y_{(\bar{i}_1,\bar{i}_2)}$ & $-\tau^z_{(\bar{i}_1,\bar{i}_2)}$ \\ [0.5ex] \hline
  \end{tabular}
\caption{Symmetry transformations of the $\tau$ spins under various microscopic symmetries, where $i=(i_1,i_2)$ and $\bar{i}_{1(2)}\equiv -i_{1(2)}$. See the corresponding text in section.~\ref{subsec_symm} for details.}
\label{table_tau_symm}
\end{center}
\end{table}

The effective low energy Hamiltonian below the $\sim |K_z|$ scale is captured by the $\tau$ spins can now be gotten using a degenerate perturbation theory with the strong coupling expansion in $1/|K_z|$. 
\subsection{The effective Hamiltonian}\label{AFM anisotropic limit}

The low energy effective Hamiltonian (up to fourth order in perturbation theory) is given by
\begin{align}\label{eq_afm_full_hamiltonian}
\mathcal{H}^{AF}_{\text{eff}}=\mathcal{H}^{AF}_{[1]}+\mathcal{H}^{AF}_{[2]}+\mathcal{H}^{AF}_{[3]}+\mathcal{H}^{AF}_{[4]}
\end{align}
where $1-4$ represents the number of spins operators involved.  While the detailed form of all these terms are relegated to Appendix \ref{appen_perturb}, it is most transparent to separate various terms in these three different limits: (i) $\Gamma=K=0$ (ii) $J=K=0$ and (iii)   $J=\Gamma=0$.

The leading contributions to the effective Hamiltonian for just the Heisenberg perturbation is given by
\begin{equation}
\begin{aligned}
\mathcal{H}^{AF}_{\Gamma=K=0} =\ 2J\sum_{i}\tau^x_i - J\sum_{\langle i,j\rangle}\tau_i^z\tau_j^z.
\end{aligned}
\label{eq_heisenberg_limit_min_afm}
\end{equation}
Higher order terms can renormalize these coefficients and also generate further neighbour interactions as has been shown in Appendix~\ref{appen_perturb}. Except for the transverse field the above leading order in nearest neighbour Ising term is exactly similar to that of the FM case of Ref.~\cite{nanda2020phases} and has similar ordering effects although the details of the magnetic patterns here are different (see below) as denoted in Fig.~\ref{fig_neel_zz}. The effect of the transverse field, for the present case, we think, does not play a major role as we discuss below, but opens up a very interesting possibility in the isotropic limit that is related to a two step transition from the QSL to the magnetic phase via an intermediate nematic \cite{Agarwala_unpublised}.

The leading pseudo-dipolar contributions, on the other hand, are of the form

\begin{equation}\label{eq_gam_limit_afm1}
\begin{aligned}
\mathcal{H}^{AF}_{J=K=0} = & \frac{\Gamma^2}{|K_z|}\sum_{i}\left(\tau^z_{i+d_1}\tau^x_{i}\tau^z_{i-d_1} + \tau^z_{i+d_2}\tau^x_{i}\tau^z_{i-d_2}\right) \\
							& + \frac{\Gamma^2}{|K_z|}\sum_{i}\left( \tau^z_{i+d_1}\tau^y_{i}\tau^z_{i-d_2}-\tau^z_{i+d_2}\tau^y_{i}\tau^z_{i-d_1}\right)
\end{aligned}
\end{equation}
whose form is drastically different from the leading transverse field term for the FM case~\cite{nanda2020phases} and is one of the central difference as we shall discuss in detail.

The pure-$\Gamma$ Hamiltonian (see Eq.~\ref{eq_gam_limit_afm1}) is a linear sum of three spin terms, separately which stabilizes stacked cluster $Z_2\times Z_2$ SPTs \cite{verresen2017one,you2018subsystem,chen2014symmetry,son2012topological,nielsen2006cluster,dubinkin2019higher} oriented in different directions. Our analysis, both numerical and based on symmetry analysis of this and unitarily related Hamiltonians (see Section~\ref{subsec_gamma_limit_afm}), shows that the pure-$\Gamma$ Hamiltonian either is a critical point or part of a gapless phase which resides proximate to gapped weak-SPT phases and this constitutes one of the central results of this work. In presence of symmetry allowed perturbations we find that this gapless phase (or the critical point) is smoothly connected to a paramagnetic phase.  Interestingly, we find that this pure-$\Gamma$ limit Hamiltonian can host novel boundary physics in terms of exact zero energy modes which, as expected is unstable to symmetry allowed perturbations.

Finally, in the limit $J=\Gamma=0$ the pure Kitaev interactions take the toric code form, albeit in Wen's representation \cite{wen2002quantum} and is given by:
\begin{align}\label{eq_tc_rot_unrot}
    \mathcal{H}^{AF}_{J=\Gamma=0}=&-J_{TC}\sum_{i} W_i
\end{align}
With $J_{TC}=\frac{K^4}{16|K_z|^3}$, and $W_i\equiv\tau^z_{i+d_1}\tau^z_{i-d_2}\tau^y_i\tau^y_{i+d_1-d_2}$ is the plaquette operator shown on the lattice (see Fig.~\ref{fig_kitaevtoric}).
which stabilises the gapped $Z_2$ QSL with bosonic Ising electric, $e$, and magnetic, $m$, charges \cite{kitaev2006anyons}. The higher order terms provide further interactions and for most part of the paper we shall neglect such interactions for analysing the leading order instability of the QSL unless stated. 

\section{Phases and phase diagram}\label{sec_phase_and_PD}

Having isolated the different terms in the effective Hamiltonian, we now discuss their effects in stabilising different phases in order to develop the theory for the associated phase transitions.

\subsection{Toric code limit \texorpdfstring{$J=\Gamma=0$}{}}\label{subsec_tc_limit_afm}

The Hamiltonian given in Eq.~\ref{eq_tc_rot_unrot} after a bond dependent unitary rotations as defined in Appendix \ref{appen_taurotations} (see also \cite{nanda2020phases,kitaev2003fault,kitaev2006anyons}) becomes 
\begin{align}\label{eq_tc_rot}
    \tilde{\mathcal{H}}^{AF}_{J=\Gamma=0}=&-J_{TC}\left[\sum_{s} A_s+ \sum_{p} B_p\right]
\end{align}
Where $A_s=\prod_{i\in s}\ttau^x_i$, $B_p=\prod_{i\in p}\ttau^z_i$.  $\ttau^\alpha$ denotes the rotated operators and  Eq.~\ref{eq_tc_rot} represents Kitaev's toric code model \cite{kitaev2003fault}.  The symmetry transformations for $\tilde{\tau}$ spins are given in the Appendix~\ref{appen_rot_ham_afm}. 

The Toric-code Hamiltonian~\cite{kitaev2003fault} is exactly solvable and stabilises a $Z_2$ QSL ground state with topological order and a four fold ground-state degeneracy on the torus with two bosonic and a fermionic-- all gapped-- excitations. The bosonic excitations are respectively the electric, $e$, and magnetic, $m$, charges of an underlying $Z_2$ gauge theory description and they have mutual semionic statistics. The fermion, on the other hand, can be thought of as a $e-m$ bound state. 

Since these gauge charges would be of central importance to the description of the phase transition out of the QSL, we briefly flesh out the well known details of the the standard mapping from the spins while all the details can be found for e.g.~,in Ref. \cite{nanda2020phases} which uses the same notations as other places \cite{PhysRevLett.98.070602,PhysRevB.91.134419}.

The electric (magnetic) charges, created by the Ising variable $\mu^x_a$ ($\tilde\mu^x_{\bar{a}}$), resides on the sites of the direct (dual) lattice and are each coupled to its own Ising gauge field  $\rho^z_{ab}$ ($\tilde{\rho}^z_{\bar{a}\bar{b}}$) that lives on the links of the direct (dual) lattice (see \cite{nanda2020phases}). Here $a\equiv (a_x,a_y)$ and $\bar{a}\equiv (\bar{a}_x,\bar{a}_y)$ denote the sites of the direct and dual lattice respectively. Therefore the Ising electric charge density is measured by $\mu^z_a= A_{a}$ with the Gauss' law constraint being given by
\begin{align}
\mu^z_a=\prod_{\bar{a}\bar{b}\in a}\tilde{\rho}^z_{\bar{a}\bar{b}}=\prod_{b\in  {a}}\rho^x_{ab}
\end{align}
where $\rho_{ab}^x$ is conjugate to $\rho_{ab}^z$. Similarly for the magnetic charge density, $\tilde{\mu}^z_{\bar{a}}=B_{\bar{a}}$, the Gauss's law is 
\begin{align}
 \tilde{\mu}^z_{\bar{a}}=\prod_{ab \in \bar{a}}\rho^z_{ab}=\prod_{\bar{b}\in {\bar{a}}}\tilde{\rho}^x_{\bar{a}\bar{b}}   
\end{align}
The two equations also encode the mutual semionic statistics between the electric and the magnetic charges.

Finally, to complete the mapping we denote the electric and magnetic charge hopping operators on the direct and dual lattices respectively and they are given by
\begin{align}\label{eq_ttau_z}
    \ttau^z_{i}=\mu^x_a\rho^z_{ab}\mu^x_b
\end{align}
and
\begin{equation}\label{eq_ttau_x}
    \ttau^x_i=\tilde{\mu}^x_{\bar{a}}\tilde{\rho}^z_{\bar{a}\bar{b}}\tilde{\mu}^x_{\bar{b}}
\end{equation}
respectively.

In the presence of the existing microscopic symmetries, the low energy anyon excitations are further enriched and this symmetry enrichment is different from that of the FM case. The (projective) symmetry transformation of the gauge charges as well as the direct and the dual Ising gauge fields are presented in Appendix~ \ref{appen_gauge_set}. Again, the complete set of transformations are different from the FM case such that in presence of these symmetries the FM and the AFM QSLs represent different symmetry enriched topologically ordered $Z_2$ QSLs. 

\subsection{Heisenberg Limit \texorpdfstring{$\Gamma=K=0$}{}}\label{subsec_heisenberg_limit_afm}

\begin{figure}
\includegraphics[]{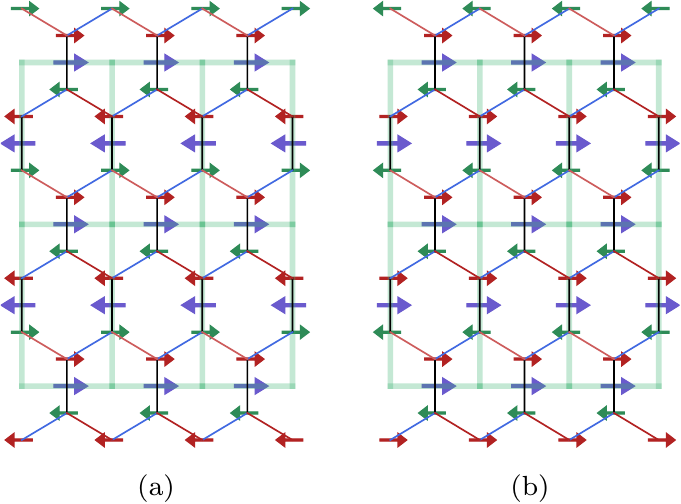}
\label{fig_ph_afm_zz}
\caption{Configuration of $\sigma$ and $\tau$ spins. (a) For $J<0,\Gamma=0, K=0$ a AFM state is realized for $\tau$ spins on $z$ links of the honeycomb lattice, which is denoted by the blue arrows. While the $\sigma$ spins are denoted by red (green) arrow for the $A~(B)$ sublattice, this corresponds to a {\it zig-zag} order for the $\sigma$ spins. Where left (right) blue arrow means $\tau^z=+1(-1)$ state, and left (right) red/green arrow means $\sigma^z=+1(-1)$ state. (b) For $J>0$ a FM state is realized for the $\tau$ spins, which corresponds to a {\it Neel} order for $\sigma$ spins.}
\label{fig_neel_zz}
\end{figure}

Deep inside the anisotropic limit, {\it i.e.} $|K_z|\rightarrow \infty$, the leading order contribution arising from the Heisenberg perturbations to the disconnected dimers is given by 
\begin{equation}
\begin{aligned}
\mathcal{H}^{AF}_{\Gamma=K=0} = - J\sum_{\langle i,j\rangle}\tau_i^z\tau_j^z + 2J\sum_{i}\tau^x_i
\end{aligned}
\label{eq_heisenberg_limit_min}
\end{equation}
where the first term is the Ising interactions that favour ferromagnetic (Neel) ordering of the $\tau^z$-spins for $J>0 (J<0)$. Qualitatively, this is similar to the FM-Kitaev case~\cite{nanda2020phases}, with an important difference in terms of the underlying $\sigma$ spins of the honeycomb magnet-- the ferromagnetic (Neel) ordering for the $\tau^z$ spins correspond to the Neel (Zig-Zag) ordering for the underlying $\sigma^z$ spins as shown in Fig.~\ref{fig_neel_zz}.


The second term, representing the transverse field in the leading order of Heisenberg coupling, however suggests a curious possibility of the Heisenberg perturbations stabilising a paramagnetic state of $\tau$-spins polarised in the $\tau^x$ direction. Very interestingly, in terms of the underlying $\sigma$-spins of the honeycomb lattice this is given by
\begin{align}
    & \ket{\psi_{+}}=\otimes_{pp^{\prime}}\left(\frac{\ket{\uparrow_p\downarrow_{p^{\prime}}}-\ket{\downarrow_p\uparrow_{p^{\prime}}}}{\sqrt{2}}\right)~;~\text{for} J>0 \\ \nonumber
    & \ket{\psi_{-}}=\otimes_{pp^{\prime}}\left(\frac{\ket{\uparrow_p\downarrow_{p^{\prime}}}+\ket{\downarrow_p\uparrow_{p^{\prime}}}}{\sqrt{2}}\right)~;~\text{for} J<0
\end{align}
which are singlet and triplet states respectively for $pp^{\prime}$ that denotes the z-bond (see Eq.~\ref{eq_tau}). Therefore, following \cite{nanda2020phases} the bond-nematic order parameter:
\begin{align}
\hat{Q}^{\alpha \beta}_{pp^{\prime}} = \left(\frac{\sigma^{\alpha}_p\sigma^{\beta}_{p^{\prime}} + \sigma^{\beta}_p\sigma^{\alpha}_{p^{\prime}}}{2} - \frac{\delta_{\alpha\beta}}{3}\boldsymbol{\sigma}_p.\boldsymbol{\sigma}_{p^{\prime}}\right)
\end{align}
is non-zero. In particular, for the $\ket{\psi_{-}}$, we have
\begin{align}
\bra{\psi_{-}}\hat{Q}^{\alpha \beta}_{pp^{\prime}}\ket{\psi_{-}}=\begin{bmatrix}
	\frac{2}{3} & 0 & 0\\
	0 & \frac{2}{3} & 0\\
	0 & 0 & -\frac{4}{3}\\
\end{bmatrix}
\label{eq_nematicop}
\end{align}

On the other hand for $\ket{\psi_{+}}$, singlet dimers are present on the $z$-bonds of the honeycomb lattice. In absence of spin-rotation symmetry, for non-Kramers doublets both these orders represent lattice nematic.

\begin{figure}
\begin{center}
\includegraphics[width=1.0\columnwidth]{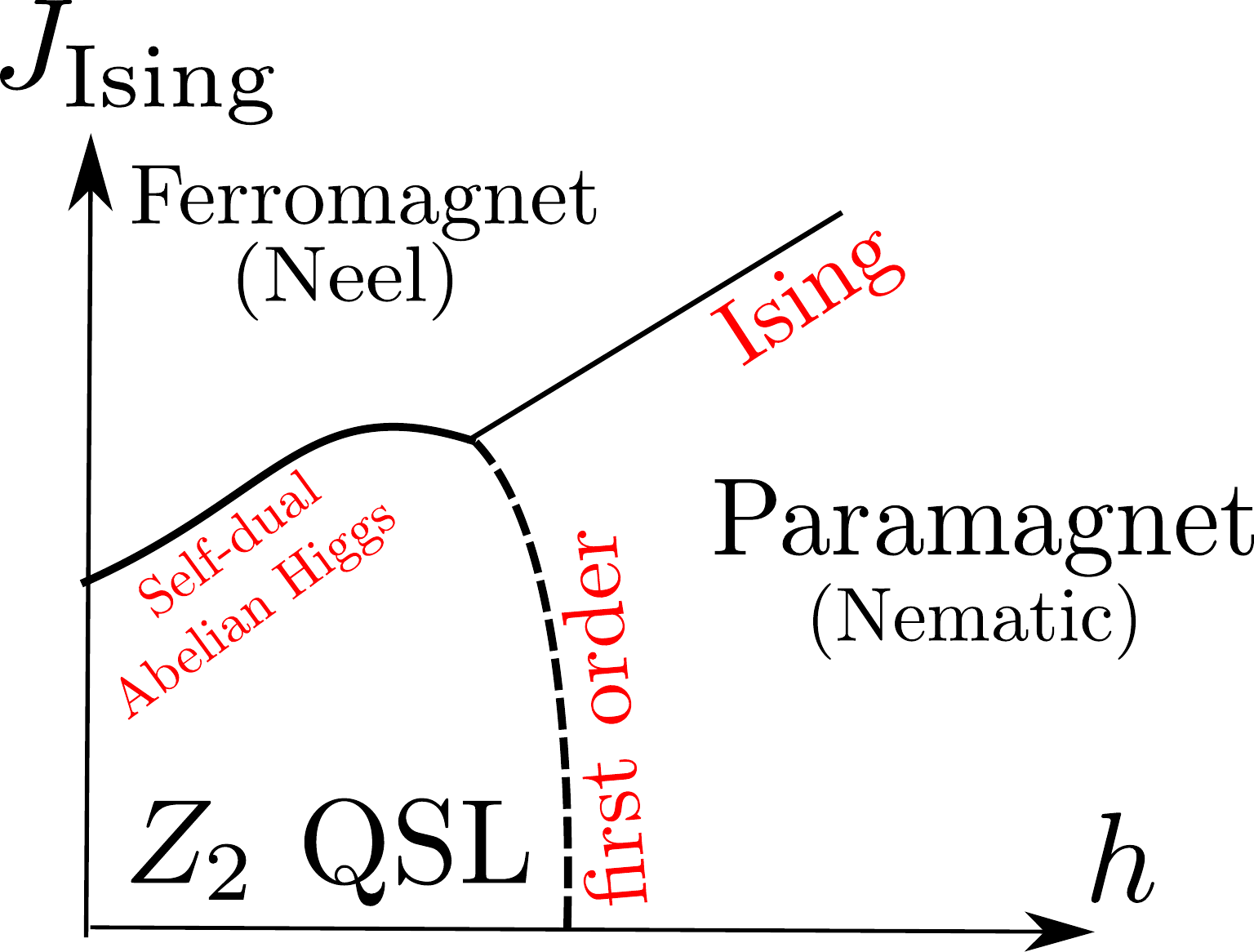}
\caption{A single or a two step transition from the $Z_2$ QSL as a function of Heisenberg coupling into a magnetically ordered state (Eq.~\ref{eq_heisenberg_limit_min_afm}). In the leading order of perturbation $h\sim 2J$ and $J_{\rm Ising}=J$ where $J$ is the strength of the Heisenberg perturbation as defined in Eq.~\ref{eq_hamiltonian}. The transition from the $Z_2$ QSL to a paramagnet in the parallel field is a first order transition\cite{Dusuel_PRL_2011} where the transition at $J_{\text{Ising}}=0$ is a self dual point. We expect this first order transition to be stable to Ising perturbation since our numerical results do not show any significant change of behavior (see  Appendix~\ref{sec_jk_ham}).}
\label{fig:SsTs}
\end{center}
\end{figure}

While in Eq.~\ref{eq_heisenberg_limit_min_afm} the couplings of the transverse field and the Ising term both are proportional to $J$, on considering higher order contributions of the perturbation theory (see Eq.~\ref{eq_afm_full_hamiltonian} and Eq.~\ref{eq_single_afm}-\ref{eq_afm_quadrapole_full}) they are differently renormalised and it is therefore useful to consider them at independent parameters and study the generalised phase diagram where  the strength of the Ising term ($\equiv J_{\rm Ising}$) and the magnetic field term ($\equiv h$) is independently varied (see Fig.~\ref{fig:SsTs}). In this generalised model for $h\rightarrow \infty$ limit we obtain the two above polarised phases for the $\tau$ spins that  correspond to a direct product state of of singlets and triplets on the $z$ bonds for the $\sigma$-spins. 

Detailed discussion regarding this model is relegated to Appendix~\ref{sec_jk_ham}, where it is shown that under unitary transformations this system is equivalent to a problem of perturbing a toric code Hamiltonian with a transverse field and a $x$-$z$ Ising term (see Eq.~\ref{eq_jk_general_pm_t1t2}). For this model, our numerical studies show three prominent phases (qualitatively shown in Fig.~\ref{fig:SsTs}) -- (i) ferromagnet (FM), (ii) paramagnet (PM) and (iii) $Z_2$ QSL. While the FM and PM are separated by an $3D$-Ising transition; the $Z_2$ QSL and the paramagnet are separated by a first order line \cite{Vidal_PRB_2009, Dusuel_PRL_2011}. The nature of transition between $Z_2$ QSL and the FM is self-dual modified Abelian Higgs transition as is discussed below.

Therefore, in the present case, in principle there can be two possible ways of destroying the $Z_2$ QSL leading to a spin-ordered phase (in the Heisenberg limit) via tuning the Heisenberg interactions-- (1) a direct second order quantum phase transition into the spin ordered phase, and (2) a two step transition where the the QSL first goes into a polarised trivial paramagnet through a first order transition and finally into the spin-ordered state via a $3D$-Ising transition.   For the purely transverse field Ising model on a square lattice (Eq.~\ref{eq_heisenberg_limit_min_afm}), existing variational and cluster Monte-Carlo calculations~\cite{Blote_PRE_2002, Albuquerque_PRB_2010,   Blass_srep_2dTFIM_2016, Huang_PRB_2020} shows the strength of the transverse field $\approx 3J$ is the critical point for the phase transition between the symmetry broken $\tau^z$ magnetically ordered state, i.e.  $\langle\tau^z_i\rangle\neq 0$ and the paramagnet state, i.e. $\langle\tau^z_i\rangle = 0$. So for Eq.~\ref{eq_heisenberg_limit_min_afm}, we expect a single step transition which is supported by our exact diagonalisation results on finite spin clusters presented in Appendix~\ref{sec_jk_ham}.

\subsection{The pseudo-dipolar limit \texorpdfstring{$J=K=0$}{}}
\label{subsec_gamma_limit_afm}

A novel and the most interesting limit of the anisotropic antiferromagnetic model is obtained when the pseudo-dipolar interactions dominate. The leading order effect of such perturbation in the $|K_z|\rightarrow\infty$ limit is given by the second order perturbation theory leading to the effective Hamiltonian (the full Hamiltonian up to fourth order perturbation is given in Eq.~\ref{eq_gammalimit}) given by Eq.~\ref{eq_gam_limit_afm1} 

Unlike the Heisenberg perturbations (Eq.~\ref{eq_heisenberg_limit_min_afm}) or the $\Gamma$ perturbations in the ferromagnetic Kitaev case~\cite{nanda2020phases}, the above term does not get contributions at the first order level. This allows for a non-trivial spin-interactions through the three spin terms. Notably, due to the unusual implementation of time reversal symmetry (see table \ref{table_tau_symm}) the above three spin terms are symmetry allowed.

\subsubsection{Stacked cluster chains}

\begin{figure}
\centering
\includegraphics[]{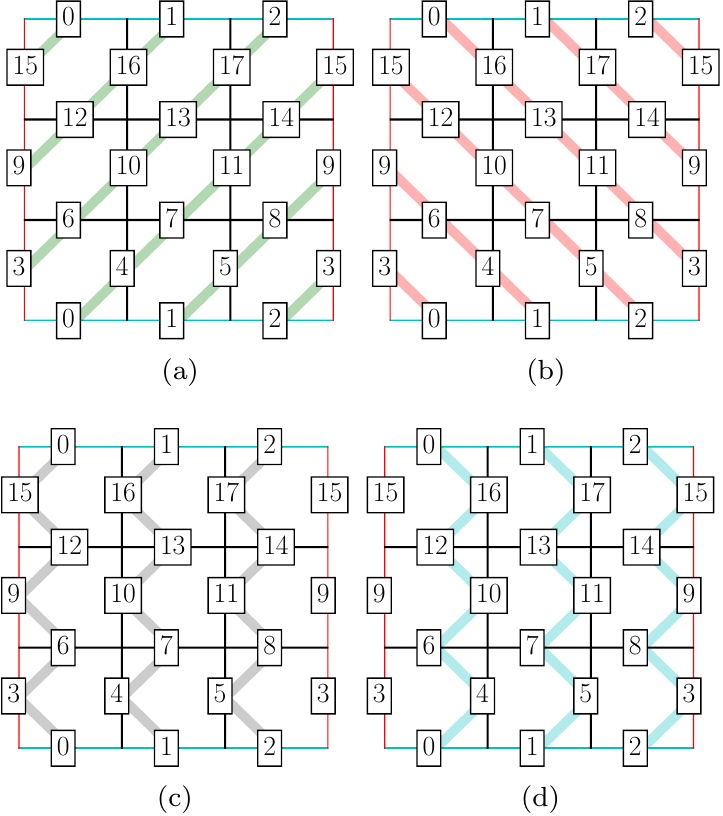}
\caption{Four stacked cluster spin chains : The panels {\bf (a)}, {\bf (b)}, {\bf (c)}, {\bf (d)} represent the four Hamiltonians in Eq.~\ref{eq_four_spt} in PBC.}
\label{fig_gam_spt_chains}
\end{figure}

We now discuss the rich structure of the Hamiltonian in Eq.~\ref{eq_gam_limit_afm1}. To this end we re-write it as
\begin{equation}
\mathcal{H}^{AF}_{J=K=0} = \frac{\Gamma^2}{|K_z|}\left(H_{1} + H_{2} + H_{3} + H_{4}\right)
\label{eq_superposegamma}
\end{equation}
with
\begin{equation}
\begin{aligned}
    H_1 &= \sum_{i} \tau^z_{i+d_1}\tau^x_{i}\tau^z_{i-d_1}  \\
    H_2 &= \sum_{i}  \tau^z_{i+d_2}\tau^x_{i}\tau^z_{i-d_2}  \\
    H_3 &= \sum_{i \in V}\left( \tau^z_{i+d_1}\tau^y_{i}\tau^z_{i-d_2}\right)-\sum_{i \in H}\left(\tau^z_{i+d_2}\tau^y_{i}\tau^z_{i-d_1}\right) \\
    H_4 &= \sum_{i \in H}\left( \tau^z_{i+d_1}\tau^y_{i}\tau^z_{i-d_2}\right)-\sum_{i \in V}\left(\tau^z_{i+d_2}\tau^y_{i}\tau^z_{i-d_1}\right)
\end{aligned}\label{eq_four_spt}
\end{equation}
where $H,V$ denote the set of sites belonging to the horizontal and vertical bonds respectively. 

We immediately note that each of these Hamiltonians represent a set of stacked one-dimensional cluster spin-1/2 chains arranged in a particular direction. This is shown in Fig.~\ref{fig_gam_spt_chains}. While $H_1$ and $H_2$ are stacked cluster chains oriented at $\frac{\pi}{4}$ and $\frac{-\pi}{4}$ in the lattice plane, $H_3$
 and $H_4$ are oriented vertically with the chains being displaced by a lattice constant with respect to each other.

If the Hamiltonians are considered independently, as discussed in Appendix \ref{appen_cluster_spt}, at this leading order each decoupled chain has an enhanced $Z_2 \times Z_2$ symmetry and stabilises a gapped symmetry protected topological (SPT) phase protected by this symmetry~\cite{you2018subsystem,chen2014symmetry,son2012topological,nielsen2006cluster,dubinkin2019higher}. As a result each chain supports a zero energy localised spin-1/2 at the edge of each chain. Each stacking pattern of these cluster Hamiltonians in Eq.~\ref{eq_four_spt}, $H_\alpha~(\alpha=1,2,3,4)$  therefore result in a weak-SPT phase~\cite{you2018subsystem} whose edge mode structure depends on the shape of the cluster chosen, as expected (see Table. \ref{table_gam_gsd_cluster} and Appendix \ref{sec:Wtrans}).

\begin{table}
\begin{center}
 \begin{tabular}{|c| c | c | c | c|} 
 \hline
 Hamiltonian & PBC & $x$-CBC & $y$-CBC & OBC  \\ [0.5ex] \hline \hline
$H_1$ & 1 & $2^{2L_x}$ & $2^{2L_y}$ & $2^{2(L_x + L_y)-2}$   \\  [0.5ex] \hline
$H_2$ & 1 & $2^{2L_x}$ & $2^{2L_y}$  &  $2^{2(L_x + L_y)-2}$  \\  [0.5ex] \hline
$H_3$ & 1 & $2^{2L_x}$ & 1 & $2^{2L_x}$ \\  [0.5ex] \hline
$H_4$ & 1 & $2^{2L_x}$ & 1 & $2^{2(L_x+L_y)-2}$  \\  [0.5ex] \hline
\end{tabular}
\caption{Ground state degeneracies for various stacked cluster Hamiltonians $H_1, H_2, H_3, H_4$ (see Eq.~\ref{eq_four_spt}) when placed under various boundary conditions. PBC (OBC) is the usual periodic (open) boundary condition on a torus, while $x$-CBC ($y$-CBC) is cylindrical boundary condition with $x(y)$ direction being periodic. The details are discussed in Appendix.~\ref{sec:Wtrans}.}
\label{table_gam_gsd_cluster}
\end{center}
\end{table}

\begin{figure}
    \centering
    \includegraphics[width=0.7\linewidth]{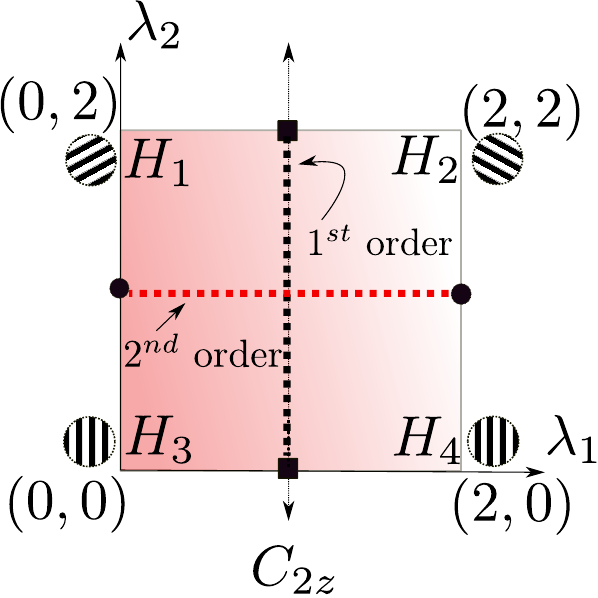}
    \caption{Anticipated phase diagram for the Hamiltonian in Eq.~\ref{eq_four_spt_interpolation}. The four corners are exact limit of the stacked cluster SPTs (given in Eq.~\ref{eq_four_spt}) which are oriented in different directions and shown in Fig.~\ref{fig_gam_spt_chains}. $C_{2z}$ symmetry transforms $H_1 \leftrightarrow H_2$ {\rm and}  $H_3 \leftrightarrow H_4$. The phase transitions at $\lambda_2=0$ ($\lambda_1=0$) as a function of $\lambda_1$ ($\lambda_2$) is a first(second) order transition which is expected to be stable when $\lambda_2 \neq 0$ ($\lambda_1 \neq 0$) (see text).}
    \label{fig_anticipatd_lamda_pd}
\end{figure}

The full Hamiltonian in the pseudo-dipolar limit (Eq.~\ref{eq_superposegamma}), however, is a equal weight superposition of the the four stackings. In order to understand this, it is useful to consider the interpolating Hamiltonian 
\begin{align}\label{eq_four_spt_interpolation}
H(\lambda_1,\lambda_2) & = \lambda_2\left((2-\lambda_1)H_1 + \lambda_1 H_2\right) \\ \nonumber
        & + (2-\lambda_2)\left((2-\lambda_1)H_3 + \lambda_1H_4\right)
\end{align}
parameterized by $\lambda_1$ and $\lambda_2$ -- such that in the  $(\lambda_1, \lambda_2)$ plane, the points $(0,2)$, $(2,2)$, $(0,0)$ and $(2,0)$ are identified with $H_1, H_2, H_3$ and $H_4$ respectively while, up to multiplicative factors, $\mathcal{H}^{AF}_{(J=K=0)}$ is given by $(1,1)$. This is illustrated in Fig.~\ref{fig_anticipatd_lamda_pd} and explained below. However, we note that on this plane the symmetry of $\pi$-rotation about the $z$-bond, $C_{2z}$ results in $H_1 \leftrightarrow H_2$ and $H_3 \leftrightarrow H_4$ and thus constraining $\lambda_1=1$ on the plane while $\lambda_2$ being free to be renormalised by higher order terms. We shall specially focus on this line while discussing the phase diagram.

The interpolating Hamiltonian of Eq.~\ref{eq_four_spt_interpolation} in the entire $(\lambda_1,\lambda_2)$ has some special symmetry and energetic features. While these properties are not stable to higher order perturbations (see Eq.~\ref{eq_gammalimit} in Appendix \ref{appen_perturb}), not only such structures are interesting in their own rights as we shall see below, but also, these {\it weakly broken} symmetries provide important insights into the nature of the phase in this pure $\Gamma$ limit. Hence, we now discuss these special symmetries.

The generic non-Kramers time-reversal symmetry is generated by (see table \ref{table_tau_symm}) the operator $\mathcal{T}=\prod_{i\in H,V}\tau^x_i \mathcal{K}$ (where $\mathcal{K}$ is the complex conjugation operator). However Eq.~\ref{eq_four_spt_interpolation} enjoys an enhanced {\it sub-lattice} time-reversal symmetry generated by the operators 
\begin{align}
    \mathcal{T}_H=\prod_{i\in H}\tau^x_i\mathcal{K}_i,\quad \mathcal{T}_V=\prod_{i\in V}\tau^x_i\mathcal{K}_i
\end{align}
where the products in the first and second expressions run over the horizontal and vertical bonds respectively. Thus this plane enjoys a global $Z_2\times Z_2$ symmetry. 

The Hamiltonian in Eq.~\ref{eq_four_spt_interpolation}, however has an even larger set of {\it sub-system symmetries} which is most apparent after a unitary rotation defined on a set of bonds, followed by a global unitary rotation. The following transformation 
\cite{plenio2007remarks,you2018subsystem,Kalis_PRA_2012, kramers1941statistics, son2011quantum}
\begin{align}
{\cal W} \equiv \prod_i U_{i,i+d_1}
\label{Wtrans}
\end{align} 
where we define a bond-dependent (direction independent) unitary operator \begin{align}
U_{ij} = \frac{1}{2} \Big(1 + \tau^z_i + \tau^z_j - \tau^z_i \tau^z_j \Big) 
\label{Uijtrans}
\end{align}
renders
\begin{align}
\mathcal{W} :~
\tau^x_i \rightarrow \tau^z_{i-d_1}\tau^x_{i}\tau^z_{i+d_1}~;~~~ \tau^z_i \rightarrow \tau^z_i .
\label{eq_kw}
\end{align}
This when followed by a global rotation
\begin{align}\label{eq_rot_local}
    \mathcal{V} :~\{\tau^x_i,\tau^y_i,\tau^z_i\}\rightarrow\{\eta^y_i,\eta^z_i,\eta^x_i\}
\end{align}
leads to
\begin{align}\label{eq_unitary_spt1_pm}
    H_\alpha\rightarrow\tilde{H}_\alpha=\left(\mathcal{V}\mathcal{W}\right)H_\alpha \left(\mathcal{V}\mathcal{W}\right)^{-1}
\end{align}
where $\eta^\alpha_i$ are the new spin degrees of freedom. Note that while the transformation $\mathcal{V}$ is not essential, as we shall see below, it simplifies parts of our analysis. 

The resultant transformed Hamiltonians are given by
\begin{align} 
    H_1 \rightarrow \tilde{H}_1 &= \sum_{i}\eta^y_i \label{eq_4_spt_rot_1} \\ 
    H_2 \rightarrow \tilde{H}_2&=\sum_i\eta^y_i\eta^x_{i+d_1}\eta^x_{i-d_1}\eta^x_{i+d_2}\eta^x_{i-d_2} \label{eq_4_spt_rot_2} \\
    H_3 \rightarrow \tilde{H}_3&=\sum_{i \in V} \eta^x_{i-d_1}\eta^z_{i}\eta^x_{i-d_2}-\sum_{i \in H} \eta^x_{i+d_2}\eta^z_{i}\eta^x_{i+d_1} \label{eq_4_spt_rot_3}  \\
    H_4 \rightarrow \tilde{H}_4&=\sum_{i \in H} \eta^x_{i-d_1}\eta^z_{i}\eta^x_{i-d_2}-\sum_{i \in V}\eta^x_{i+d_2}\eta^z_{i}\eta^x_{i+d_1} \label{eq_4_spt_rot_4}
\end{align}
Therefore under this particular transformation, the four differently stacked weak cluster SPTs respectively get mapped to a $y$-paramagnet (PM) ($\tilde{H}_1$), strong sub-system  symmetry protected topological phase (SSPT) ($\tilde{H}_2$) of the topological plaquette Ising model~\cite{PhysRevLett.86.5188,you2018subsystem}, and two horizontally stacked weak cluster SPTs ($\tilde{H}_3$ and $\tilde{H}_4$). We have explicitly checked that the transformation when defined for an open system restores the correct number of zero modes in both $H_\alpha$ and $\tilde{H_\alpha}$. A discussion about the transformation ${\cal W}$ (Eq.~\ref{Wtrans}) and the way it acts on the boundary Hamiltonians in an open system see Appendix~\ref{sec:Wtrans}. The cluster SPT is briefly discussed in Appendix \ref{appen_cluster_spt}.

\begin{figure}
\centering
\includegraphics[]{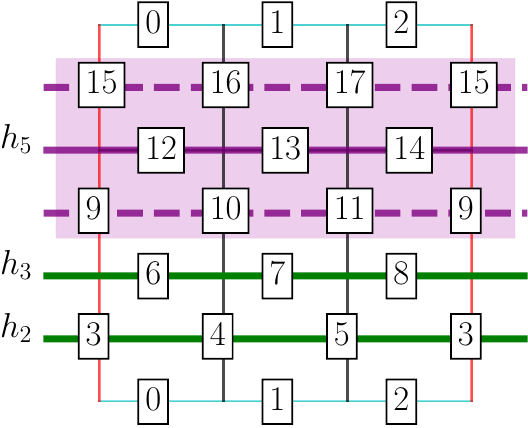}
\caption{The anti-unitary symmetry operator in Eq.~\ref{eq_sspt_hor_symm} is shown as a single green continuous line ($h_n$), for example $h_3$ passes through the horizontal bonds (spins 6, 7, 8) and $h_2$ passes through the vertical bonds (spins 3, 4, 5). The symmetry operator in the Eq.~\ref{eq_spt_hor_symm_original} is shown using the three magenta lines. Dashed (continuous) line shows the unitary (anti-unitary) operation.}
\label{fig_gam_spt_chains_trans}
\end{figure}

In the transformed basis, the Hamiltonian (Eq.~\ref{eq_four_spt_interpolation}) is invariant under the following set of {\it anti-unitary subsystem symmetries} that are generated by
\begin{equation}\label{eq_sspt_hor_symm}
    \tilde{PT}_{h_n} = \prod_{i\in n^{th}\text{ Hor. line}} \eta^z_i \mathcal{K}_{i}
\end{equation}
\begin{equation}\label{eq_sspt_ver_symm}
    \tilde{PT}_{v_n} = \prod_{i\in n^{th} \text{Vert. line}}  \eta^x_{i-d_1}\mathcal{K}_{i-d_1} \eta^z_i  \mathcal{K}_i  \eta^x_{i+d_1} \mathcal{K}_{i+d_1}
\end{equation}
where in Eq.~\ref{eq_sspt_hor_symm} (\ref{eq_sspt_ver_symm}), $h_n (v_n)$ denotes the n$^{th}$ horizontal (vertical) line which either can pass through the horizontal (vertical) bonds or cut through the vertical (horizontal) bonds of the square lattice (see Fig.~\ref{fig_gam_spt_chains_trans}) and $\mathcal{K}_i$ is the local complex conjugation operation which acts on site $i$. 

In terms of the untransformed basis (by Eq.~\ref{eq_unitary_spt1_pm}), Eq.~\ref{eq_sspt_hor_symm} and \ref{eq_sspt_ver_symm} can be obtained from Eqs. \ref{eq_kw} and \ref{eq_rot_local} and are given respectively by 
\begin{equation}\label{eq_spt_hor_symm_original}
    PT_{h_n} = \prod_{i\in n^{th} \text{Hor. line}} \tau^z_{i+d_1}\tau^x_i\tau^z_{i-d_1} \mathcal{K}_i
\end{equation}
\begin{equation}\label{eq_spt_ver_symm_original}
    PT_{v_n} = \prod_{i\in n^{th} \text{Vert. line}} \tau^z_{i+d_1}\mathcal{K}_{i+d_1}\tau^x_i\mathcal{K}_i\tau^z_{i-d_1}\mathcal{K}_{i-d_1}
\end{equation}
Note that both Eq.~\ref{eq_spt_hor_symm_original} and \ref{eq_spt_ver_symm_original} involves the same  transformation on horizontal/vertical stacks of three consecutive spins separated by, $T_{d_1}$, {\it i.e.}  translation along ${\bf d}_1$. However, while for the horizontal stacking in Eq.~\ref{eq_spt_hor_symm_original}, the conjugation operator acts only on the spin in the middle, for the vertical stacking in Eq.~\ref{eq_spt_ver_symm_original} they act on all the spins involved.

In addition, along the $\lambda_2=2$ line the system has another set of subsystem symmetries generated by : 
\begin{equation}
    {\tilde{PT}}'_{v_n} = \prod_{i\in n^{th} \text{Vert. line}} \eta^z_i \mathcal{K}_i
    \label{eq_sspt_ver_symm_2}
\end{equation}
Similar to the Eq.~\ref{eq_spt_hor_symm_original}, we can write this symmetry in the original spin basis of Eq.~\ref{eq_four_spt} as a combination of unitary and anti-unitary symmetry, now in the vertical direction which is
\begin{equation}\label{eq_spt_ver_symm_original2}
    {PT'}_{v_n} = \prod_{i\in n^{th} \text{Ver. line}} \tau^z_{i+d_1}\tau^x_i\tau^z_{i-d_1} \mathcal{K}_i
\end{equation}

We shall later return to the constraints imposed by these sub-system symmetries. However, as briefly discussed in Appendix \ref{sec:excitations}, due to the particular non-Kramers nature of the time reversal symmetry the above subsystem symmetries do not constrain the dispersion of excitations unlike fractons \cite{Pretko_IJMPA_2020,Nandkishore_ARCMP_2019}.

\begin{figure}
    \centering
    \includegraphics[]{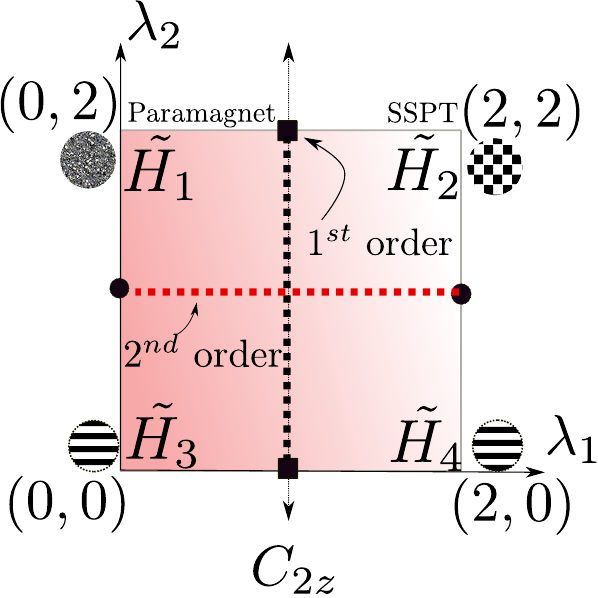}
    \caption{Phase diagram corresponding to the Hamiltonian given in Eq.~\ref{eq_four_spt_interpolation} after the transformation defined in Eq.~\ref{eq_unitary_spt1_pm}, also see Fig.~\ref{fig_anticipatd_lamda_pd}.}
    \label{fig_anticipatd_transform_lamda_pd}
\end{figure}

The above transformation (Eq.~\ref{eq_unitary_spt1_pm}) allows for new insights into the phase diagram of the pure $\Gamma$ Hamiltonian given by Eq.~\ref{eq_superposegamma}. In particular the transitions along the four boundaries, as shown in Fig.~\ref{fig_anticipatd_transform_lamda_pd}, can be immediately read off from from existing literature. These are as follows :

\begin{itemize}
    \item The transition between $\tilde{H}_1$ and $\tilde{H}_2$ along the $\lambda_2=2$ line is between a trivial paramagnet and a two dimensional SSPT respectively. This transition is known to be first order~\cite{Kalis_PRA_2012, Orus_PRA_2013} and occurs at $\lambda_1=1$. In the un-transformed basis, we note that this represents a transition between two stacked cluster models, $H_1$ and $H_2$. Remarkably, the effective dimensional reduction at the critical point is far from apparent in this un-transformed basis.
    
    There exists a transformation similar to Eq.~\ref{eq_unitary_spt1_pm} which transforms, on the  $\lambda_2=0$ line, $H_3$ to a trivial transverse field paramagnet and $H_4$ to an SSPT. The discussion of the above paragraph then can be immediately applied to the $\lambda_2=0$ line. (Notably, such a transformation map $H_1$ and $H_2$ to weak cluster SPTs.)
    
    Therefore at $\lambda_1=1$, both $\lambda_2=0,2$ are first order transition points. This implies that the phase diagram in $(\lambda_1, \lambda_2)$ phase has a reflection symmetry about $\lambda_2=1$ line.

    \item The transition from $\tilde{H}_1$ to $\tilde{H}_3$ along the  $\lambda_1=0$ line is between a trivial paramagnet and decoupled one dimensional cluster chains. This is a self dual transition at $\lambda_2=1$ that is described by a $SO(2)_1$ conformal field theory (CFT) with central charge, $c=1$~\cite{PhysRevLett.115.237203,verresen2017one}. Given the  the existence of sub-system symmetry operators it may seem that dynamics of the excitations from the $\tilde{H_3}$ state is constrained. As is discussed in Appendix \ref{sec:excitations} we show that the antiunitary character of these subsystem symmetries effectively renders the dynamics to be free especially on the $\lambda_1=0$ line. Again, as above, in the un-transformed basis,  the above transition is between two stacked cluster models, $H_1$ and $H_3$, again, with non-obvious effective dimensional reduction at the critical point.
    
    A yet third set of transformations similar to Eq.~\ref{eq_unitary_spt1_pm} transforms $H_2$ to a transverse paramagnet and $H_4$ to a stacked cluster SPT. This immediately allows us to import the above physics of $\lambda_1=0$ and apply it to the case of $\lambda_1=2$ line. Further the rotation about the $z$-bond ($C_{2z}$ symmetry, see table \ref{table_tau_symm}) leads to $(\lambda_1,\lambda_2)\rightarrow (2-\lambda_1,\lambda_2)$ which also leads to the same conclusion regarding the phases and phase transitions.
\end{itemize}

\begin{figure}
\includegraphics[]{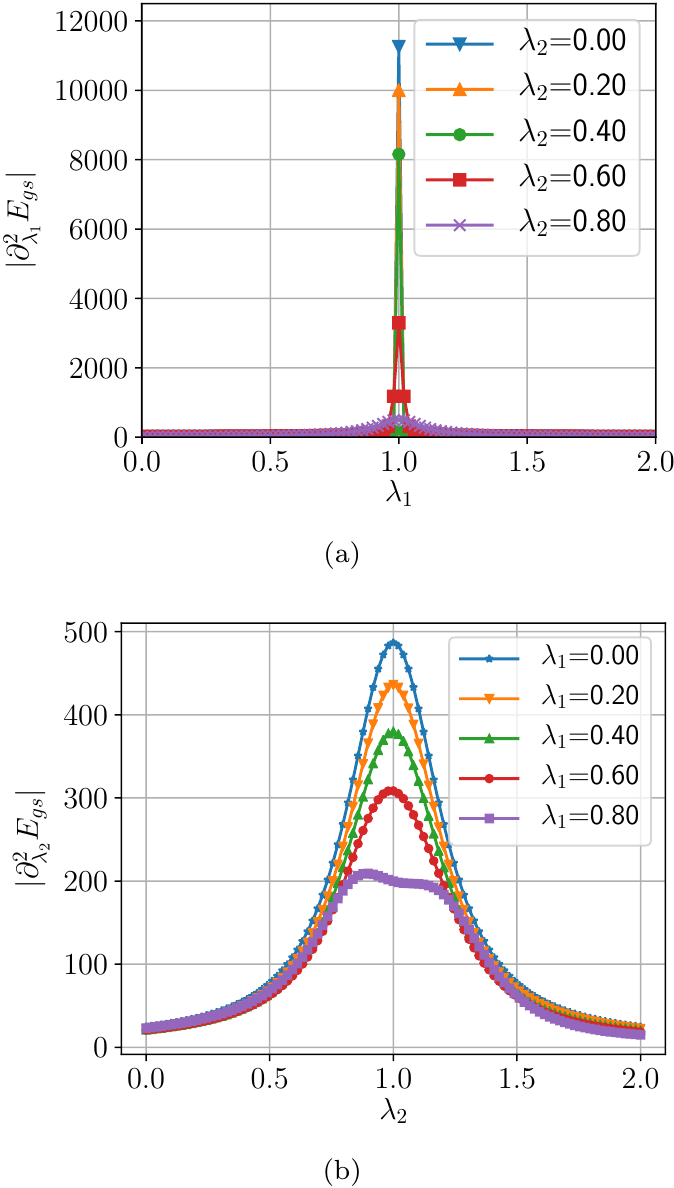}
\caption{(a) Ground state (GS) susceptibility as the absolute value of second derivative of the GS energy ($E_{gs}$) with respect to $\lambda_1$ for constant $\lambda_2$ values for the Hamiltonian given in Eq.~\ref{eq_four_spt_interpolation} (b) Ground state susceptibility along $\lambda_2$ for constant $\lambda_1$ values. Both results are for a 18 spin cluster with $L_x =3$ and $L_y=3$ and PBC geometry.}
\label{fig_sus_lam_scan}
\end{figure}

The entire $(\lambda_1, \lambda_2)$ plane respects the sub-system symmetries protecting the above SPTs and the associated phase transitions. Hence we expect that the continuous (discontinuous) transitions at $\lambda_1=0,2$ ($\lambda_2=0,2$) are perturbatively stable away from these lines. To investigate this we perform numerical Exact Diagonalisation (ED) on small spin clusters for various system sizes of up to 32 spins using QuSpin~\cite{weinberg2017quspin,weinberg2019quspin}. In addition to the bulk excitation gap, we calculate the ground state fidelity susceptibility whose peaks locate the bulk gap closing phase transitions in the $(\lambda_1,\lambda_2)$ plane.

In the Fig.~\ref{fig_sus_lam_scan}(a) we show the ground state (GS) susceptibility \cite{yu2009fidelity} discontinuous peak along the $\lambda_1$ for various constant $\lambda_2$ values between 0 and 0.8. We can clearly see for $\lambda_2=0$ line the sudden jump in the susceptibility at $\lambda_1=1$ is indicating a first order phase transition \cite{Kalis_PRA_2012,Orus_PRA_2013} expected  between the SSPT and the trivial paramagnet. On departing from the $\lambda_2=0$ line, the weight of the discontinuous peak monotonically comes down as we approach $\lambda_2=1$ indicating that the discontinuous nature of the transition weakens as we approach $\lambda_2=1$ and disappears at this point. However our present calculations cannot discern if the discontinuity persists all the way to $\lambda_2=1$. Similar physics is observed coming down from the $\lambda_2=2$ line (not shown). It is pertinent to point out that given the limited system sizes accessible in ED, there are significant even-odd (commensurability) effects in all regions of the phase diagram. This therefore makes the role of symmetries and various transformations, even more crucial to understand the nature of the phases. 

The above first order transition is in stark contrast with the transition obtained by tuning $\lambda_2$ as shown in Fig.~\ref{fig_sus_lam_scan}(b). Here the susceptibility shows a peak without a shoulder ({\it i.e.}, a sudden jump) indicative of a continuous transition at $\lambda_2=1$. Indeed for $\lambda_1=0$, this  transition originates from a stack of cluster chains and is described by decoupled $(1+1)$ dimensional critical point of $SO_1(2)$ CFT \cite{PhysRevLett.115.237203,verresen2017one} with a description in terms of Majorana fermions (see Appendix \ref{appen_cluster_spt}). The continuous nature of the transition persists for larger values of $\lambda_1$ until close to the $\lambda_1=1$ whence the peak bifurcates indicating the possibility of opening up of an intermediate phase in the vicinity of $\lambda_1=1$. However our present numerical calculations are limited by system size to probe this aspect. However, as we discuss below, we expect that this intermediate phase, even if it exists, to be very fragile due to the large number of special symmetries (see the discussion above) in the $(\lambda_1, \lambda_2)$ plane. Again we find a similar picture on the $\lambda_1>1$ region due to the $\pi$-rotation about the z-bond symmetry $C_{2z}$, (see table \ref{table_tau_symm}).

\begin{figure}
\includegraphics[]{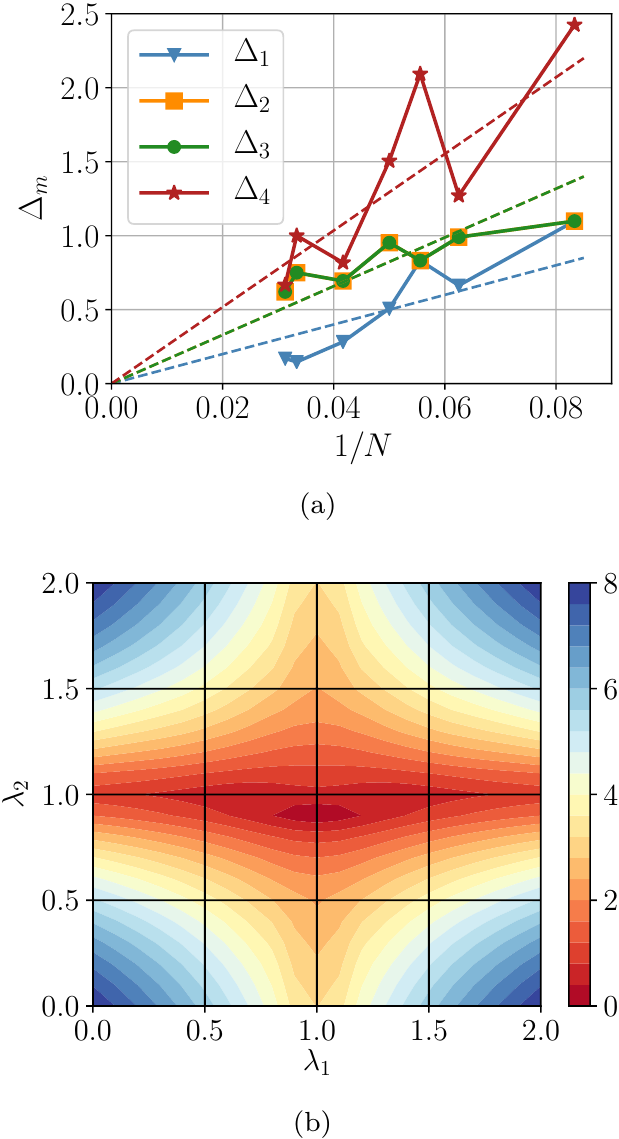}
\caption{(a) Scaling of energy gaps ($\Delta_m$) to  $m^{th}$ excited state as a function of inverse system size ($N = 2(L_x \times L_y)$) at pure $\Gamma$ limit, i.e. $(\lambda_1,\lambda_2)=(1,1)$ (see Eq.~\ref{eq_four_spt_interpolation}) for systems sizes involving $N=12$ to $N=32$ spins. The dashed lines are guide to eye. (b) The gap to the first excited state in the complete $(\lambda_1,\lambda_2)$ plane for a $4 \times 2$ cluster. Both results are for a PBC geometry.}
\label{fig_gap_lamda_pd}
\end{figure}

Right at the point $\lambda_1=\lambda_2=1$, our present ED calculations reveal a bulk gapless phase. This is shown in Fig. Fig.~\ref{fig_gap_lamda_pd}(a) where we plot the bulk gap to the four lowest excitations as a function of the few system sizes to indicate that the gap to these excitations vanish almost linearly in inverse system size. The contour plot of the bulk gap to the first excited state in the entire $(\lambda_1, \lambda_2)$ plane is shown in Fig.~\ref{fig_gap_lamda_pd}(b). This shows that gap indeed closes along the $\lambda_1=1$ and $\lambda_2=1$ lines with the former leading to a first order transition and the later leading to second order transition. This separates the plane into four phases as shown in Fig.~\ref{fig_anticipatd_lamda_pd} and \ref{fig_anticipatd_transform_lamda_pd}. 

As indicated above, the first order transitions weaken near the $\lambda_1=\lambda_2=1$ and possibly leading to a bulk gapless phase right at that point. Remarkably, our ED calculations on systems with open boundary conditions show that at this point, in addition to the gapless bulk modes the system has an additional $2^{2L_x}$ exact zero energy Ising boundary modes on the top and bottom boundary  which do not hybridize with the bulk gapless modes due to special subsystem symmetries (Eq.~\ref{eq_spt_ver_symm_original}). A detailed discussion on the anomalous character of these symmetry operations in the $(\lambda_1,\lambda_2)$ plane is discussed in Appendix~\ref{subsubsec_disc_bound_gamma}. Such gapless phases with boundary modes have recently being discussed in context of symmetry enriched criticality in one dimension  \cite{Scaffifi_PRX_2017, Verresen_PRL_2018}
 and more recently for related two dimensional phases\cite{Thorngren_arXiv_2020}, however to best of our knowledge none of the these phases lie in the interjection of such weak SPTs as here.

We now turn to the important question regarding the nature of the possible gapless phase at $\lambda_1=\lambda_2=1$ with extra sub-system symmetry-protected zero energy boundary modes. At the outset such a gapless phase is rather remarkable in a system with no continuous symmetries and hence would be rather novel if found to be stable. As noted above, whether such a gapless phase is limited to the only the single point or extends over a finite region is not clear from our present ED calculations due to severe finite size effects, however as we shall discuss now, we think it is the former and this gapless point is rather fragile. 

The first clue to the fragility of this gapless point comes from the rather fine tuned nature of the Hamiltonian in Eq.~\ref{eq_four_spt_interpolation} which allows for a whole class of sub-system symmetries not present in the microscopic Hamiltonian and  are an  artifact of keeping just the second order terms in $\Gamma$. For example, on considering the higher order ($\mathcal{O}(\Gamma^3/|K_z|^3)$ in perturbation theory) term for the $\Gamma$-Hamiltonian (see Eq.~\ref{eq_gammalimit}), such sub-system symmetries are explicitly broken. However they serve as important approximate symmetries in discerning the general structure of the phase diagram in the $(\lambda_1,\lambda_2)$ plane-- particularly the gapped part of the phase diagram. However for the gapless part of the phase diagram the absence of these sub-system symmetries are rather subtle. Indeed the boundary modes are susceptible to symmetry breaking perturbations or to boundary interactions which can lead to spontaneous symmetry breaking at the boundary. A discussion of such symmetry breaking terms on the boundary Hamiltonian of the large $\Gamma$ phase is shown in section~\ref{subsubsec_disc_bound_gamma}.  

To check the stability of the gapless point at $\lambda_1=\lambda_2=1$, we added simple perturbations that explicitly break the sub-system symmetries, but are still allowed by the microscopic symmetries and studied the fate of such a Hamiltonian. In particular, we performed ED on 
\begin{equation}\label{eq_gamma_zz_x}
\begin{aligned}
    H(\delta_1,\delta_2)= &(1-\delta_1)(1-\delta_1)H(1,1) \\
    &-\delta_1(1-\delta_2)\sum_{i}\tau^x_i -\delta_2(1-\delta_1)\sum_{\langle ij\rangle}\tau^z_i\tau^z_j
\end{aligned}
\end{equation}

where $H(1,1)$ is the Hamiltonian which belongs to the general Hamiltonian given in Eq.~\ref{eq_four_spt_interpolation} with $(\lambda_1,\lambda_2)=(1,1)$, the second term represents a $x$-field and the third term is a nearest neighbour Ising exchange in the $z$ direction, both of which are allowed within the microscopic symmetries (see table \ref{table_tau_symm}).

\begin{figure}
\centering
\includegraphics[]{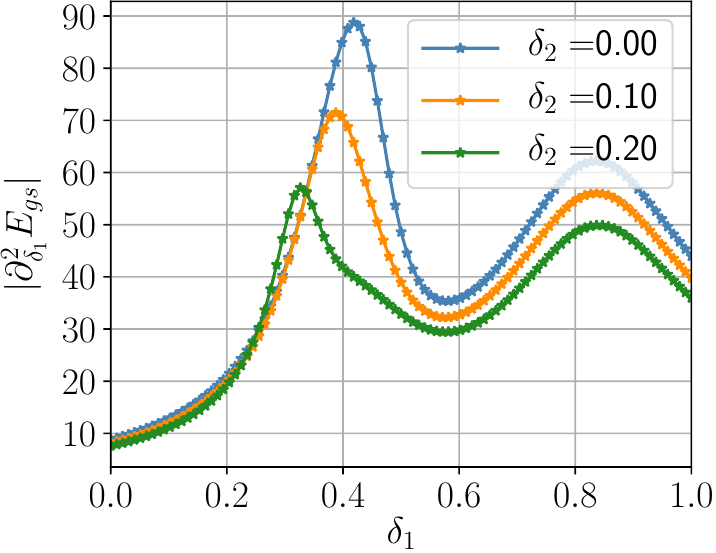}
\caption{Behavior of GS susceptibility as the large $\Gamma$ phase is tuned to a $x$ paramagnet in presence of varying strengths of Ising perturbation($\sim \delta_2$), see Eq.~\ref{eq_gamma_zz_x}. The results are for a
system size $N=16, L_x=2, L_y=4$ with PBC geometry.}
\label{fig_sus_gamma_zz_x}
\end{figure}

Fig.~\ref{fig_sus_gamma_zz_x} shows the ground state fidelity as a function of the two interpolating parameters, where we find that while a finite size system shows a GS susceptibility ($|\frac{\partial^2{E_{gs}}}{\partial{\delta_1}^2}|$) peak suggesting a phase transition -- the peak falls significantly with inclusion of an Ising coupling suggesting that the large $\Gamma$ phase is indeed smoothly connected to a $x$-paramagnet {\it without} any intervening phase transition within the symmetry allowed parameter space. A further insight into the nature of the phase is-- as more systematically discussed in the next section we also find that the phase has no topological entropy content and is short range entangled (see Fig.~\ref{fig_topEE}). Interestingly as the system is tuned to a paramagnet this topological entropy content continues to remain zero showing that the phase is smoothly connected to a trivial state. The behavior of the energy gaps $(\Delta_m)$ as well as the topological entanglement entropy (see Eq.~\ref{eq_ent_entropy}) are shown in Fig.~\ref{fig_gap_gamma_zz_x} in Appendix~\ref{sec_gamma_lim}.

Together the above signatures of the $\Gamma$ phase, we conclude that the $\lambda_1=\lambda_2=1$ is a fine tuned point which even while it is itself gapless, gets gapped out immediately by generic microscopic symmetry allowed perturbations and the resultant gapped phase is continuously connected to a trivial paramagnet. This insight as will discuss later will guide both the nature of the phase and their nature of transitions in the complete $KJ\Gamma$ parameter space. Before going into the field theoretic discussion, we numerically study the complete $KJ\Gamma$ parameter space within exact diagonalization studies.

\section{Phase diagram : Exact diagonalisations}\label{sec_numerics_afm}

Having discussed the phases in the different limits, we now study the phase boundaries via ED on finite spin cluster. For this we use the interpolating Hamiltonian : 
\begin{align}\label{eq_def_full_ham_num}
\mathcal{H}(t_1,t_2) &= (1-t_1)(1-t_2)\mathcal{H}^{\prime}_{(J=\Gamma=0)} \\ \nonumber
& + t_2(1-t_1)\mathcal{H}^{\prime}_{(\Gamma=K=0)} 
+ t_1(1-t_2)\mathcal{H}^{\prime}_{(J=K=0)}
\end{align}
Where $\mathcal{H}^{\prime}_{X}$ is defined as $\mathcal{H}_{X}$ with a unit energy scale. The explicit forms of the Hamiltonians($\mathcal{H}_{X}$) are given in Eqs.~\ref{eq_heisenberg_limit_min_afm}-\ref{eq_tc_rot_unrot}. The rescaled parameters are:
\begin{equation}\label{eq_def_t1_t2_afm}
 t_1=\frac{\Gamma^2/|K_z|}{J_{TC}+\Gamma^2/|K_z|}~;~t_2=\frac{J}{J_{TC}+|J|}
\end{equation}

In this parameter space, at the points $(t_1,t_2)=(0,0),(0,1),(1,0)$ the $\mathcal{H}(t_1,t_2)$ are Toric code, the Heisenberg and the pseudo-dipolar limit respectively. We perform ED for system sizes of up to 32 spins with periodic boundary conditions (PBC). We calculate the following quantities to estimate the phase boundaries as well as the nature of the phases-- (1) {\it Ground state fidelity susceptibility}, (2) {\it Spectral gaps}, (3) {\it Topological entanglement entropy}, (4) {\it Plaquette expectation}, (5) {\it Magnetization}, and, (6) {\it Spin-spin correlation}.

\begin{figure}
\centering
\includegraphics[]{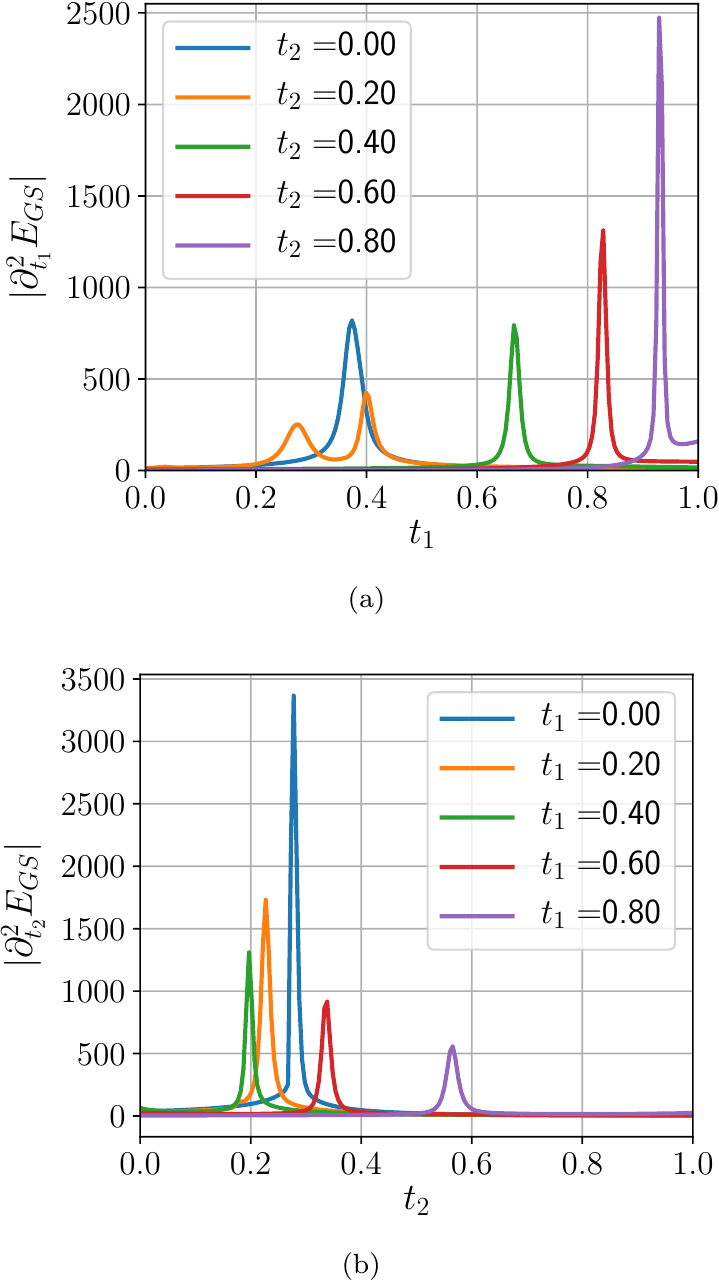}
\caption{Behavior of ground state susceptibility for Eq.~\ref{eq_def_full_ham_num}. (a) $\chi_1$  along the $t_1$ direction for constant values of $t_2$. (b) $\chi_2$ along the $t_2$ direction for various values of $t_1$. ($N=2L_xL_y$ for $L_x\times L_y=4\times 3$)}
\label{fig_full_pd_33_sus}
\end{figure}

\begin{figure}
\centering
\includegraphics[]{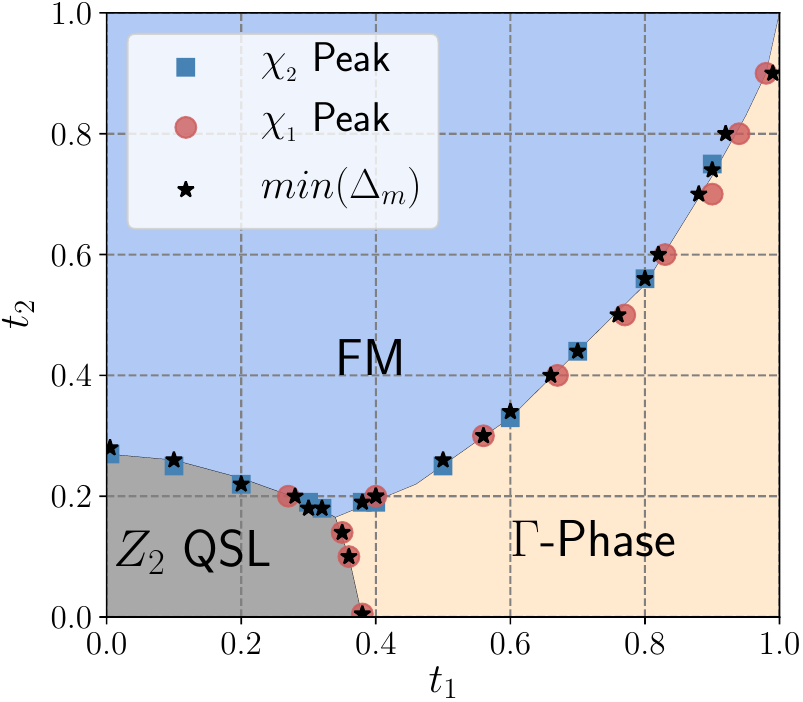}
\caption{Phase diagram of Eq.~\ref{eq_def_full_ham_num} for $t_1,t_2\in[0,1]$ for the complete KJ$\Gamma$ Hamiltonian. The phase boundaries are obtained by analysing ground state fidelity susceptibility and values where the gaps to the bulk excited states ( $\equiv \Delta_m$) (see text) takes the minimum value ($min(\Delta_1)$) of a 24 spin ($L_x=4, L_y=3$) cluster.}
\label{fig_full_pd_43}
\end{figure}

\paragraph*{1. \underline{Ground state fidelity susceptibility} ($\chi_1, \chi_2$) :} 
As introduced above, this is the double derivative of the ground state energy $E_{GS}$ as a function of any of the parameters $t_1$  and $t_2$ : $\chi_1= |\frac{\partial^2 E_{GS}}{\partial t^2_1}|$ and $\chi_2= |\frac{\partial ^2 E_{GS}}{\partial t^2_2}|$.  The behavior of the fidelity susceptibility for fixed values of $t_1$ as a function of $t_2$ and vice-versa shows pronounced peaks (see Fig.~\ref{fig_full_pd_33_sus}) showing transitions between the $Z_2$ QSL (stabilized by 
$\tilde{\mathcal{H}}_{(J=\Gamma=0)}$), the ferromagnet (stabilized by $\mathcal{\tilde{H}}^{AF}_{(\Gamma=K=0)}$ ) and the large $\Gamma$ phase stabilized by $\mathcal{\tilde{H}}^{AF}_{(J=K=0)}$. The position of these peaks is plotted in Fig.~\ref{fig_full_pd_43} to demarcate the phase boundaries.  

\begin{figure}
\centering
\includegraphics[]{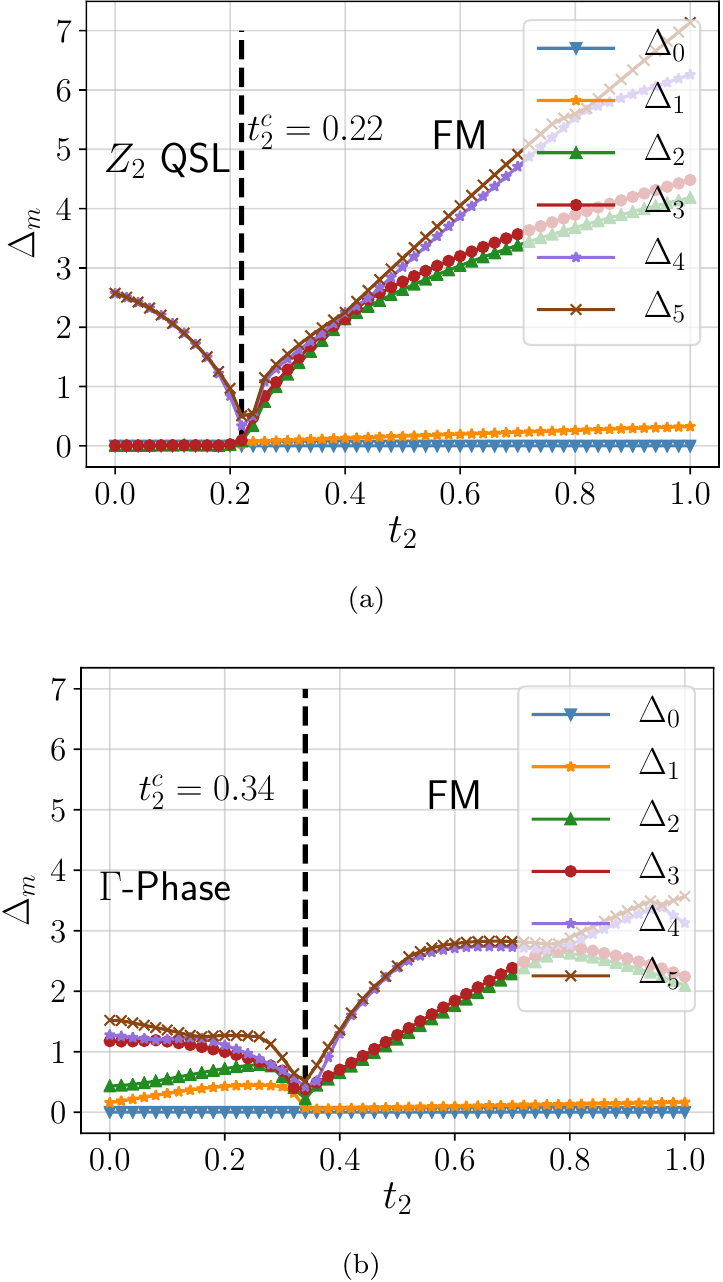}
\caption{Low energy spectra to excited states from the ground state where $\Delta_m$ defines the gap between the  $m^{th}$ excited state and the ground state is shown as a function of $t_2$ for constant values of $t_1$ (see Eq.~\ref{eq_def_full_ham_num}). In (a) $t_1=0.2$ and (b) $t_1=0.6$. The system size is $(L_x,L_y)=(4,3)$ with $N=2L_xL_y$ spins.}
\label{fig_full_pd_33_gap}
\end{figure}

\paragraph*{2. \underline{ Spectral gap} ($\Delta_m$):} Further insights into the nature of phases and phase boundaries are obtained from the bulk spectral gap of the low lying energy eigenstates, $\Delta_m$,-- the gap between the  $m^{th}$ excited state and the ground state. For instance in the FM state $\Delta_1 \sim $ zero given the expected two fold degenerate ground states (pertaining to two symmetry broken states in the thermodynamic limit), while in the $Z_2$ QSL one expects $\Delta_1-\Delta_3 \sim$ zero, since the latter has a 4 fold topological degeneracy on a torus. One expects no such degeneracy for the large $\Gamma$ phase since it is a gapless point where the bulk states would show gaps due to finite size effects. All these expectations are correctly borne out in our numerical results shown in Fig.~\ref{fig_full_pd_33_gap}, where the behavior $\Delta_1 - \Delta_5$ helps to demarcate the various phases.

Further the minimum of bulk gap (min($\Delta_m$)) coincides with the susceptibility peaks (see Fig.\ref{fig_full_pd_43}) which serves as a self consistent check for the phase boundaries for our finite spin clusters.

\begin{figure}
\centering
\includegraphics[]{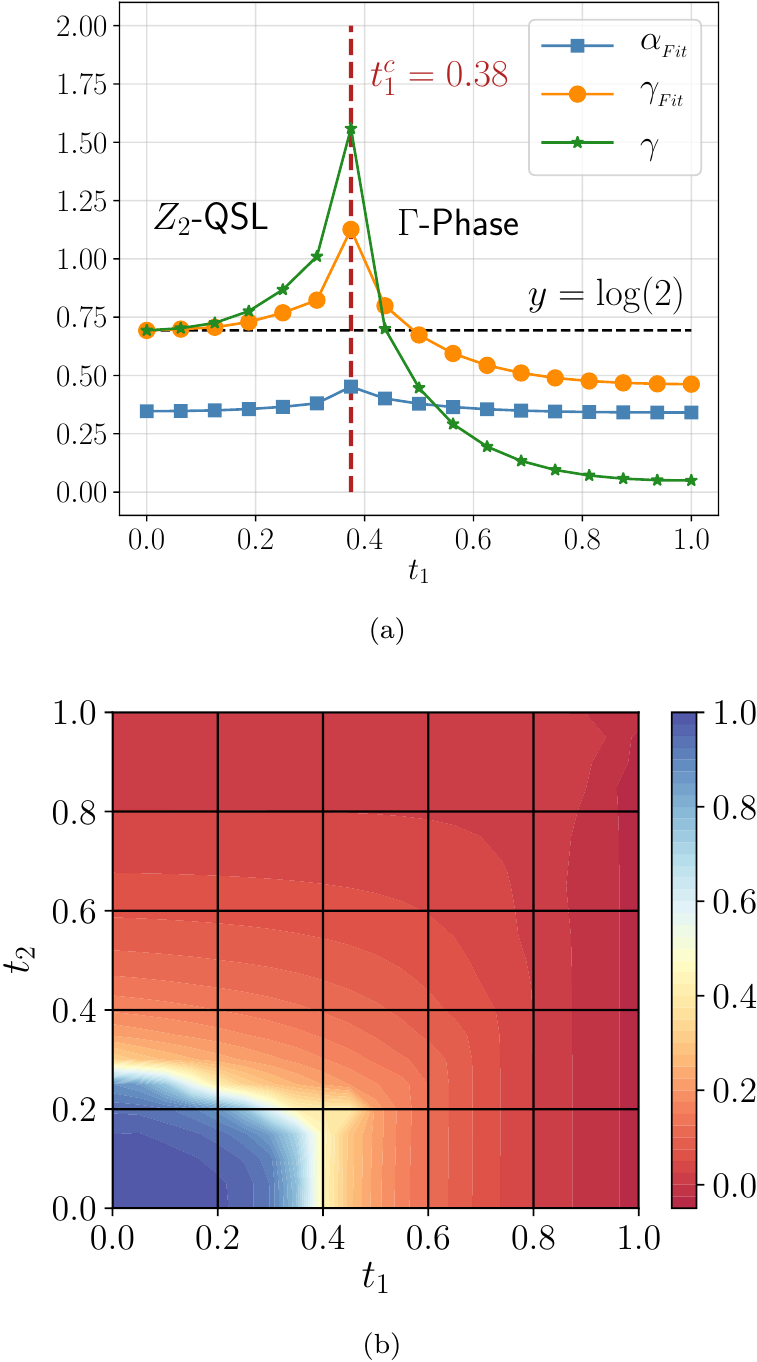}
\caption{(a)The bipartite entanglement entropy in the $\Gamma K$ direction ($t_2=0$ line in Eq.~\ref{eq_def_full_ham_num}) follows an area law, i.e. $S_A(L)=\alpha L$, however, in the $Z_2$-QSL phase this is supplemented by a topological correction ($\gamma$). Calculations done on a ($(L_x,L_y)=(5,3)$ cluster). (b) Average of the plaquette ($W_i$ in Eq.~\ref{eq_tc_rot_unrot}) expectation value for the ground state of Eq.~\ref{eq_def_full_ham_num} in the $(t_1,t_2)$ plane (for a ($(L_x,L_y)=(3,3)$) spin cluster).}
\label{fig_topEE}
\end{figure}

\paragraph*{3. \underline{Topological entanglement entropy} ($\gamma$):} The non-trivial entanglement of the gapped $Z_2$ QSL can be captured via the topological entanglement entropy ($\equiv \gamma$).  In order to distill this it is useful to employ the Kitaev-Preskil prescription\cite{nanda2020phases,kitaev2006topological,levin2006detecting} where the area law contributions cancel perfectly. The behavior of $\gamma$ as a function of $t_1$ for $t_2=0$ is shown in Fig.~\ref{fig_topEE}(a). One finds that while  $\gamma \sim \log(2)$ in the $Z_2$ QSL, $\gamma\sim 0$ in the $\Gamma$ phase reflecting that the latter has no topological order of a  gapped spin liquid state. To investigate the area law contributions in the various phases it useful to calculate, for a given a spin cluster,  the bipartite entanglement entropy($S_{A}(L)$) of any sub part of volume $A$ with a linear boundary of size $L$ and fit it to this functional form 
\begin{equation}\label{eq_ent_entropy}
S_{A}(L)=\alpha L -\gamma + O(1/L)
\end{equation}
where $\alpha,~\gamma$ are the coefficients of the area law entanglement, and the topological entanglement entropy respectively \cite{eisert2010area,kitaev2006topological,levin2006detecting}. The behavior of $\alpha$ is also shown in Fig.~\ref{fig_topEE}(a) reflecting that both $Z_2$ QSL and large $\Gamma$ phase has finite area law contributions. It is worthwhile to point out that $\gamma$ obtained by fitting Eq.~\ref{eq_ent_entropy} ($\equiv \gamma_{Fit}$) seems to show a finite value in the large $\Gamma$ phase, this however is a spurious artifact of the fitting scheme as has been pointed out in \cite{Zou_PRB_2016} for stacked/cluster SPT like states. It is pertinent to point out that in the large $\Gamma$ phase we often find a curvature in the behavior of $S$ as a function of $L$ which may suggest a logarithmic correction \cite{luo2021gapless}. However, in our limited ED calculations it is hard to separate out if this due to the gapless nature of the $(\lambda_1, \lambda_2)$ point or due to a finite correlation length in the large $\Gamma$ phase. Some additional results in other parameter regimes are discussed in Appendix~\ref{sec_JKGnumerical}.

\paragraph*{4. \underline{Plaquette expectation} ($w$):} The non-trivial topological entanglement entropy of the QSL is closely related to the type of topological order realised. As discussed above in the section \ref{subsec_tc_limit_afm} the low energy excitations of the QSL are gapped bosonic Ising electric and magnetic charges \cite{kitaev2003fault,kitaev2006anyons} whose density are encoded by the plaquette spin operators $W_i \equiv\tau^z_{i+d_1}\tau^z_{i-d_2}\tau^y_i\tau^y_{i+d_1-d_2}$ (see Eq.\ref{eq_tc_rot_unrot}). We plot the expectation value of such average charge density $w = \sum_i \frac{1}{N} \langle W_i \rangle$ in Fig.~\ref{fig_topEE}(b) for the entire $t_1, t_2$ plane (where the expectation value is taken over the ground state). Clearly in the QSL the ground state does not contain any charges resulting in $w\approx 1$ which gives away to $w\approx 0$ in both the spin ordered as well as the large $\Gamma$ phase showing that in the ground states of these phases the charges proliferate. 

This provides an important clue into the mechanism of the phase transitions out of the QSL via the proliferation and condensation of the gauge charges. We use these soft modes to construct our critical theory for the phase transition in the next section.

\begin{figure}
\centering
\includegraphics[]{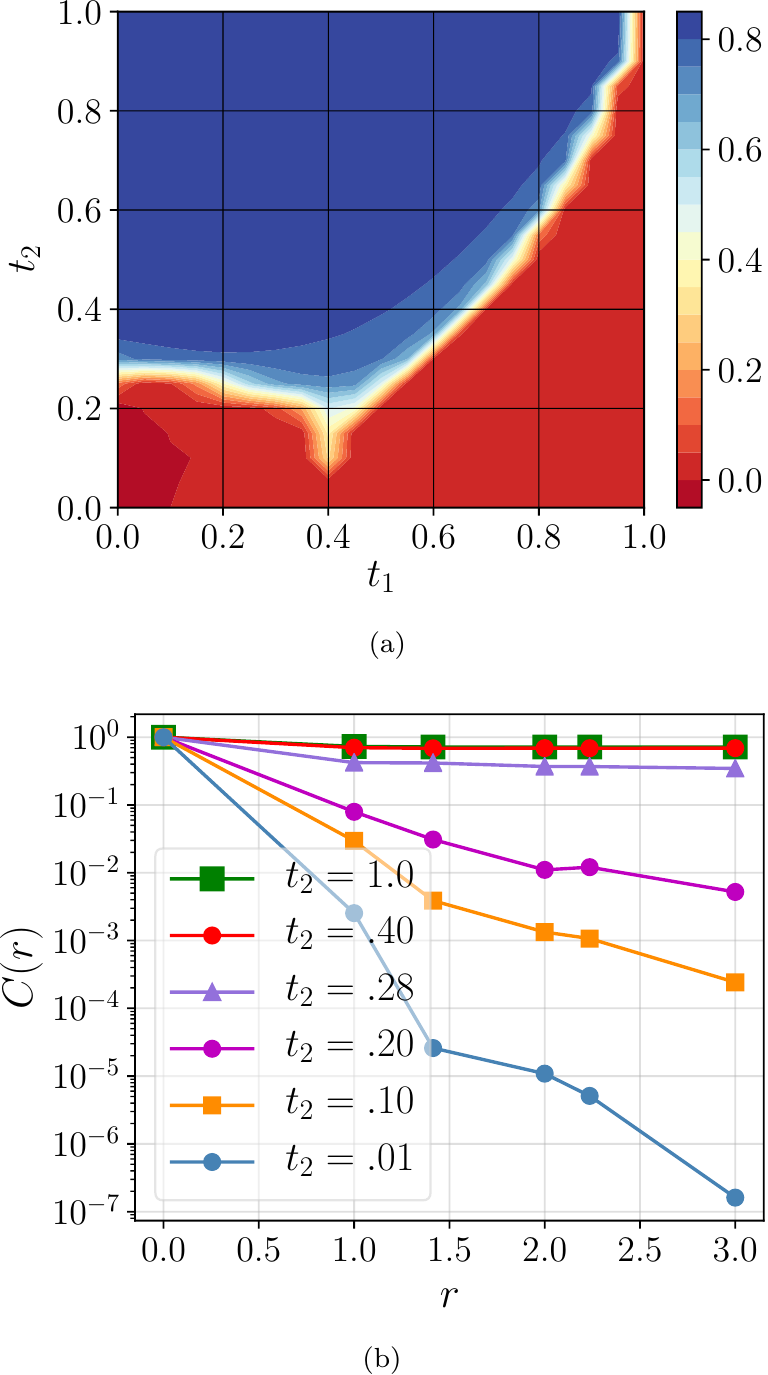}
\caption{(a) Behavior of $M_z =  \frac{1}{N} \sum_i \langle \tau^z_i \rangle   $ in the $t_1,t_2$ parameter regime of Eq.~\ref{eq_def_full_ham_num}. To characterize the ferromagnet state we apply a weak symmetry breaking perturbation ($\sim \frac{t_2(1-t_1)}{100} \sum_i \tau^z_i$). (b) Behavior of connected correlator $C(r)=\langle\tau^z_i\tau^z_{i+r}\rangle-\langle\tau^z_i\rangle\langle\tau^z_{i+r}\rangle$ over the ground state of Eq.~\ref{eq_def_full_ham_num} for different values of $t_2$ for $t_1=0$. (System size, $L_x=3,L_y=3$)}
\label{fig_full_pd_33_exp}
\end{figure}

\paragraph*{5. \underline{Magnetization} ($M_z$):} While the QSL does not break any symmetry spontaneously, the spin-ordered phase on the other hand, is characterised by symmetry breaking captured by a finite magnetisation $M_z= \frac{1}{N} \sum_i \langle \tau_z \rangle$ which is calculated in presence of a small symmetry breaking field ($\sim \frac{t_2(1-t_1)}{100} \sum_i \tau^z_i$). The resultant plot is shown in  Fig.~\ref{fig_full_pd_33_exp}(a). Clearly the complete FM region shows a finite $M_z$ while both the $Z_2$ liquid and the large $\Gamma$ phase shows no such feature. Thus we expect that this region spontaneously break symmetry in the thermodynamic limit as the symmetry breaking field is taken to zero as our calculation of the spin-spin correlations (below) indicate.
 
\paragraph*{6. \underline{Spin-spin correlation:}} 
To further characterize the ferromagnet, connected correlator $C(r)=\langle\tau^z_i\tau^z_{i+r}\rangle-\langle\tau^z_i\rangle\langle\tau^z_{i+r}\rangle$ is evaluated over the ground state in {\it absence} of any perturbing field.  An exponentially falling correlation signals no magnetic order while a long range ordered state will show that $C(r)$ takes a finite value. The behavior of $C(r)$ as a function of $r$ is shown in Fig.~\ref{fig_full_pd_33_exp}(b) for different values of $t_2$ with $t_1$ being zero showing the systems realizes a long-range magnetic order in the FM state.

The above numerical results, when taken together, lead to the phase diagram as shown in Fig.~\ref{fig_full_pd_43} which illustrates the three phases and the intervening transitions.  In the rest of the paper we investigate the nature of the intervening phase transitions and develop its field theory.  

\section{Theory of Phase transitions}
\label{sec_phasetrans}

Our numerical studies leading to the phase diagram of Fig.~\ref{fig_full_pd_43} shows that the phase transitions out of the QSL are brought about by the condensation of the Ising electric and magnetic charges. We now build on the above observation to develop the field theories for the phase transitions. 

\subsection{Phase transition between QSL and the spin ordered phase}\label{sec_phase_tran_tc_heisenberg}

Along the $\Gamma=0$ line ($t_1=0$ line in Fig.~\ref{fig_full_pd_43}),  there are two competing phases-- the $Z_2$ QSL for  $J\sim 0$ and the spin ordered phase in the Heisenberg limit, $J/|K|\gg 1$.  To understand the phase transition between them,  it is convenient to start with the QSL and obtain the description of the transition in terms of the soft electric and magnetic modes similar to Ref.~\onlinecite{nanda2020phases}, as a function of $J$,  of its excitations-- the $e$ and $m$ charges. To the leading order in $J$ the Hamiltonian is given by Eq.~\ref{eq_heisenberg_limit_min_afm},  where, we neglect the higher order terms in $J$. Since at large $J$ (depending on the sign) the system goes into an ferromagnet or anti-ferromagnet state for $\tau$-spins we for now ignore the transverse field term and look at the effect of the Ising exchange term on the Toric code Hamiltonian.

In terms of the gauge charges of Eq.~\ref{eq_ttau_z}, the Hamiltonian in Eq.~\ref{eq_afm_full_hamiltonian} in the limit $\Gamma=0$ becomes:
\begin{equation}
\begin{aligned}
& \tilde{\mathcal{H}}^{AF}_{\Gamma=0}={-J}\sum_{\langle ab\rangle \in H; \langle bc\rangle\in V}\left[\mu^x_a\rho^z_{ab}\mu^x_b\right]\left[\rho^x_{bc}\right] \\
& ~~~~~~~~~-J_{TC}\sum_a\mu^z_a-J_{TC}\sum_p\prod_{\langle ab\rangle\in p}\rho^z_{ab}
\label{eq_leading}
\end{aligned}
\end{equation}
Where $a,~b,~c$ are the square lattice vertices (see Fig \ref{fig_kitaevtoric}). Similar to Ref. \cite{nanda2020phases} we identify the soft modes within a gauge mean field analysis (also see Appendix \ref{subsec_gmft}) appropriately modified to the present symmetry considerations. As in the ferromagnetic case~\cite{nanda2020phases} (see Appendix \ref{sec:smAFM}), we get two soft modes for each of electric and magnetic sectors \cite{lannert2001quantum,xu2009global,PhysRevB.84.104430}:
\begin{align}
\Psi_e({\bf r},\tau)=\phi_e^{(1)}({\bf r},\tau)~\hat\nu_e^{(1)}+\phi_e^{(2)}({\bf r},\tau)~\hat\nu_e^{(2)} \\
\Psi_m({\bf r},\tau)=\phi_m^{(1)}({\bf r},\tau)~\hat\nu_m^{(1)}+\phi_m^{(2)}({\bf r},\tau)~\hat\nu_m^{(2)}
\label{eq_esm}
\end{align}
where $(\phi_e^{(1)}({\bf r},\tau), \phi_e^{(2)}({\bf r},\tau))$ (and $(\phi_m^{(1)}({\bf r},\tau), \phi_m^{(2)}({\bf r},\tau))$) are real fields that represents amplitudes of the electric (magnetic) soft modes. Defining complex variables 
\begin{align}
    \Phi_e=\phi_e^{(1)}+i\phi_e^{(2)}=|\Phi_e|e^{i\theta^e}
    \label{eq_soft_e}
\end{align}
and 
\begin{align}
    \Phi_m=\phi_m^{(1)}+i\phi_m^{(2)}=|\Phi_m|e^{i\theta^m}
    \label{eq_soft_m}
\end{align}
provides us the fields using which the critical field theory is formulated. The symmetry transformation of these fields are given in Eq.~\ref{eq_emme}, we notice that the transformation rules for the symmetries $\sigma^v$ and $R_{\pi}$ are different from the FM case \cite{nanda2020phases} given the different implementation of microscopic symmetries.


Given the electric and magnetic charges see each other as source of mutual $\pi$-flux due to their statistics such long range statistical interactions need to be accounted for through an appropriate Chern-Simons term. In Ref. \cite{nanda2020phases}, we presented a mutual $U(1)$ gauge theory to account for this long range statistical interactions. The same techniques can be applied to the present case as shown below. However, here we employ a somewhat more microscopic formulation using a mutual $Z_2$ gauge theory formalism to obtain the same critical field theory. We discuss them in turn.

\subsubsection{Mutual \texorpdfstring{$U(1)$}{} Chern Simons theory}

The semionic statistics between the Ising electric and magnetic charges can be captured \cite{kitaev2006anyons} using a mutual $U(1)$ Chern-Simons term \cite{PhysRevB.78.155134,xu2009global,PhysRevB.80.125101}
\begin{align}
\mathcal{S}^{U(1)}_{CS}=\frac{i}{\pi}\int d^2{\bf r}d\tau~\epsilon^{\mu\nu\lambda} A_\mu\partial_\nu B_\lambda
\label{eq_u1cs}
\end{align}
where $\mu,\nu,\lambda=x,y,\tau$ and $A_\mu$ and $B_\mu$ are $U(1)$ gauge fields that couple to the electric and magnetic soft modes respectively. The symmetry transformations for the gauge fields $A_{\mu}~(B_{\mu})$ are given in Eq.~\ref{eq_abba}.

The critical theory is identical to the FM case and is given by
\begin{align}
\mathcal{S}_c=\int d^2{\bf r}d\tau~\mathcal{L} +\mathcal{S}^{U(1)}_{CS}
\label{eq_softct}
\end{align}
where $\mathcal{S}_{CS}$ is given by Eq.~\ref{eq_u1cs} and
\begin{align}
\mathcal{L}=\mathcal{L}_e +\mathcal{L}_m+\mathcal{L}_{em}
\label{eq_spinct}
\end{align}
with
\begin{align}
\mathcal{L}_e=|(\partial_\mu-i A_\mu)\Phi_e|^2+& u|\Phi_e|^2+v|\Phi_e|^4\nonumber\\
&-\lambda\left[(\Phi_e)^4+(\Phi_e^*)^4\right]
\label{eq_e4}
\end{align}
\begin{align}
\mathcal{L}_m=|(\partial_\mu-i B_\mu)\Phi_m|^2+& u|\Phi_m|^2+v|\Phi_m|^4\nonumber\\
&-\lambda\left[(\Phi_m)^4+(\Phi_m^*)^4\right]
\label{eq_m4}
\end{align}
\begin{align}
\mathcal{L}_{em}=w\left[(\Phi_e\Phi_m)^2+(\Phi_e\Phi_m^*)^2+{\rm c.c.}\right]
\label{eq_emint}
\end{align}
Similar to FM case~\cite{nanda2020phases} (see details in Appendix~\ref{sec:phases}) we find that while $u>0$ signifies the $Z_2$ spin liquid state, $(u<0, \lambda<0, w \neq 0 \propto \text{sgn}(J))$ specifies the spin symmetry broken ordered state, where the ordered states correspond to FM (AFM) for $J>0~(J<0)$ in Eq.~\ref{eq_heisenberg_limit_min_afm} for the $\tau$ spins state, which translates into Neel (zig-zag) order for underlying $\sigma$ spins (see Fig.~\ref{fig_neel_zz}).

\subsubsection{The Mutual \texorpdfstring{$Z_2$}{} gauge theory}

The soft modes of the Ising electric and magnetic charges in Eq.~\ref{eq_soft_e} and \ref{eq_soft_m} respectively are charges under a $Z_2$ gauge field and hence their mutual semionic statistics are naturally captured by a mutual $Z_2$  CS theory as we describe below~\cite{PhysRevB.62.7850,PhysRevB.84.104430}. This provides for connecting the more prevalent mutual $U(1)$ approach described above \cite{Xu_PRB_2009} with a systematic $Z_2$ approach. Indeed, the latter approach is generically more suited to faithfully capture the nature of phase transitions~\cite{abhishodh}. However, in the present case we obtain the same continuum theory for the transition.

The starting point of the mutual $Z_2$ formalism is obtaining a lattice version of the soft mode theory since the $Z_2$ gauge fields are naturally formulated on the lattice. Hence using Eq.~\ref{eq_soft_e} and \ref{eq_soft_m}, we write the lattice low energy action as~\cite{PhysRevD.19.3682}
\begin{align}
    \mathcal{S}=\mathcal{S}_e+\mathcal{S}_m+\mathcal{S}_{CS}
    \label{eq_mutualz2action}
\end{align}
where
\begin{align}
    \mathcal{S}_e=-t\sum_{ab}\rho_{ab}\cos(\theta^e_a-\theta^e_b)+\cdots
\end{align}
is the electric action defined on the direct square lattice with $\rho_{ab}$ being the $Z_2$ link field  with which it is minimally coupled,
\begin{align}
    \mathcal{S}_m=-t\sum_{\bar{a}\bar{b}}\tilde{\rho}_{\bar{a}\bar{b}}\cos(\theta^m_{\bar{a}}-\theta^m_{\bar{b}})+\cdots
\end{align}
is the magnetic action defined on the dual square lattice with $\tilde{\rho}_{\bar{a}\bar{b}}$  being the dual $Z_2$ gauge field and 
\begin{align}
    \mathcal{S}_{CS}=i\frac{\pi}{4}\sum_{ab\in \Box}(1-\rho_{ab})\left(1-\prod_{\bar{a}\bar{b}\in\Box}\tilde{\rho}_{\bar{a}\bar{b}}\right)
    \label{eq_z2cs}
\end{align}
is the Ising Chern-Simons action that implements the mutual semionic statistics between the electric and the magnetic charges\cite{kitaev2006anyons}.

Note that the hopping amplitude for both the electric and magnetic charges are fixed to be the same (denoted by $t$ above) by the self-dual structure of the action since the electric and magnetic soft modes transform into each other under unit lattice translation (see \cite{nanda2020phases} and the discussion near eqns.~\ref{eq_emme}). Similarly $(\cdots)$ represents higher order interaction terms that are highly constrained by the self dual structure of the theory. We shall consider such interaction terms soon.

To proceed further we seek to dualise either the electric or the magnetic sectors both of which are XY fields and hence can be dualised using the particle-vortex duality \cite{Dasgupta_PRL_1981,Fisher_PRB_1989}. We choose to dualise the electric sector.

To this end, we re-write the electric action 
\begin{align}
    \mathcal{S}_e&=-t\sum_{ab}\cos\left(\theta_a^e-\theta_b^e+\frac{\pi}{2}(1-\rho_{ab})\right)
    \label{eq_electricz2}
\end{align}
using Villain approximation \cite{Villain_1977} to obtain (the details are given in Appendix \ref{appen_z2})
\begin{align}
    \mathcal{S}'_e&=\frac{1}{2t}\sum_{ab} L_{ab}^2+iL_{ab}\left(\theta_a^e-\theta_b^e+\frac{\pi}{2}(1-\rho_{ab})\right)
    \label{eq_intermediatese}
\end{align}
where $L_{ab}$ is an integer value link field. Further integration over $\theta^e_a$ gives rise to the zero divergence (on a lattice) constraint on them, {\it i.e.},
\begin{align}
    \nabla_j L_{ab}=0
\end{align}
which is solved by defining an integer field $C_{\bar{a}\bar{b}}$ on the dual lattice through a lattice curl
\begin{align}
    L_{ab}=\nabla\times C_{\bar{a}\bar{b}}
    \label{eq_latticecurl}
\end{align}
Putting this together with $\mathcal{S}_{CS}$ (Eq.~\ref{eq_z2cs}), we have
\begin{align}
    \mathcal{S}_e+\mathcal{S}_{CS}=&\sum_{\bar{a}\bar{b}}\frac{\left(\nabla\times C_{\bar{a}\bar{b}}\right)^2}{2t}\nonumber\\
    &+i\frac{\pi}{2}\sum_{ab}[1-\rho_{ab}]\left[\nabla\times C_{\bar{a}\bar{b}}+\frac{1-\prod_{\Box}\tilde{\rho}_{\bar{a}\bar{b}}}{2}\right]
\end{align}
such that on integrating over $\rho_{ab}$ we get the constraint
which gives rise to
\begin{align}
    \prod_{\bar{a}\bar{b}\in\Box}\tilde{\rho}_{\bar{a}\bar{b}}=(-1)^{(\nabla\times C_{\bar{a}\bar{b}})}
\end{align}
which can be solved by dividing $C_{\bar{a}\bar{b}}$ into an even and an odd part as
\begin{align}
    C_{\bar{a}\bar{b}}=2A_{\bar{a}\bar{b}}+\eta_{\bar{a}\bar{b}}
\end{align}
where $\eta_{\bar{a}\bar{b}}=0,1$ and $A_{\bar{a}\bar{b}}\in \mathbb{Z}$, such that
\begin{align}
    \tau_{\bar{a}\bar{b}}=1-2\eta_{\bar{a}\bar{b}}
\end{align}
In continuation with our soft mode treatment, we now implement the integer constraint on $A_{\bar{a}\bar{b}}$ softly through the potential
\begin{align}
    -w\cos(2\pi A)~~~~~~~~~(w>0)
\end{align}
such that the whole action (Eq.~\ref{eq_mutualz2action}) becomes
\begin{align}
    \mathcal{S}=&\sum_{\bar{a}\bar{b}}\frac{\left(\nabla\times C_{\bar{a}\bar{b}}\right)^2}{2t\pi^2}\nonumber\\
    &-\sum_{\bar{a}\bar{b}}\tilde{\rho}_{\bar{a}\bar{b}}\left[w\cos(C_{\bar{a}\bar{b}}+\vartheta_{\bar{a}}-\vartheta_{\bar{b}})+t~\cos(\theta^m_{\bar{a}}-\theta^m_{\bar{b}})\right]
\end{align}
where we have re-scaled $C\rightarrow \pi C$ and have separated out a vortex field $\vartheta_{\bar a}$ through a gauge choice~\cite{Senthil_PRB_2006,Senthil_2001,Bhattacharjee_PRB_2011}: $\nabla\cdot C=0$. Integrating out $\tilde{\rho}$, we get, to the leading order

\begin{align}
    \mathcal{S}=&\frac{1}{2t\pi^2}\sum_{\bar{a}\bar{b}}\left(\nabla\times C_{\bar{a}\bar{b}}\right)^2\nonumber\\
    &+\frac{t^2}{4}\sum_{\bar{a}\bar{b}}~\cos[2(\xi_{\bar{a}}-\xi_{\bar{b}})-2(\vartheta_{\bar{a}}-\vartheta_{\bar{b}})]\nonumber\\
    &+\frac{w^2}{4}\sum_{\bar{a}\bar{b}}~\cos[2(\vartheta_{\bar{a}}-\vartheta_{\bar{b}}-C_{\bar{a}\bar{b}})]\nonumber\\
    &+\frac{tw}{2}\sum_{\bar{a}\bar{b}}~\cos[\xi_{\bar{a}}-\xi_{\bar{b}}-C_{\bar{a}\bar{b}}]
\end{align}

where we have defined 
\begin{align}
    \xi=\theta^m+\vartheta
\end{align}
The continuum limit can be obtained by introducing bosonic fields
\begin{align}
    \varphi=e^{i2\vartheta},~~~~~~\chi=e^{-i\xi}
\end{align}
to get
\begin{align}
    \mathcal{S}=&\frac{1}{2t\pi^2}\sum_{\bar{a}\bar{b}}\left(\nabla\times C_{\bar{a}\bar{b}}\right)^2 + tw\sum_{\bar{a}\bar{b}}~\chi_{\bar{a}}^*e^{-i C_{\bar{a}\bar{b}}}\chi_{\bar{b}}\nonumber\\
    &+\frac{w^2}{2}\sum_{IJ}~\varphi_{\bar{a}}^*~e^{i2C_{\bar{a}\bar{b}}}~\varphi_{\bar{b}} + \frac{t^2}{2}\sum_{\bar{a}\bar{b}}~(\chi_{\bar{a}}^*\chi_{\bar{b}})^2(\varphi_{\bar{a}}^*\varphi_{\bar{b}})
\end{align}
such that the continuum action is given by
\begin{align}
    \mathcal{S}_{cont}=\int d^2{\bf x}d\tau~\mathcal{L}_{cont}
\end{align}
where
\begin{align}
\mathcal{L}_{cont}=&|(\partial_\mu-iC_\mu)\chi|^2+|(\partial_\mu+i2C_\mu)\varphi|^2+ V\left[\chi,\varphi\right]\nonumber\\
&+g(\epsilon_{\mu\nu\lambda}\partial_\nu C_\lambda)^2
\end{align}
where $V\left[\chi,\varphi\right]$ denotes the interactions between the modes that are allowed by symmetry. The above critical theory is exactly dual to Eq.~\ref{eq_softct}. Indeed starting with Eq.~\ref{eq_softct}, we can dualise the electric charges to get the above field theory as was shown in Ref. \cite{nanda2020phases}. Similarly, based on the symmetry transformations of the soft modes and in particular the permutation of the electric and the magnetic soft modes under translation, we have:
\begin{align}
    V\left[\chi,\varphi\right]=u\left(|\chi|^2-|\varphi|^2\right)+v\left(|\chi|^4+|\varphi|^2\right)+\tilde{w}|\chi|^4|\varphi|^2+\cdots
\end{align}
where the relative negative sign for the quadratic term is obtained by noting that $\varphi$ is dual to the electric soft mode. Thus the transition belongs to a self-dual modified Abelian Higg's theory. This concludes our discussion of the deconfined critical point describing the quantum phase transition between the $Z_2$ QSL and the spin-ordered phase. For a detailed discussion on this critical theory we refer to \cite{nanda2020phases}.

\subsubsection{A two-step or a single step transition}

In the above discussion we have presently ignored the transverse field term (see Eq.~\ref{eq_heisenberg_limit_min_afm}) which occurs with a strength of $2J$ may potentially open up an intermediate phase as $J$ is increased (see  Fig.~\ref{fig:SsTs}). We now focus the viability of such a scenario.

The inclusion of Heisenberg term leads to a perturbation of both an Ising term and a transverse field to the parent Toric code Hamiltonian in the strong anisotropic limit. In the  complete parameter space, therefore we clearly have three phases (i) The $Z_2$ QSL for the Toric code. (ii) The Ising ordered phase which breaks a $Z_2$ symmetry stabilizing a Neel order for the original $\sigma$ spins. (iii) A $x$ paramagnet (in $\tau$s)(see  Fig.~\ref{fig:SsTs} and Fig.~\ref{fig_sus_mag_tcm_zz_y_t1_scan}).

Even while for $\tau$ spins the the paramagnet is may seem featureless and trivially $x$ polarized, in terms of underlying $\sigma$ spins its an intriguing state given the eigenstates correspond to $\tau^x$ are the essentially a singlet or a triplet bond 
\begin{equation}
|\pm \rangle_x = \frac{1}{\sqrt{2}} \Big( | \uparrow \downarrow \rangle \pm | \downarrow \uparrow \rangle \Big)    
\end{equation}
ordered state on every $z$ bond of the underlying honeycomb lattice. A polarized state in the $\tau$ spins therefore corresponds to a direct product state of singlets on all $z$ bonds which in turn corresponds to a lattice nematic state for the $\sigma$ spins (see discussion in section~\ref{subsec_heisenberg_limit_afm}). The analysis  already provides some interesting insights. This present study in the anisotropic limit already leads to the fact that the transition from the Neel state to the lattice nematic phase is essentially an Ising transition. The transition from the $Z_2$ liquid to the Neel phase is the self-dual modified Abelian Higgs transition. Now at infinite $J$ we know the system enters a Neel phase -- this can either occur directly through a single step transition or the route may entail an intermediate paramagnet phase which could then imply a two step transition (see Fig.~\ref{fig:SsTs}). A detailed numerical study of the Toric code Hamiltonian with a generalized Ising perturbation and a transverse field is given in section~\ref{sec_jk_ham}. We find that in general a Heisenberg perturbation in this strong anisotropic limit is in fact a single-step transition where the $Z_2$ QSL undergoes a self dual modified Abelian Higgs transition to a ferromagnet state.

\subsection{Transition between large $\Gamma$ phase and $Z_2$ QSL}

While the nature of the transition from a Toric code to the Neel state is captured in the above discussed framework - the transition from the $Z_2$-QSL to the paramagnet is quite interesting and we now discuss this transition. The $Z_2$ QSL for the $\tau$ spins is in Wen's representation \cite{wen2002quantum} while the paramagnet it transits to is $x$-polarized which is adiabatically connected to the large-$\Gamma$ phase (see discussion above). Under a unitary rotation (see Eq.~\ref{eq_rot_ver}) while the QSL can be exactly mapped to the Kitaev's Toric code ground state (see Eq.~\ref{eq_tc_rot}), the paramagnet gets converted to $y$-polarized state. The nature of transition from a Toric code QSL to a transverse field in $y$ direction is known to be a first order transition \cite{Vidal_PRB_2009, Dusuel_PRL_2011}. Given the first order nature of this transition we do not expect any universal physics, except noting that this transition has a fundamentally different character from our related FM study \cite{nanda2020phases} where the transition between the QSL and large $\Gamma$ phase was a second order transition.  

\subsection{Transition between spin-ordered phase and large \texorpdfstring{$\Gamma$}{} phase}

This leaves us with the transition between the FM and the large $\Gamma$ phase. Given the large $\Gamma$ phase contains all the microscopic symmetries, we expect the transition from the large $\Gamma$ to the FM transition to be of the Ising kind where a symmetry breaking order gets develop at a critical value of Ising coupling. Our numerical estimation of the phase boundary shows that the transition from the large $\Gamma$ phase to the FM occurs along the $t_2=\frac{t_1}{3 -2t_1}$ curve. This corresponds to a critical value of $J_c$ which quadratically increases with the strength of the $\Gamma$ coupling strength ($\Gamma^2/J_c|K_z| \sim 3$).

This completes our discussion of the phase transitions.
\section{Summary and outlook}
\label{sec_summary}

We now summarise our results. In this follow up (to Ref. \cite{nanda2020phases}) work, we have investigated the Heisenberg-Kitaev-$\Gamma$ model in the anisotropic limit with Kitaev interactions being antiferromagnetic. This leads to important difference in the symmetry transformation of the low energy degrees of freedom-- the non-Kramers doublets which is manifested in the nature of the phases stabilised. In particular the large $\Gamma$ limit appears to be proximate to equal superposition of stacked $Z_2\times Z_2$ spin SPT phases where the symmetries protecting the SPTs are only weakly broken leading by small higher order terms. Our numerical studies on small spin clusters reveal the general structure of the phase diagram indicating that the $Z_2$ QSL is destroyed via proliferation and condensation of its gauge charges-- both electric and magnetic. While the transition to the paramagnetic phase in the large $\Gamma$ limit turns out to be discontinuous, for the continuous transition to the spin-ordered state (from the QSL) we construct a critical continuum field theory in terms of the soft modes of the electric and magnetic charges via a mutual $Z_2$ CS theory and show that the results are exactly with the mutual $U(1)$ CS theory used by us in Ref. \cite{nanda2020phases}. This leads us to conclude that the direct transition between the QSL and the spin-ordered phase is described by a self-dual modified Abelian Higgs field theory. 

\begin{figure}
\centering
\includegraphics[width=1.0\columnwidth]{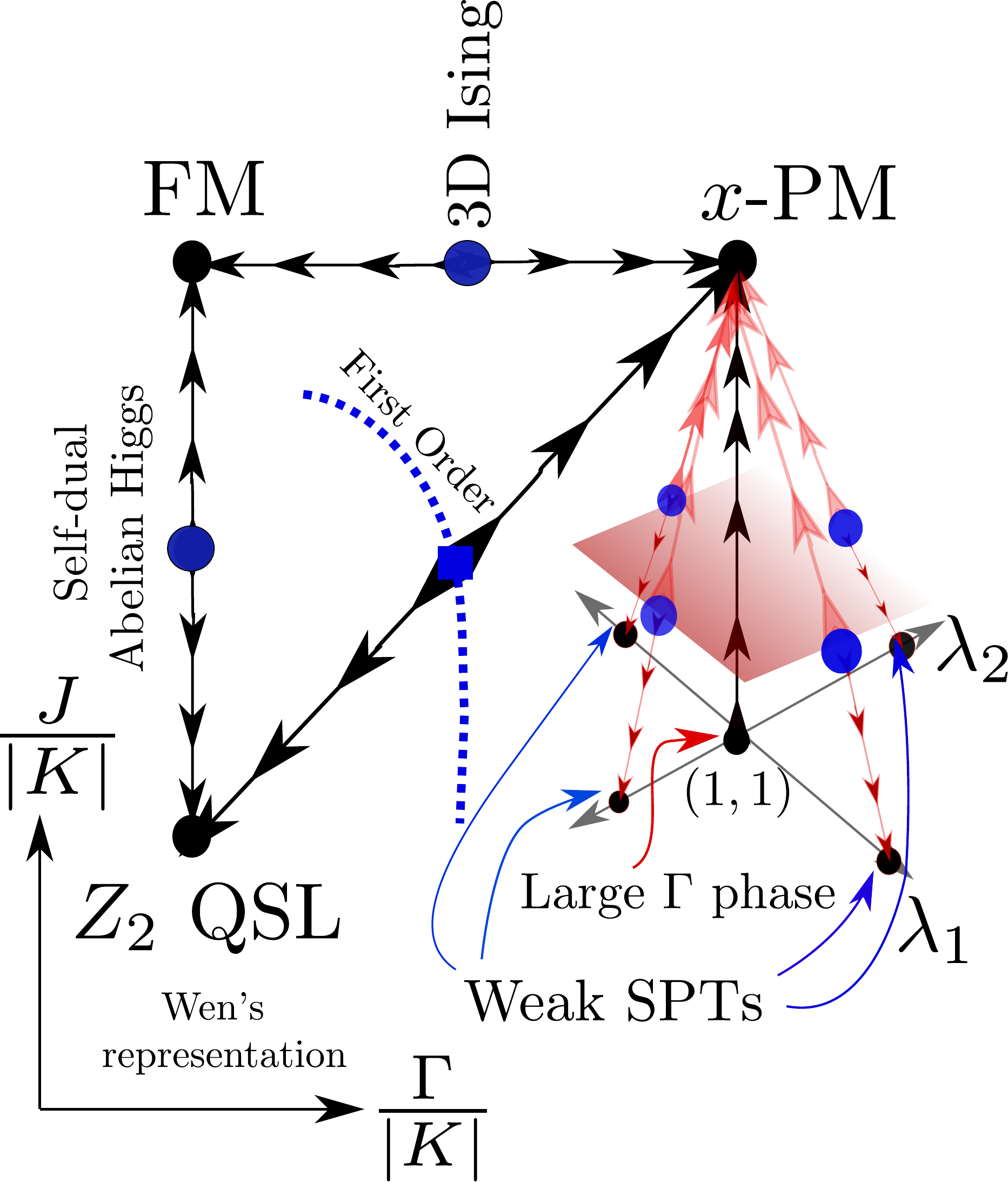}
\caption{Schematic depiction of the phases and phase transitions which are accessible within the parameter space of the complete $KJ\Gamma$ Hamiltonian (see Eq.~\ref{eq_afm_full_hamiltonian}) in the anisotropic limit of anti-ferromagnetic Kitaev model.}
\label{fig_full_pd_schematic}
\end{figure}

The overall summary of our phase diagram is then illustrated in Fig.~\ref{fig_full_pd_schematic} where the following scenario emerges for the $\tau$ spins. There are three phases (i) $Z_2$ QSL, (ii) the Ising FM and (iii) $\Gamma$ phase. The leading order Hamiltonian for the phase in large $\Gamma$ limit is described by a fine tuned point in the $(\lambda_1, \lambda_2)$ plane which interpolate between differently stacked weak SPTs where a gapless phase with boundary modes appear and possibly belongs to a critical point which is trivially gapped out immediately by local transverse field perturbations. The Heisenberg coupling, on the other hand,  drives the $Z_2$ QSL to the spin ordered phase via a deconfined critical point.

The present work, along with Ref. \cite{nanda2020phases}, therefore completes the understanding of the physics of the anisotropic Heisenberg-Kitaev-$\Gamma$ system. While not directly relevant to the present set of experimentally relevant materials, we think our results some shed light on the nature of the soft modes and the phases proximate to the Kitaev QSL on the isotropic honeycomb lattice. 

We would like to end with a few comments about the critical point describing the gapless phase obtained in the large $\Gamma$ limit at the leading order of the Hamiltonian. In this limit, our detailed symmetry and numerical analysis (see Appendix.~\ref{subsubsec_boun_pert_gamma}) shows that the boundary modes are fragile to microscopic symmetry allowed perturbations. Within our exact diagonalisation results this critical point has zero topological entropy consistent with a gapless or a short range entangled gapped phase. Indeed our numerical calculations seem to indicate that the large $\Gamma$ phase is continuously connected to a trivial gapped paramagnet. All these indicate that the gapless point in the large $\Gamma$ limit is indeed critical and unstable. However further detailed understanding of the Hamiltonian in the large $\Gamma$ limit (Eq.~\ref{eq_four_spt_interpolation}) is needed to understand the nature of the phase realised by superposing stacked SPTs. In this regard a naive Majorana mean field theory of the Hamiltonian given by Eq.~\ref{eq_tc_rot_unrot} and the stacked cluster SPT $H_1$, given by Eq.~\ref{eq_four_spt} reveals a generic intermediate gapless phase between the two limits (see Appendix~\ref{appen_majoranamft}). 

\section{Acknowledgements}
The authors would like to thank K. Damle, Y. B. Kim, R. Moessner, A. Prakash, A.V. Mallik, V. B. Shenoy and V. Tripathi for various enlightening discussions and collaborations in related topics. We acknowledge use of open-source QuSpin\cite{weinberg2017quspin,weinberg2019quspin} for exact diagonalisation calculations.  SB and AA acknowledges financial support through Max Planck partner group on strongly correlated systems at ICTS. We acknowledge support of SERB-DST (Govt. of India) early career research grant (No.~ECR/2017/000504) and the Department of Atomic Energy, Government of India, under project no. RTI4001. Computations were performed at the ICTS clusters {\it boson} and {\it boson1}.

\appendix
\section{The perturbation theory for the anisotropic limit}
\label{appen_perturb}

In the strong anisostropic limit ($K_z \gg J, \Gamma, K$) a perturbation series for the Hamiltonian in $\tau$ spins can be obtained. Various terms can be arranged in terms of the number of spins they entail. We perform an analysis upto four spin terms and present the results below. 

The  {\it single spin} terms are given by 
\begin{equation}\label{eq_single_afm}
\mathcal{H}^{AF}_{[1]} = 2J\left(1-\frac{\Gamma^2}{\Delta^2}\right) \sum_{i}\tau^x_i
\end{equation}
where  $\delta=J+|K|,\Delta=|K_z|+J$. 
\begin{widetext}
{\small
{\it Two-spin} contributions are:
\begin{equation}\label{eq_double_afm}
\begin{aligned}
	& \mathcal{H}^{AF}_{[2]} = -\left[J + \frac{J\delta}{2\Delta} + \frac{\delta^3J+\delta J^3}{8\Delta^3}\right]\sum_{\langle i,j\rangle}\tau_i^z\tau_j^z  + \frac{J^2\delta^2}{2\Delta^3}\sum_{i}\tau^z_i\tau^z_{i+d_1-d_2} - \left[\frac{2\Gamma^3}{\Delta^2}+\frac{J^2\delta^2}{4\Delta^3}\right]\sum_{i}\tau^z_{i+d_1}\tau^z_{i-d_2} \\
	& -\frac{J^2\delta^2}{8\Delta^3}\sum_i\left(\tau^z_{i+d_1}\tau^z_{i-d_1} + \tau^z_{i+d_2}\tau^z_{i-d_2}\right) + \frac{\delta J\Gamma^2}{\Delta^3}\sum_{i}\left(\tau^x_i\tau^y_{i+d_1-d_2}-\tau^y_i\tau^x_{i+d_1-d_2}\right)  -\frac{5J^2\delta^2}{8\Delta^3}\sum_{i}\left(\tau^x_i\tau^x_{i+d_1-d_2}+\tau^y_i\tau^y_{i+d_1-d_2}\right)
\end{aligned}
\end{equation}

{\it Three spin} contributions are:
\begin{equation}\label{eq_triple_afm}
\begin{aligned}
	& \mathcal{H}^{AF}_{[3]} = \sum_i \Big( \left[\frac{\Gamma^2}{\Delta}-\frac{(\Gamma)^4}{\Delta^3} + \frac{7\Gamma^2\delta^2}{4\Delta^3}\right]\left(\tau^z_{i+d_1}\tau^x_{i}\tau^z_{i-d_1} + \tau^z_{i+d_2}\tau^x_{i}\tau^z_{i-d_2}\right) 
	  + \left[\frac{\Gamma^2}{\Delta} - \frac{4\Gamma^4+J^2\Gamma^2}{4\Delta^3} + \frac{3\Gamma^2\delta^2}{2\Delta^3}\right]\left( \tau^z_{i+d_1}\tau^y_{i}\tau^z_{i-d_2} - \tau^z_{i+d_2}\tau^y_{i}\tau^z_{i-d_1}\right) \Big) \\
	&  -\left[\frac{\Gamma^4}{2\Delta^3} + \frac{3\Gamma^2\delta^2}{2\Delta^3}\right]\sum_{i}\left(\tau^z_i\tau^z_{i-d_2}\tau^x_{i+d_1-d_2} + \tau^z_{i+d_1-d_2}\tau^z_{i+d_1}\tau^x_i -\tau^z_i\tau^z_{i+d_1}\tau^x_{i+d_1-d_2} - \tau^z_{i+d_1-d_2}\tau^z_{i-d_2}\tau^x_i\right) \\
	& + \frac{J\Gamma^2}{\Delta^2}\sum_{i}\left(\tau^z_i\tau^z_{i+d_1}\tau^y_{i+d_1-d_2}-\tau^z_{i+d_1-d_2}\tau^z_{i-d_2}\tau^y_i + \tau^z_i\tau^z_{i-d_2}\tau^y_{i+d_1-d_2}-\tau^z_{i+d_1-d_2}\tau^z_{i+d_1}\tau^y_i\right)  + \frac{\Gamma^3}{\Delta^2}\sum_{i}\left(\tau^z_{i+d_1}\tau^x_i\tau^z_{i+d_2} + \tau^z_{i-d_1}\tau^x_i\tau^z_{i-d_2}\right)
\end{aligned} 
\end{equation}

{\it Four spins} contributions are:
\begin{equation}\label{eq_afm_quadrapole_full}
\begin{aligned}
	& \mathcal{H}^{AF}_{[4]} = -\left[\frac{\delta^4}{16\Delta^3}+\frac{J^4}{16\Delta^3}\right]\sum_{i}\tau^z_{i+d_1}\tau^z_{i-d_2}\tau^y_i\tau^y_{i+d_1-d_2}  - \frac{J^2\delta^2}{8\Delta^3}\sum_{i}\tau^z_{i+d_1}\tau^z_{i-d_2}\tau^x_i\tau^x_{i+d_1-d_2}
\end{aligned}
\end{equation}.

In the pseudo-dipolar limit, {\it i.e.} for $J=K=0$, the effective Hamiltonian is:
\begin{equation}
\begin{aligned}\label{eq_gammalimit}
\mathcal{H}^{AF}_{J=K=0} =
 & \left[\frac{\Gamma^2}{\Delta}-\frac{\Gamma^4}{\Delta^3}\right]\sum_{i}\left(\tau^z_{i+d_1}\tau^x_{i}\tau^z_{i-d_1} + \tau^z_{i+d_2}\tau^x_{i}\tau^z_{i-d_2}\right)  	  + \left[\frac{\Gamma^2}{\Delta} - \frac{\Gamma^4}{\Delta^3}\right]\sum_{i}\left( \tau^z_{i+d_1}\tau^y_{i}\tau^z_{i-d_2} - \tau^z_{i+d_2}\tau^y_{i}\tau^z_{i-d_1}\right) \\ &- \left[\frac{2\Gamma^3}{\Delta^2}\right]\sum_{i}\tau^z_{i+d_1}\tau^z_{i-d_2} -\left[\frac{\Gamma^4}{2\Delta^3}\right]\sum_{i}\left(\tau^z_i\tau^z_{i-d_2}\tau^x_{i+d_1-d_2} + \tau^z_{i+d_1-d_2}\tau^z_{i+d_1}\tau^x_i -\tau^z_i\tau^z_{i+d_1}\tau^x_{i+d_1-d_2} - \tau^z_{i+d_1-d_2}\tau^z_{i-d_2}\tau^x_i\right) \\
	& + \frac{\Gamma^3}{\Delta^2}\sum_{i}\left(\tau^z_{i+d_1}\tau^x_i\tau^z_{i+d_2} + \tau^z_{i-d_1}\tau^x_i\tau^z_{i-d_2}\right)
\end{aligned}
\end{equation}
}
\end{widetext}

\section{The effective Hamiltonian in rotated basis of the AFM limit}\label{appen_rot_ham_afm}

\subsection{Rotation to the \texorpdfstring{$\tilde{\tau}$}{}-basis}
\label{appen_taurotations}

Form the Eq.~\ref{eq_tc_rot_unrot}, to bring the TC model in it's usual form in Eq.~\ref{eq_tc_rot}, we use the bond dependent unitary rotation:

\begin{align}
&\{\tau_i^x,\tau_i^y,\tau_i^z\}\rightarrow\{-\ttau_i^y,\ttau_i^x,\ttau_i^z\}
~~\forall i\in {\rm horizontal.~bonds}\nonumber\\
&\{\tau_i^x,\tau_i^y,\tau_i^z\}\rightarrow\{\ttau_i^y,\ttau_i^z,\ttau_i^x\}
~~\forall i\in {\rm vertical~bonds}
\label{eq_rot_ver}
\end{align}

\subsection{Symmetry transformations}

The symmetry transformations for the $\ttau$-spins is obtained from table \ref{table_tau_symm} and is given in table \ref{table_tau_symm_afm} where $H~(V)$ denotes the horizontal (vertical) links of the square lattice (see Fig.~\ref{fig_kitaevtoric}).


\begin{table}
\begin{center}
 \begin{tabular}{|c| c c c|c c c |} 
 \hline
 Symmetry & $\ttau^x_{h}$ & $\ttau^y_{h}$ & $\ttau^z_{h}$ & $\ttau^x_{v}$ & $\ttau^y_{v}$ & $\ttau^z_{v}$  \\ [0.5ex] 
 \hline\hline
 $\mathcal{T}$ & $\ttau^x_{h}$ & $\ttau^y_{h}$ & $-\ttau^z_{h}$ & $-\ttau^x_{v}$ & $\ttau^y_{v}$ & $\ttau^z_{v}$ \\ \hline
 $\sigma_v$ & $\ttau^x_{h^{\prime}}$ & $\ttau^y_{h^{\prime}}$ & $\ttau^z_{h^{\prime}}$ & $\ttau^x_{v^{\prime}}$ & $\ttau^y_{v^{\prime}}$ & $\ttau^z_{v^{\prime}}$ \\ \hline
 $C_{2z}$ & $-\ttau^x_{h^{\prime}}$ & $\ttau^y_{h^{\prime}}$ & $-\ttau^z_{h^{\prime}}$ & $-\ttau^x_{v^{\prime}}$ & $\ttau^y_{v^{\prime}}$ & $-\ttau^z_{v^{\prime}}$\\ \hline
 $R_{\pi}$ & $-\ttau^x_{h^{\prime}}$ & $\ttau^y_{h^{\prime}}$ & $-\ttau^z_{h^{\prime}}$ & $-\ttau^x_{v^{\prime}}$ & $\ttau^y_{v^{\prime}}$ & $-\ttau^z_{v^{\prime}}$\\ \hline
 $T_{d_j}$ & $\ttau^z_{h^{\prime}}$ & $-\ttau^y_{h^{\prime}}$ & $\ttau^x_{h^{\prime}}$ & $\ttau^z_{v^{\prime}}$ & $-\ttau^y_{v^{\prime}}$ & $\ttau^x_{v^{\prime}}$\\ \hline
  
\end{tabular}
\caption{Symmetry transformation of the $\ttau$ spins on the horizontal ($h$) and vertical ($v$) links of the AFM anisotropic limit (see Fig.~\ref{fig_kitaevtoric}). Where $v^{\prime}~\&~h^{\prime}$ denotes the lattice points transformation, $h^{\prime}\equiv\mathcal{S}(h)~\&~v^{\prime}\equiv\mathcal{S}(v)$ for $\mathcal{S}\equiv \{\mathcal{T},\sigma_v,C_{2z},R_{\pi},T_{d_{1(2)}}\}$.}
\label{table_tau_symm_afm}
\end{center}
\end{table}

\subsection{Action of the symmetries on the gauge charges and the gauge fields}
\label{appen_gauge_set}

Following the symmetry transformation of the $\tilde{\tau}$-spins in table \ref{table_tau_symm_afm} we will now discuss the transformation rules for the gauge charges and gauge fields.

\paragraph{\underline{Lattice Translations } :}\label{para_gauge_symm_tran} 
Under both the translations, along the directions ${\bf d}_1$ and ${\bf d}_2$ (see Fig.~\ref{fig_kitaevtoric}), the plaquettes and the vertices are interchanged. Hence the $e$ and $m$ charges are interchanged.
\begin{equation}
\begin{aligned}
 T_{\bf d_j}: \begin{array}{l} \{\mu^x,\mu^z\}_{a} \rightarrow \{\tilde{\mu}^x,\tilde{\mu}^z\}_{T_{d_j}({a})} \\
  \{\tilde{\mu}^x,\tilde{\mu}^z\}_{\bar{a}} \rightarrow \{\mu^x,\mu^z\}_{T_{d_j}({\bar{a}})}  \\
\{\rho^x,\rho^z\}_{ab} \rightarrow \{\tilde{\rho}^x,\tilde{\rho}^z\}_{T_{d_j}({ab})} \\
\{\tilde{\rho}^x,\tilde{\rho}^z\}_{\bar{a}\bar{b}} \rightarrow \{\rho^x,\rho^z\}_{T_{d_j}({\bar{a}\bar{b}})}
\end{array}
\end{aligned}\label{eq_gauge_trans_transl}
\end{equation}

For translation along the cartesian axes, the lattice vectors are given by $\hat x={\bf d_1-d_2}$ and $\hat y={\bf d_1+d_2}$. Under this, the gauge charges and potentials transform as
\begin{equation}
\begin{aligned}
T_{\hat x(\hat y)} : \begin{array}{l}\{\mu^x,\mu^z\}_{a} \rightarrow \{\mu^x, \mu^z\}_{a+\hat x(\hat y)} \\
\{\tilde{\mu}^x,\tilde{\mu}^z\}_{\bar{a}} \rightarrow \{\tilde{\mu}^x,\tilde{\mu}^z\}_{\bar{a}+\hat x(\hat y)}  \\
\{\rho^x,\rho^z\}_{\bar{a}\bar{b}} \rightarrow \{\rho^x,\rho^z\}_{\bar{a}+\hat x(\hat y),\bar{b}+\hat x(\hat y)} \\
\{\tilde{\rho}^x,\tilde{\rho}^z\}_{\bar{a}\bar{b}} \rightarrow \{\tilde{\rho}^x,\tilde{\rho}^z\}_{\bar{a}+\hat x(\hat y),\bar{b}+\hat x(\hat y)}
\end{array}
\end{aligned}
\end{equation}

\paragraph{\underline{Time Reversal} :}\label{para_gauge_symm_tr} Due the bond dependent nature of the $\ttau$ transformation the gauge degrees of freedoms transform as:
\begin{align}
\mathcal{T}:  \begin{array}{l}
\{\mu^x,\mu^z\}_{a}\rightarrow \{\mu^x,\mu^z\}_{a} \\
\{\tilde\mu^x,\tilde\mu^z\}_{\bar{a}}\rightarrow \{\tilde\mu^x,\tilde\mu^z\}_{\bar{a}}  \\
\{\rho^x,\rho^z\}_{ab} \rightarrow \{(-1)^{a_y+b_y}\rho^x,(-1)^{a_x+b_x}\rho^z\}_{ab} \\
\{\tilde{\rho}^x,\tilde{\rho}^z\}_{\bar{a}\bar{b}} \rightarrow \{(-1)^{\bar{a}_y+\bar{b}_y}\tilde{\rho}^x,(-1)^{\bar{a}_x+\bar{b}_x}\tilde{\rho}^z\}_{\bar{a}\bar{b}}\\
\end{array}
\label{eq_gauge_trans_tr}
\end{align}

\paragraph{\underline{Reflections about $z$ bond, $\sigma_v$} :}
This transformation is different compared the ferromagnetic case:
{
\begin{equation}
\begin{aligned}
\sigma_v: \begin{array}{l}
\{\mu^x,\mu^z\}_{a} \rightarrow \{\mu^x,\mu^z\}_{\sigma_v({a})} \\
\{\tilde{\mu}^x,\tilde{\mu}^z\}_{a} \rightarrow \{\tilde{\mu}^x,\tilde{\mu}^z\}_{\sigma_v({a})} \\
\{\rho^x,\rho^z\}_{ab} \rightarrow \{\rho^x,\rho^z\}_{\sigma_v({ab})} \\
 \{\tilde{\rho}^x,\tilde{\rho}^z\}_{\bar{a}\bar{b}} \rightarrow \{\tilde{\rho}^x,\tilde{\rho}^z\}_{\sigma_v({\bar{a}\bar{b}})}\\
 \end{array}
\end{aligned}\label{eq_gauge_trans_sigv}
\end{equation}
}

\paragraph{\underline{$\pi$-rotation about the $z$-bond, $C_{2z}$} :}  
This transformation is also different compared the ferromagnetic case:
{
\begin{equation}
\begin{aligned}
C_{2z}: \begin{array}{l}\{\mu^x,\mu^z\}_{\bf a} \rightarrow \{\mu^x,\mu^z\}_{C_{2z}({a})} \\
\{\tilde{\mu}^x,\tilde{\mu}^z\}_{\bar{a}} \rightarrow \{\tilde{\mu}^x,\tilde{\mu}^z\}_{C_{2z}({\bar{a}})} \\
\{\rho^x,\rho^z\}_{ab} \rightarrow \{-\rho^x,-\rho^z\}_{C_{2z}(ab)} \\
\{\tilde{\rho}^x,\tilde{\rho}^z\}_{\bar{a}\bar{b}} \rightarrow \{-\tilde{\rho}^x,-\tilde{\rho}^z\}_{C_{2z}({\bar{a}\bar{b}})}
\end{array}
\end{aligned}\label{eq_gauge_trans_c2z}
\end{equation}
}

\paragraph{\underline{$\pi$-rotation about honeycomb lattice centre, $R_{\pi}$} :}
We can obtain the transformation rules from the Eq.~\ref{eq_gauge_trans_sigv} and \ref{eq_gauge_trans_c2z}
{
\begin{equation}
\begin{aligned}
R_{\pi}: \begin{array}{l}\{\mu^x,\mu^z\}_{a} \rightarrow \{\mu^x,\mu^z\}_{R_{\pi}({a})} \\
\{\tilde{\mu}^x,\tilde{\mu}^z\}_{\bar{a}} \rightarrow \{\tilde{\mu}^x,\tilde{\mu}^z\}_{R_{\pi}({\bar{a}})} \\
\{\rho^x,\rho^z\}_{ab} \rightarrow \{-\rho^x,-\rho^z\}_{R_{\pi}({ab})} \\
\{\tilde{\rho}^x,\tilde{\rho}^z\}_{\bar{a}\bar{b}} \rightarrow \{-\tilde{\rho}^x,-\tilde{\rho}^z\}_{R_{\pi}({\bar{a}\bar{b}})}
\end{array}
\end{aligned}\label{eq_gauge_trans_rpi}
\end{equation}
}


\section{$J-K$ Hamiltonian}
\label{sec_jk_ham}

\begin{figure}
\includegraphics[]{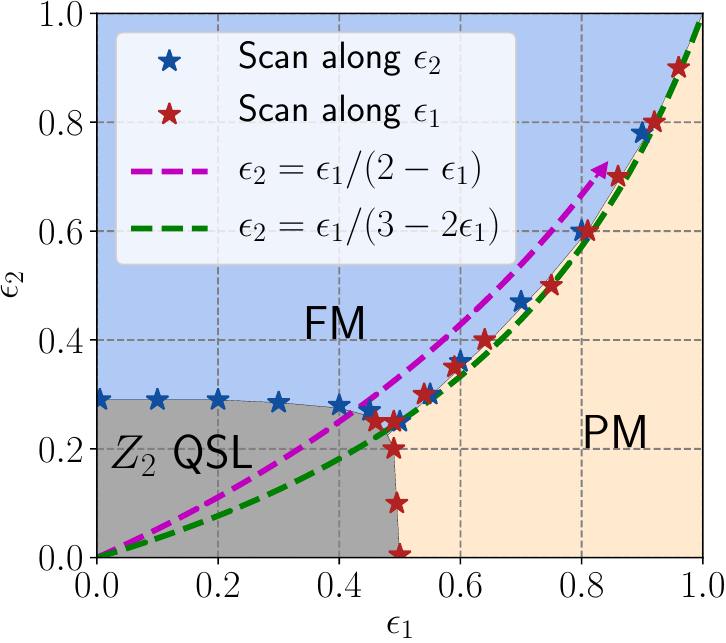}
\caption{Phase diagram of Eq.~\ref{eq_jk_general_pm_t1t2} where we consider a general paramagnetic field along with the toric code and Ising contribution. The three phases are (i) ferromagnet (FM), (ii) Paramagnet (PM) and (iii) toric code spin liquid ($Z_2$ QSL). The green dashed line indicates the expected phase boundary between ferromagnetic and the paramagnetic phase in absence of the toric code contribution (see text).  The magenta dashed shows the effect of Heisenberg coupling (J) on the anisotropic anti-ferromagnetic Kitaev model (see Eq.~\ref{eq_heisenberg_limit_min}).}
\label{fig_33_tcm_zx_y}
\end{figure}

The generalisation of the Hamiltonian for the antiferromagnetic Kitaev model in the strong anisotropic limit with the Heisenberg term (Eqs. \ref{eq_tc_rot_unrot} and \ref{eq_heisenberg_limit_min_afm}) is given by
\begin{equation}\label{eq_JKoriginal}
\begin{aligned}
\mathcal{H}^{AF}_{\Gamma=0} = & \ h_{\text{eff}}\sum_{i}\tau^x_i - J_{\text{eff}}\sum_{\langle i,j\rangle}\tau_i^z\tau_j^z \\
& - J^{\text{TC}}_{\text{eff}}\sum_{i}\tau^z_{i+d_1}\tau^z_{i-d_2}\tau^y_i\tau^y_{i+d_1-d_2}
\end{aligned}
\end{equation}
where $J_{\text{eff}}$, $h_{\text{eff}}$ and $J^{\text{TC}}_{\text{eff}}$ are the strengths of the Ising term, magnetic field  and of the quartic term respectively. On transforming the above Hamiltonian via a unitary rotation in Eq.~\ref{eq_rot_ver} followed by $\tilde{\tau}^y_i \rightarrow -\tilde{\tau}^y_i$ on the horizontal bonds, we get
\begin{equation}\label{eq_jk_general_pm}
\begin{aligned}
\mathcal{H}^{AF}_{\Gamma=0} = & \ h_{\text{eff}}\sum_{i}\ttau^y_i - J_{\text{eff}}\sum_{\langle i,j\rangle, i \in H, j\in V}\ttau_i^z\ttau_j^x \\
& - J^{\text{TC}}_{\text{eff}} \Big( \sum_s A_s  + \sum_p B_p \Big)
\end{aligned}
\end{equation}
which now takes the form the toric code Hamiltonian when perturbed by a {\it transverse} magnetic field and an Ising perturbation, although of a $\ttau^z\ttau^x$ kind. This Hamiltonian, in parts, has been a subject of recent numerical studies \cite{Vidal_PRB_2009, Dusuel_PRL_2011}; and we now investigate it further to develop a field theoretic understanding of the phases and intervening phase transitions. To understand this phase diagram numerically we define two interpolating parameters: $\epsilon_1$ and $\epsilon_2$, and study the following Hamiltonian

\begin{align}\label{eq_jk_general_pm_t1t2}
\mathcal{H}^{\prime} = & \ \epsilon_1(1-\epsilon_2)\sum_{i}\ttau^y_i - \epsilon_2(1-\epsilon_1)\sum_{\langle i,j\rangle, i \in H, j\in V}\ttau_i^z\ttau_j^x \nonumber \\
& - (1-\epsilon_1)(1-\epsilon_2)\left(\sum_s A_s + \sum_p B_p\right)
\end{align}
which interpolates between the exact toric code Hamiltonian ($\epsilon_1=\epsilon_2=0$), a $z-x$ ferromagnet ($\epsilon_1=0,\epsilon_2=1$) and a $y$ paramagnet ($\epsilon_1=1,\epsilon_2=0$). We perform exact diagonalization (ED) studies on a $18$ spin (3$\times$3) periodic cluster and track the ground state fidelity and other observables to identify  the phase boundaries. The numerically obtained phase diagram is shown in Fig.~\ref{fig_33_tcm_zx_y}. 

\begin{figure}
\centering
\includegraphics[]{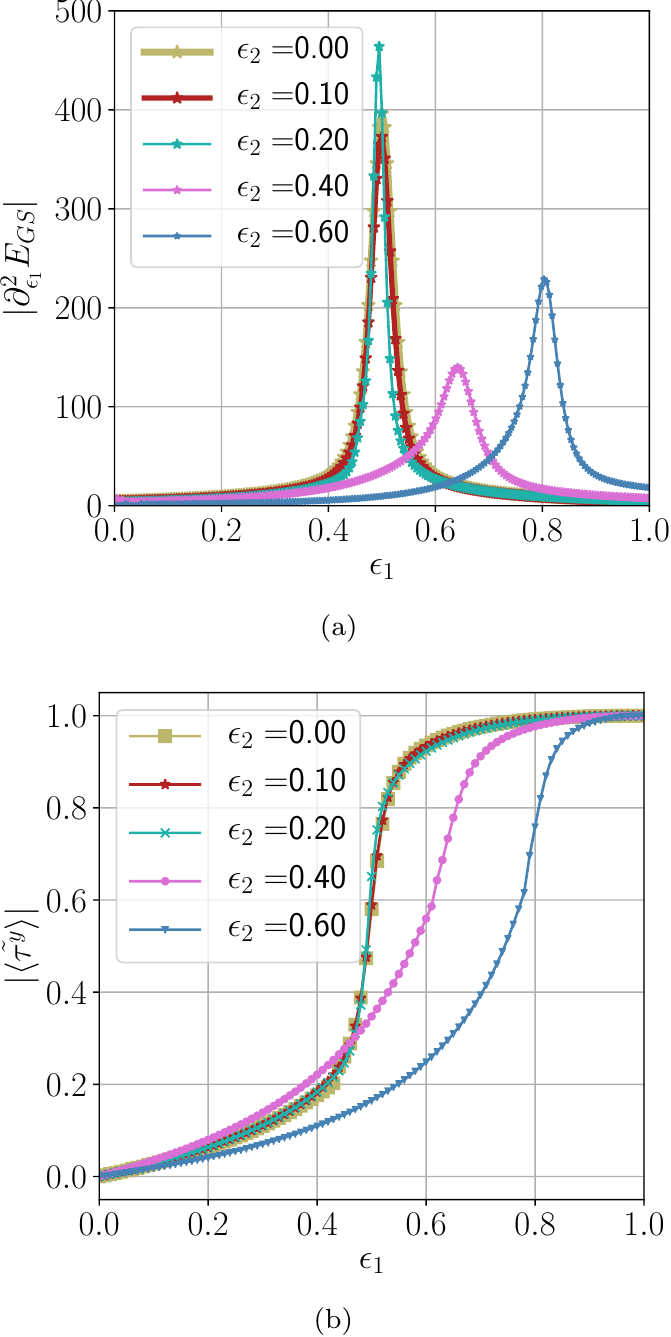}
\caption{The absolute value of ground state (GS) susceptibility ($\frac{\partial^2 E_{GS}}{\partial \epsilon^2}|_{\epsilon_2}$) (a) and the absolute value of $\ttau^y$-magnetization (b) as a function of $t_1$ for constant values of $t_2$ are shown for the Hamiltonian given in Eq.~\ref{eq_jk_general_pm_t1t2}.}
\label{fig_sus_mag_tcm_zz_y_t1_scan}
\end{figure}

In absence of the Ising term, i.e.~the toric code Hamiltonian with a transverse field, is self dual under ($h_{\text{eff}} \leftrightarrow  J^{TC}_{\text{eff}}$) which is known to be a first order transition at $\epsilon_1=0.5$ \cite{Vidal_PRB_2009, Dusuel_PRL_2011}.  Here in our finite sized system this transition shows up as a peak in the fidelity susceptibility which does not change remarkably with increasing $\epsilon_2$ ($\sim J$) suggesting that the transition is stable with increasing $\epsilon_2$ (see Fig.\ref{fig_sus_mag_tcm_zz_y_t1_scan} (a)). However, strictly our finite size numerics cannot distinguish the order of transition when transiting to either the large $\Gamma$ phase or the FM. The transition is concomitant with a finite magnetization (along the transverse field) signaling a transition to a paramagnetic (polarized) phase (see Fig.\ref{fig_sus_mag_tcm_zz_y_t1_scan} (b)). In absence of the toric code term, the Ising magnet to paramagnet transition is governed by the Ising transition and occurs at $h_{\text{eff}} \sim 3 J_{\text{eff}}$ \cite{Blote_PRE_2002, Albuquerque_PRB_2010,   Blass_srep_2dTFIM_2016, Huang_PRB_2020} where the ordered phase spontaneously breaks a $Z_2$ Ising symmetry operator given by $\prod_i \ttau^y_i$.  This corresponds to $\epsilon_2 = \frac{\epsilon_1}{3-2\epsilon_1}$ (dashed green) line in the $\epsilon_1-\epsilon_2$ phase diagram (see Fig.~\ref{fig_33_tcm_zx_y}). Clearly the numerically obtained phase boundary follows this quite closely specially when the toric code term is small ($\epsilon_1, \epsilon_2 >0.5$). We find that this second order line and the (expected) first order line (separating the $Z_2$ QSL and the paramagnet) meet at $\epsilon_1\sim 0.5, \epsilon_2 \sim 0.3$, potentially a multicritical point. The phase boundary between the $Z_2$ QSL and the Ising ferromagnet (in absence of any magnetic field) \cite{nanda2020phases} (see section \ref{sec_jk_no_pm_ft}) is argued to be a 3D-Higgs transition with mutual Chern Simons term. We find that this transition with increasing $\epsilon_1$ remains stable and meets both the first order line and the second order Ising transition line again at $\epsilon_1\sim 0.5, \epsilon_2 \sim 0.3$.

Having discussed the phase diagram of the generalized $J-K$ Hamiltonian (see Eq.~\ref{eq_jk_general_pm}) we now specify which intervening phases to expect as we increase the Heisenberg coupling in the anisotropic limit. Given the form of the effective Hamiltonian (see Eq.~\ref{eq_heisenberg_limit_min}) we find that $h_{\text{eff}} = 2 J_{\text{eff}}$ which corresponds to $\epsilon_2 = \frac{\epsilon_1}{2-\epsilon_1}$ line (shown in a magenta dashed line with an arrow) in Fig.~\ref{fig_33_tcm_zx_y} suggesting a single step transition. 

\section{Summary of the 1D cluster phase \texorpdfstring{($Z_2\times Z_2$ SPT)}{}}
\label{appen_cluster_spt}

Here we briefly summarise the essential results for one dimensional cluster model for completion. The one dimensional (on an open chain) cluster model Hamiltonian is given by~\cite{verresen2017one,you2018subsystem,chen2014symmetry,son2012topological,nielsen2006cluster,dubinkin2019higher}
\begin{align}
    H_{1d}=\sum_{i=2}^{N-1} \mathcal{U}_i
    \label{eq_clusterham}
\end{align}
where $\mathcal{U}_i=\tau^z_{i-1}\tau^x_i\tau^z_{i+1}$ and we consider $N\in {\rm Even}$. The Hamiltonian, in particular, is symmetric under a $Z_2\times Z_2$ transformation generated by
\begin{align}
 P_1 &= \prod_{i=2}^{N/2} \tau^x_{2i-1}=\tau^x_1\tau^z_2\left(\prod_{i=2}^{N/2}\mathcal{U}_{2i-1}\right)\tau^z_{N}\\
P_2 &= \prod_{i=1}^{N/2} \tau^x_{2i}=\tau^z_1\left(\prod_{i=1}^{N/2-1}\mathcal{U}_{2i}\right)\tau^z_{N-1}\tau^x_N
 \label{clus1D}
\end{align}
The Hamiltonian in Eq.~\ref{eq_clusterham} is exactly solvable since $[\mathcal{U}_i,\mathcal{U}_j]=0~~~\forall~i,j$. Since $\mathcal{U}_i^2=1$, the ground state, $|\psi_g\rangle$, satisfies
\begin{align}
    \mathcal{U}_i|\psi_g\rangle=-|\psi_g\rangle~~~~~~~\forall~i
    \label{eq_stab_1dcluster}
\end{align}
and can be obtained explicitly as 
\begin{equation}\label{eq_spt_1_gs}
    \ket{\psi_g}=\prod_{i}\left[\frac{1-\mathcal{U}_{2i-1}}{2}\right] \ket{\tau^x_{2i}=-1}\ket{\tau^z_{2i\pm 1}=1} 
\end{equation}

Therefore for the ground state on the open chain
\begin{align}
    P_1|\Psi_g\rangle&=(-1)^{N/2-1}\tau^x_1\tau^z_2\tau^z_N|\Psi_g\rangle\\
    P_2|\Psi_g\rangle&=(-1)^{N/2-1}\tau^z_1\tau^z_{N-1}\tau^x_N|\Psi_g\rangle
\end{align}
Assuming that $(N/2-1)\in~{\rm Even}$, We find that the two conserved operators $P_1$ and $P_2$ have non-trivial structure at the two edges of the open chain, {\it i.e.},
\begin{align}
    P_{1L}=\tau^x_1\tau^z_2~~~~~~~~P_{2L}=\tau^z_1
    \label{eq_ledge}
\end{align}
for the left edge and
\begin{align}
    P_{1R}=\tau^z_N~~~~~~~~P_{2R}=\tau^z_{N-1}\tau^x_N
    \label{eq_redge}
\end{align}
for the right edge such that the edge operators anti-commute on the same edge leading to a four dimensional representation of ground state manifold generated by
\begin{align}
    |P_{1L}=\pm1,P_{1R}=\pm1\rangle,
\end{align}
with each edge supporting a zero energy spin-1/2 or equivalently a complex fermion mode that transforms under a projective representation of the above $Z_2\times Z_2$ symmetry. In fact due to exact solvability, each energy eigenstate is four-fold degenerate on the open chain~\cite{fendley2016strong} The edge modes are characteristic signature of the one dimensional $Z_2\times Z_2$ SPT.

Since the Hamiltonian in Eq.~\ref{clus1D} is invariant under the global spin-flip generated by $P_1P_2=\prod\tau^x_i$, we can map it to a fermionic Hamiltonian via the following one dimensional Jordan-Wigner transformations~\cite{verresen2017one} :
\begin{align}
    \gamma_i=\left(\prod_{j=1}^{i-1}\tau^x_j\right)\tau^z_i,~~~~~~~~~~\tilde\gamma_i=\left(\prod_{j=1}^{i-1}\tau^x_j\right)\tau^y_i
\label{eq_JWmajorana}
\end{align}
into the Majorana fermions $\gamma_i$ and $\tilde\gamma_i$ whence we get $\mathcal{U}_i=i\tilde\gamma_{i-1}\gamma_{i+1}$ such that Eq.~\ref{clus1D} becomes
\begin{align}
    H_{1d}=\sum_{j=2}(i\tilde\gamma_{j-1}\gamma_{j+1})
    \label{eq_fermioncluster}
\end{align}
which is nothing but two stacked Kitaev superconducting chains \cite{kitaev2001unpaired} with a complex fermionic mode at each boundary which are annihilated respectively on the left and right edge by
\begin{align}
    c_L=(\gamma_1+i\gamma_2)/2\quad{\rm and}\quad c_R=(\tilde\gamma_{N-1}+i\tilde\gamma_N)/2
    \label{eq_fermionedge}
\end{align}

The generator of the spin-flips is local under the Jordan-Wigner Transformation, {\it i.e}
\begin{align}
    \tau^x_i=-i\tilde\gamma_i\gamma_i
\end{align}
and is related to the fermion parity operator. Therefore the generators of the $Z_2\times Z_2$ symmetry becomes
\begin{align}
    \mathcal{P}_1=\prod_{j=2}^{N/2}\left(-i\tilde\gamma_{2j-1}\gamma_{2j-1}\right),\quad \mathcal{P}_2=\prod_{j=1}^{N/2}\left(-i\tilde\gamma_{2j}\gamma_{2j}\right)
\end{align}
which shows that the parity of the even sites and the odd sites are separately preserved. Now following arguments similar to those given above we can find the edge representations of the symmetry in terms of the complex fermions given by Eq.~\ref{eq_fermionedge}.

Remarkably, the representation in terms of the majorana fermions reveal further rich symmetry structures of the cluster Hamiltonian through its fermionic form~\cite{verresen2017one} which usefully connects to the microscopic symmetries in the our case. This is seen by noticing that the fermionic representation of the cluster Hamiltonian in Eq.~\ref{eq_fermioncluster} is invariant under the following anti-unitary transformations :
\begin{align}\label{1DclusSyms}
    V_1=P_1\prod_{j=2}^{N/2}\mathcal{K}_{2j-1}:&\left\{\begin{array}{ll}
    \{\tilde\gamma_{2j},\gamma_{2j}\}&\rightarrow \{\tilde\gamma_{2j},\gamma_{2j}\}\\
    \{\tilde\gamma_{2j-1},\gamma_{2j-1}\}&\rightarrow \{\tilde\gamma_{2j-1},-\gamma_{2j-1}\}
    \end{array}\right.\\
    V_2=P_2\prod_{j=1}^{N/2}\mathcal{K}_{2j}:&\left\{\begin{array}{ll}
    \{\tilde\gamma_{2j},\gamma_{2j}\}&\rightarrow \{\tilde\gamma_{2j},-\gamma_{2j}\}\\
    \{\tilde\gamma_{2j-1},\gamma_{2j-1}\}&\rightarrow \{\tilde\gamma_{2j-1},\gamma_{2j-1}\}
    \end{array}\right.\\
    V_3=\prod_{j=2}^{N/2}\mathcal{K}_{2j-1}:&\left\{\begin{array}{ll}
    \{\tilde\gamma_{2j},\gamma_{2j}\}&\rightarrow \{\tilde\gamma_{2j},\gamma_{2j}\}\\
    \{\tilde\gamma_{2j-1},\gamma_{2j-1}\}&\rightarrow \{-\tilde\gamma_{2j-1},\gamma_{2j-1}\}
    \end{array}\right.\\
    V_4=\prod_{j=1}^{N/2}\mathcal{K}_{2j}:&\left\{\begin{array}{ll}
    \{\tilde\gamma_{2j},\gamma_{2j}\}&\rightarrow \{-\tilde\gamma_{2j},\gamma_{2j}\}\\
    \{\tilde\gamma_{2j-1},\gamma_{2j-1}\}&\rightarrow \{\tilde\gamma_{2j-1},\gamma_{2j-1}\}
    \end{array}\right.
    \end{align}
where $\mathcal{K}_j$ is the complex conjugation operator at site $j$. Clearly the four transformations are related to the microscopic symmetries and the $Z_2\times Z_2$ spin-flip symmetries as follows :
\begin{align}
    &P_1=V_1V_3;\quad\quad P_2=V_2V_4\nonumber\\
    &\mathcal{T}=V_1V_2;~\quad\quad \mathcal{K}=V_3V_4
\end{align}
where $\mathcal{T}$ is the global non-Kramers time reversal defined in Table. \ref{table_tau_symm} and $\mathcal{K}$ is the global complex conjugation operator. Depending on convenience, we can either use $(P_1,P_2)$ or $(\mathcal{T},\mathcal{K})$ to understand the properties of the $Z_2\times Z_2$ SPT and the edge modes. However the flexibility allows us to study the fate of perturbations. 

Clearly a transverse field term of the form $h\sum_i\tau^x_i$ is invariant under the $Z_2\times Z_2$ symmetry and hence the SPT is perturbatively stable to it and gives away to a trivial paramagnet polarised in the $\tau^x$ direction through a quantum phase transition at $|h|=1$~\cite{verresen2017one}. This transition is described by a $SO(2)_1$ conformal field theory (CFT) with central charge, $c=1$~\cite{PhysRevLett.115.237203}. 

A transverse field perturbation along $\tau^y$, {\it i.e.} $h\sum_i\tau^x_i$, however it naively appears that the above $Z_2\times Z_2$ symmetry is broken. To be precise, we consider the  $(\mathcal{T},\mathcal{K})$ implementation of the symmetries. While the above term is invariant under $\mathcal{T}$, it changes sign under $\mathcal{K}$. However such change in sign can be rectified by applying unitary global spin-flip $P_1P_2$ and thus rendering the above perturbation invariant under the $Z_2\times Z_2$ symmetry. Indeed the SPT is perturbatively stable under the above transverse field and gives away to the trivial $\tau^y$-polarised phase through the similar critical point as for the $\tau^x$ case above. 

\section{${\cal W}$ transformation}
\label{sec:Wtrans}

\begin{figure}
\centering
\includegraphics[]{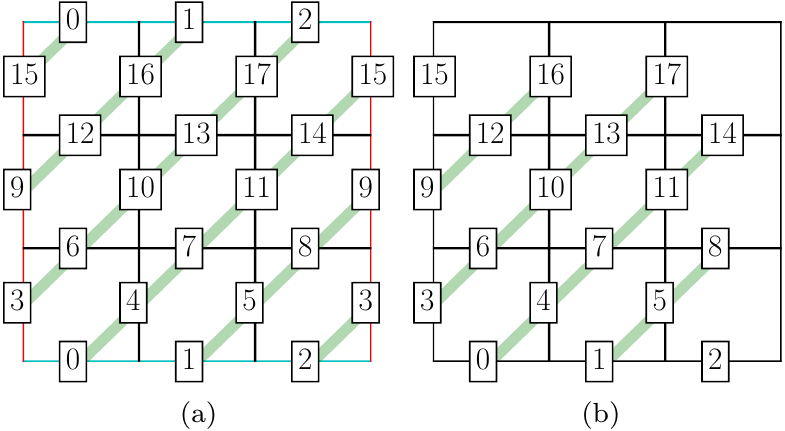}
\caption{$U_{ij}$ are transformations on the bonds (see Eq.~\ref{Uijtrans}) connected shown in the cluster for (a) PBC, and (b) OBC.}
\label{fig_33_string}
\end{figure}

In order to build intuition for the phase diagram in the $(\lambda_1,\lambda_2)$ plane (Figs. \ref{fig_anticipatd_lamda_pd} and \ref{fig_anticipatd_transform_lamda_pd})  and the nature of transitions we summarise the effect of the unitary transformation, $\mathcal{W}$ (Eq.~\ref{Wtrans}), applied on the Hamiltonian (see Eq.\ref{eq_four_spt_interpolation}). The transformation, defined on a bond $ij$, follows $U_{ij} = U_{ji} = U_{ij}^\dagger$. The bonds involved in a periodic and open system are shown in Fig.~\ref{fig_33_string}. While it is straightforward to see how periodic Hamiltonian then transforms from to $H_\alpha$ to $\tilde{H_\alpha}$, we discuss the same physics in an open system below to understand the intricacies of the boundary modes.

Consider spins $i$ where $i \in {\cal B}$, $i \in {\bf t}$ , $i \in {\bf b}$, $i \in {\bf l}$, $i \in {\bf r}$ and $i \in {\bf c}$ represents bulk, top, bottom, left, right boundary and corner of the cluster respectively. For e.g~, in the cluster shown in Fig.~\ref{fig_33_string}(b), 
${\cal B} = \{4,5,6,7,10,11,12,13\}, {\bf t} = \{16,17\},{\bf b} = \{0,1\} , {\bf r} = \{14,8\}, {\bf l} = \{9,3\}, {\bf c} = \{15,2\}$. When the transformation ${\cal W}$ is performed on a open problem one obtains (for Hamiltonian $H$ in Eq.~\ref{eq_four_spt_interpolation} in OBC) $\tilde{H}$ as 

\begin{align}
\tilde{H}_1&= \sum_{i \in {\cal B}} \tau^x_i    \\
\tilde{H}_2&= \sum_{i \in {\cal B}} \tau^x_i \tau^z_{i+d_1} \tau^z_{i-d_1} \tau^z_{i+d_2} \tau^z_{i-d_2}     \\
\tilde{H}_3&= \sum_{i \in V,{\cal B} } \tau^z_{i-d_1}\tau^y_{i}\tau^z_{i-d_2}-\sum_{i \in H,{\cal B}} \tau^z_{i+d_2}\tau^y_{i}\tau^z_{i+d_1}
 \notag \\ &+  \sum_{i \in {\bf l}}  \tau^y_i \tau^z_{i-d_2} +\sum_{i \in {\bf r}}  \tau^y_i \tau^z_{i+d_2}   \\
\tilde{H}_4 &=  \sum_{i \in H,{\cal B} } \tau^z_{i-d_1}\tau^y_{i}\tau^z_{i-d_2}-\sum_{i \in V, {\cal B}}\tau^z_{i+d_2}\tau^y_{i}\tau^z_{i+d_1}
\end{align}
$H_1$ has a set of $2^{2(L_x + L_y-1)}$ degeneracy which is reflected in the fact that $\tilde{H_1}$ has no terms which involve boundary spins. $H_2$ has a set of $2^{2(L_x + L_y-1)}$ degeneracy given it is a SSPT on an open system. $\tilde{H_3}$ has free spins on top and bottom boundaries while symmetry breaking terms on left and right boundary. This leads to a degeneracy of $2^{2L_x}$. Since $\tilde{H}_4$ has free spins on boundaries it again as $2^{2(L_x+L_y-1)}$ degeneracy. These are the exact ground state degeneracies for $H_1,H_2,H_3, H_4$ when placed in an open system. The analysis therefore shows that ${\cal W}$ is suitably defined for both open and periodic systems.

\section{Excitations and their dynamics in the pure $\Gamma$ limit}
\label{sec:excitations}

\begin{figure}
    \centering
    \includegraphics[]{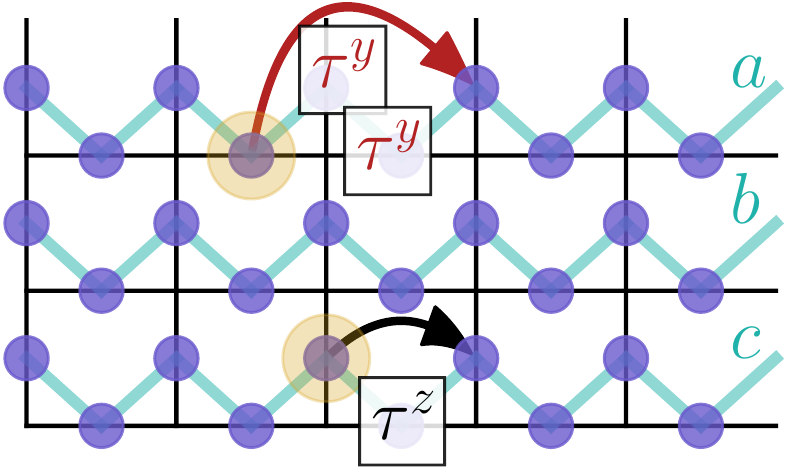}
    \caption{Excitations of the cluster state and its dynamics in presence of magnetic fields. Three chains labeled by $a,b,c$ are shown.  The orange blob shows the initial position of the $\langle \tau^x_{i-d_1}\tau^z_{i}\tau^x_{i-d_2} \rangle = 1$  excitations (on chain $a$ and $c$) which hops under the action $\tau^y (\tau^z)$ fields to third (first) neighbour at the second (first) order of perturbation theory.}
    \label{fig:XZXpm}
\end{figure}

Consider the Hamiltonian
\begin{align}
 H &= \sum_{i \in V} \tau^x_{i-d_1}\tau^z_{i}\tau^x_{i-d_2}+\sum_{i \in H}\tau^x_{i+d_2}\tau^z_{i}\tau^x_{i+d_1} \nonumber\\ & +  h_y \sum_{i \in H,V} \tau^y_i +   h_z \sum_{i \in H,V} \tau^z_i \label{eq_1Dcluster}
\end{align}
which perturbs the cluster Hamiltonian (similar to Eq.~\ref{eq_clusterham}) with a magnetic field in $y$ direction ($\equiv h_y$) and $z$ direction ($\equiv h_z$). Both these are symmetry allowed and in either field there exists a second order transition with $c=1$. Note that in Eq.~\ref{eq_4_spt_rot_4} for $\lambda_1=\lambda_2=0$ the cluster Hamiltonian in $\tilde{H_3}$ is of the above form where perturbations along $\tilde{H_1}$ direction is essentially that of a $y$-field.

Here we explore the properties of the low energy excitations of the cluster state as the magnetic field is tuned to understand their role in the eventual transition to the trivial paramagnet. The ground state in absence of any fields is characterized by  $\langle \tau^x_{i-d_1}\tau^z_{i}\tau^x_{i-d_2} \rangle = -1$ for every $i$, where an excitation with energy gap $=2$ localized at particular site is given by $\langle \tau^x_{i-d_1}\tau^z_{i}\tau^x_{i-d_2} \rangle = 1$. A $y$ field can effectively hop a charge by three lattice constants at quadratic order, but a $z$ field hops it by two lattice constants at linear order (see Fig.~\ref{fig:XZXpm}).

\begin{figure}
    \centering
    \includegraphics[]{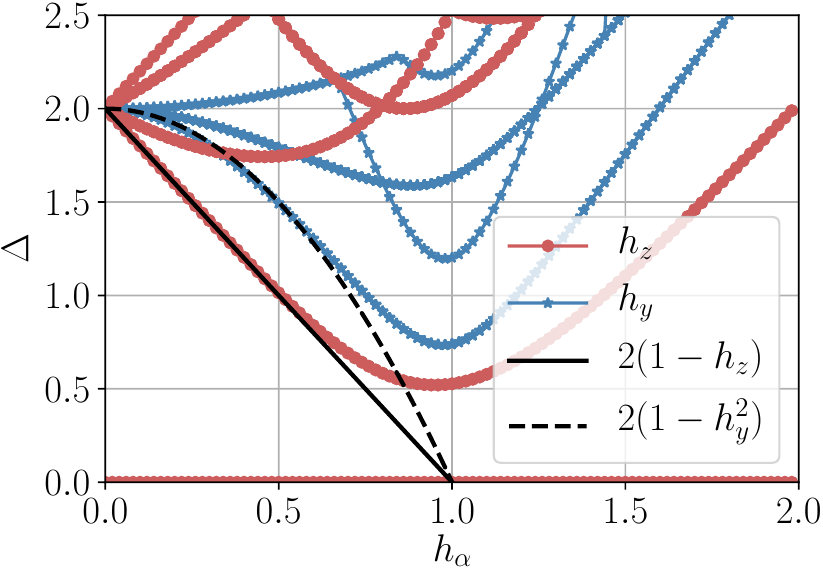}
    \caption{The low energy spectrum of a one dimensional cluster state in presence of magnetic fields ($h_\alpha= \{h_x, h_z\}$) (see Eq.~\ref{eq_1Dcluster}) for a 16 site system.}
    \label{fig:XZXpmED}
\end{figure}

Therefore the charge gap behaves, for small fields, as $\Delta \sim 2 - 2 h_z$ or $\Delta \sim 2 - 2 h^2_y$ depending on the field direction both going to zero at $h_z$ (or $h_y$) $=1$  signaling that the Ising transition (with $c=1$) can be understood as the condensation of these excitations. The exact diagonalization spectrum and how the low energy spectrum behaves is shown in Fig.~\ref{fig:XZXpmED}. 

Under a unitary rotation ($\{\tau^x,\tau^y,\tau^z\} \rightarrow \{-\tau^z,\tau^y,\tau^x \}$) where the cluster Hamiltonian gets mapped to Eq.~\ref{eq_clusterham} and perturbation $h_z (h_y)$ leads to a $x (y)$ polarized state. Using transformation to Majorana operators (see Eq.~\ref{eq_JWmajorana}) and defining bond complex fermion operators through 
\begin{align}
    c_i &= \frac{1}{2}(\gamma_{i-1} + i \tilde{\gamma}_{i+1})
\end{align}
Eq.~\ref{eq_1Dcluster} becomes
\begin{align}
    H&=\sum_i (2n_i-1) \\ &+ h_z\sum_i\Big( c_{i+1}c_{i-1}+c^{\dagger}_{i+1}c_{i-1} + c^{\dagger}_{i-1}c_{i+1} + c^{\dagger}_{i+1}c^{\dagger}_{i-1} \Big) \notag \\
    &- h_y \sum_i \prod^{i-1}_{j} \Big(i \gamma_j \tilde{\gamma_j}\Big) \Big[i(c_{i-1}-c^\dagger_{i-1})\Big]
\end{align}
where $n_i=c_i^\dagger c_i$. Therefore the ground state of the cluster Hamiltonian ($h_y=h_z=0$) is given by $\langle n_i \rangle =0 ~~\forall i$ while the excitations are given by particles at site $i$ with $\langle n_i\rangle= +1$.  Using this fermionic description it is easily seen that $\tau^x_i$ leads to a hopping process by two lattice sites in the single excitation sector; while a $\tau^y_i$ operator changes the parity sector (along with a string) leading to creation of charges. A quadratic action of $\tau^y$ brings it to the same excitation sector leading to an effective hopping by three lattice sites.

While the magnetic field terms above are ultra local and cannot lead to any dispersion of a single excitations in the vertical direction for the stacked system (Eq.~\ref{eq_four_spt_interpolation}) -- no interchain couplings of the kind mediated by ($\tilde{H_4}$) or by ($\tilde{H_2}$) can lead to any vertical dispersion for these single excitations. This leads to the fact that the $\lambda_2=1$ (even at a non-zero $\lambda_1$) transition is extremely anisotropic in character where the spin-spin correlations are expected to be power law only in the $x$ direction, while continues to remain short ranged in the $y$-direction. At $\lambda_2=0$, $\lambda_1$ direction creates no dynamics in the single excitation sector, but perturbatively brings down the two-excitation sector. However before the gap to the two-excitation sector closes, a level crossing mediated by an excited state with a host of excitations leads to a first order transition at $\lambda_2=1$.

To understand the role of the subsystem symmetries on the dynamics, we consider the cluster term of the Hamiltonian in Eq. ~\ref{eq_1Dcluster} and revisit the above discussion in light of the sub-system symmetries. Consider the Hamiltonian given in Eq.~\ref{eq_1Dcluster} which has two equivalent way of considering the sub-symmetries which protect the SPT order.
$(i)~~P_1 = \prod_i \tau^z_{2i-1}, P_2 = \prod_i \tau^z_{2i}$ and $(ii) P_1 = \prod_i \tau^z_{2i-1} {\cal K}_{2i-1}, P_2 = \prod_i \tau^z_{2i} {\cal K}_{2i}$. A $z$ perturbation preserves both pairs of symmetries (i) and (ii), leading to a sublattice preservation of the excitations. On the other hand a $y$ perturbation, given the way time-reversal symmetry behaves in this system, continues to preserve (ii), and does not change the eigenvalues of the horizontal sub-system symmetries with anti-unitary character. It is the same way that the $y$-perturbations don't change the eigenvalue of vertical subsystem symmetries (see Eq.~\ref{eq_sspt_ver_symm}). Therefore, excitations over the cluster state as generated by the $y$ field are not constrained by horizontal and vertical sub-system symmetries. 

\begin{figure}
\centering
\includegraphics[]{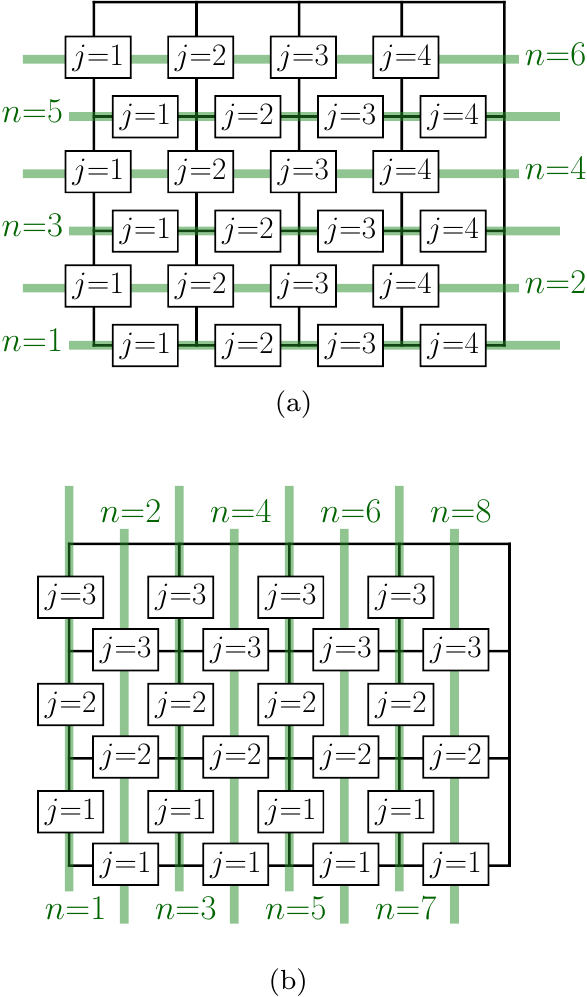}
\caption{The horizontal (in (a)) and vertical (in (b)) subsystems symmetries, in an open $(L_x,L_y)=(4,3)$ system.}
\label{fig_43_sub_sym}
\end{figure}

\section{Boundary modes of large \texorpdfstring{$\Gamma$}{} Hamiltonian}\label{subsubsec_disc_bound_gamma}

Our discussion of the large $\Gamma$ phase in $(\lambda_1, \lambda_2)$ plane in section~\ref{subsec_gamma_limit_afm} focused on the bulk system, the corresponding phases and tentative phase diagram of the same. We now discuss the nature of boundary modes in this system and which symmetries protect them. 

We investigate how the subsystem symmetries (see Eq.~\ref{eq_spt_ver_symm_original}) act on the boundaries in the $(\lambda_1,\lambda_2)= (1,1)$ point. It is easier to start from the $\lambda_2=0$ line where we have a set of stacked cluster phases. When we are in one of the vertical SPTs (say $H_3$). Interestingly one finds that both in the ground states of $H_3$ and $H_4$,
\begin{equation}
\{ PT_{v_n}, PT_{v_{n+1}} \} = 0   
\end{equation}
on the top and bottom boundaries leading to $2^{2L_x}$ degeneracies where $PT_{v_n}$ symmetries are given in Eq.~\ref{eq_spt_ver_symm_original} (see Fig~\ref{fig_43_sub_sym}) . The corresponding horizontal subsystem symmetries given in Eq.~\ref{eq_spt_hor_symm_original}
commutes leading to no protected boundary modes on the right and left boundaries for $\tilde{H_3}$.

Given the $H_1$ and $H_2$ perturbations respect the symmetries given in Eq.~\ref{eq_spt_ver_symm_original}, the boundary modes on the top and bottom boundaries remain stable in all of $(\lambda_1, \lambda_2)$ plane as is found leading to $2^{2L_x}$ degeneracy even at the $(1,1)$ point. Interestingly given the energetics at the $\lambda_2=2$ line, one gets additional boundary modes on the left and right boundaries which increasing the degeneracy to $2^{2(L_x+L_y)}$.

We now investigate how the symmetries protect the boundary modes at $\lambda_2=2$ line where we have boundary modes on all the four boundaries. Here again for $H_1$ (see Eq.~\ref{eq_superposegamma}) each of the vertical and horizontal sub-system symmetries (see Eq.~\ref{eq_spt_ver_symm_original}) can be written as a product of stabilizers where it acts anomalously on the boundaries. For instance the horizontal subsystem symmetries behave as 
\begin{align}
    PT_{h_{n}}^L &= \tau^z_{j=1+d_2} \qquad \qquad \forall  ~n\in \text{odd} \\
    PT_{h_{n}}^R &= \tau^x_{j=L_x}\tau^z_{j=L_x-d_1} {\cal K}_{L_x}  \qquad \forall ~n\in \text{odd} \\
    PT_{h_{n}}^L &= \tau^x_{j=1}\tau^z_{j=1+d_1} {\cal K}_{j=1} \qquad \forall ~n\in \text{even} \\
    PT_{h_{n}}^R &= \tau^z_{j=L_x-d_2} \qquad \qquad \forall ~n\in \text{odd}
\end{align}
It is easy to see that these anti-commutes on the left and right boundaries. The vertical subsystem symmetries are given by
\begin{align}
    PT_{v_{n}}^B &= \tau^z_{j=1-d_2} \qquad \qquad \forall  ~n\in \text{odd} \\
    PT_{v_{n}}^T &= \tau^x_{j=L_y}\tau^z_{j=L_y-d_1} {\cal K}_{j=L_y} {\cal K}_{j=L_y-d_1}  ~~ \forall ~n\in \text{odd} \\
    PT_{v_{n}}^B &= \tau^x_{j=1}\tau^z_{j=1+d_1} {\cal K}_{j=1} {\cal K}_{j=1+d_1} \qquad \forall ~n\in \text{even} \\
    PT_{v_{n}}^T &= \tau^z_{j=L_y+d_2} {\cal K}_{j=L_y+d_2} \qquad \qquad \forall ~n\in \text{odd}
\end{align}
Since these anticommute on the top and bottom boundaries they again lead to the $2^{2L_x}$ degeneracy. This shows why $H_1$ has a $2^{2L_x+2L_y-2}$ degeneracy in the system.  Similar analysis for $H_2$ shows the same degeneracy count. Therefore on the $\lambda_2=2$ line the subs-system symmetries (see Eq.~\ref{eq_spt_ver_symm_original}) protect the boundary modes on all the boundaries.  Introduction of $H_3$ and $H_4$ even while they do not break the symmetries interfere with the anomalous character of the symmetry operators since the way they behave in the bulk is dependent on the stabilizers. Since $H_3$ and $H_4$ couple spins in the vertical direction, they immediately hybridize the free spins which lie on the left and right boundaries leading to the removal of degeneracies stabilized in the $\lambda_2=2$ limit. On the other hand these same vertical SPTs stabilize free spins on the top and bottom boundaries, as discussed before, and hence do not disturb the degeneracies there. Hence the complete $(\lambda_1,\lambda_2)$ has exact degeneracies on the top and bottom boundaries. Given this degeneracies are independent on the any finite size (or therefore even when the bulk gap is dominated by Kubo gaps), these are stable and occurs in all of $(\lambda_1, \lambda_2)$ plane.


\subsection{Effect of perturbations on the boundary  modes}\label{subsubsec_boun_pert_gamma}

The large $\Gamma$ phase (see Eq.~\ref{eq_four_spt_interpolation}),  at $(\lambda_1,\lambda_2)=(1,1)$) has a degeneracy of $2^{2L_x}$ in an OBC geometry, see Fig.~\ref{fig_33_string}(b), where $L_x$ is the length of the top and bottom boundaries. We now study the effect of various perturbations on this ground state degenerate manifold. 

\paragraph{\underline{Symmetry allowed Ising perturbation:}} When ferromagnetic Ising interactions among the boundary spins are introduced, which are allowed by the microscopic symmetries (see table \ref{table_tau_symm}), we find that the top and the bottom boundaries behave as one-dimensional Ising Hamiltonians which spontaneously break time-reversal symmetry to order in the $z$ direction. 

\begin{figure}
\centering
{\includegraphics[]{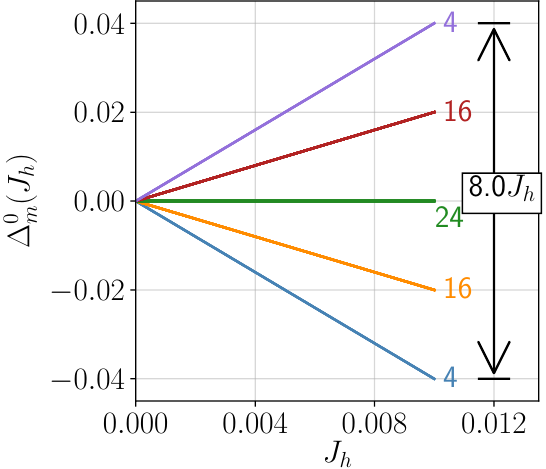}}
\caption{Evolution of the ground state degeneracy splits in presence of an Ising exchange perturbation, see Eq.~\ref{eq_pure_gamma_ising_boun}. Along the y-axis we plot the energy of the $m^{th}$ state after subtracting the $J_h=0$ GS energy, {\it i.e.} $\Delta^0_m(J_h)=E_m(J_h)-E_0(J_h=0)$.}
\label{fig_pure_gamma_boun_ising_pert}
\end{figure}

More concretely, in a $3\times 3$ cluster (see Fig.~\ref{fig_33_string}(b)) whose resulting Hamiltonian is 
\begin{equation}\label{eq_pure_gamma_ising_boun}
    H(J_{h})=H(1,1)-J_{h}\left(\tau^z_0\tau^z_1+\tau^z_1\tau^z_2+\tau^z_{15}\tau^z_{16}+\tau^z_{16}\tau^z_{17}\right)
\end{equation}
the splitting of the ground state degenerate manifold is shown in Fig.~\ref{fig_pure_gamma_boun_ising_pert}. Clearly among the $2^{2L_x}|_{L_x=3}=64$ degenerate states (at $J_h=0$), four unique states are chosen which correspond to the two Ising symmetry broken states in the top and bottom boundaries.

\paragraph{\underline{Bulk $\tau^{x(z)}$ field}:}
Next we apply the symmetry allowed (breaking) bulk $\tau^{x(z)}$-field to the $(\lambda_1,\lambda_2)=(1,1)$ point of the Hamiltonian in Eq.~\ref{eq_four_spt_interpolation} in an OBC geometry. We consider the Hamiltonian:

\begin{equation}\label{eq_bulk_xz_pert}
    H(h_{x(z)})=H(1,1)-h_{x(z)}\sum_i\tau^{x(z)}_i
\end{equation}

\begin{figure}
\centering
\includegraphics[]{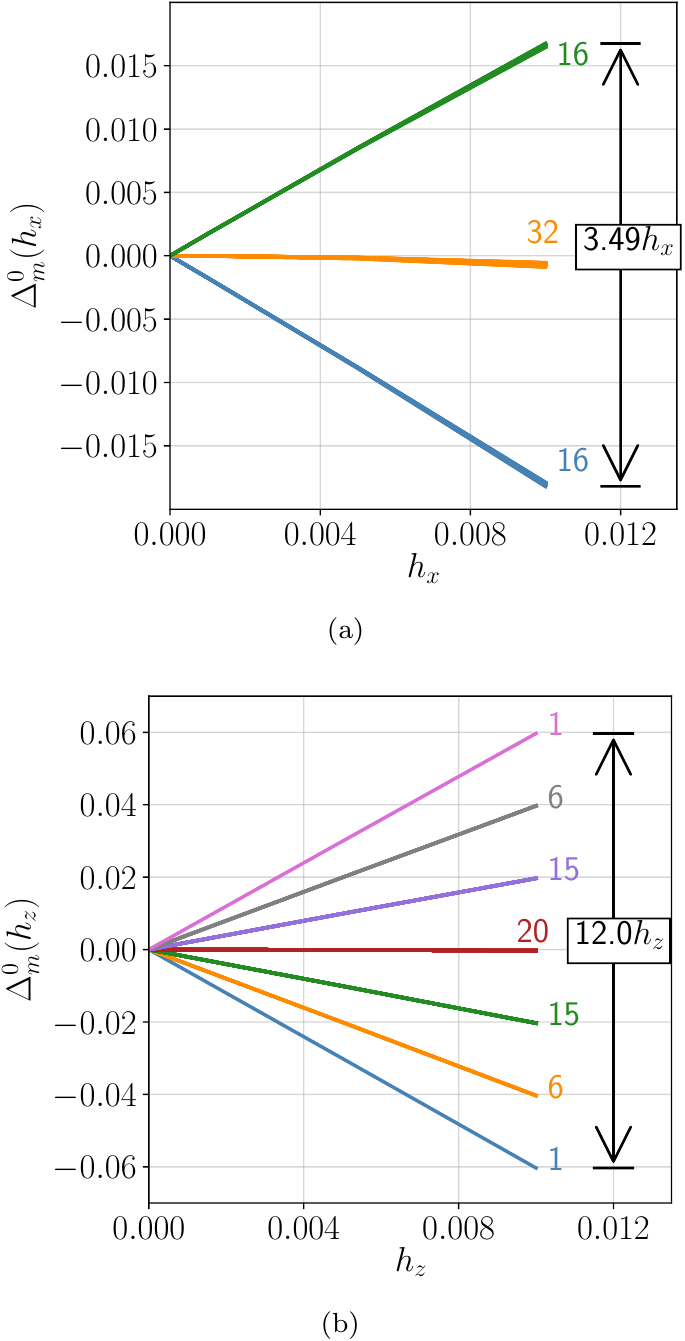}
\caption{Splitting of the degeneracy in an OBC geometry of the $H(1,1)$ (in Eq.~\ref{eq_four_spt_interpolation}) in the presence of $\tau^{x(z)}$-field as a function of the field strength ($h_{x(z)}$), also see Eq.~\ref{eq_bulk_xz_pert}.}
\label{fig_pure_gamma_bulk_field_pert}
\end{figure}

In Fig.~\ref{fig_pure_gamma_bulk_field_pert}(a) and \ref{fig_pure_gamma_bulk_field_pert}(b) we show the corresponding results. Even while a symmetry allowed $x$-field splits the 64-fold degeneracy of the same $3\times 3$ cluster (see Fig.~\ref{fig_33_string}(b)) into sub-branches, a time-reversal symmetry breaking $\tau^z$ field immediately polarizes the boundary spins into a unique ground state.

The above analysis shows that even while the large $\Gamma$ phase has a set of boundary modes (given its proximity to weak SPTs), these modes are extremely susceptible to both symmetry preserving and symmetry breaking perturbations, thereby reflecting their fragile character.

\section{Additional numerical results for the $\Gamma$ limit}
\label{sec_gamma_lim}

\begin{figure}
\includegraphics[]{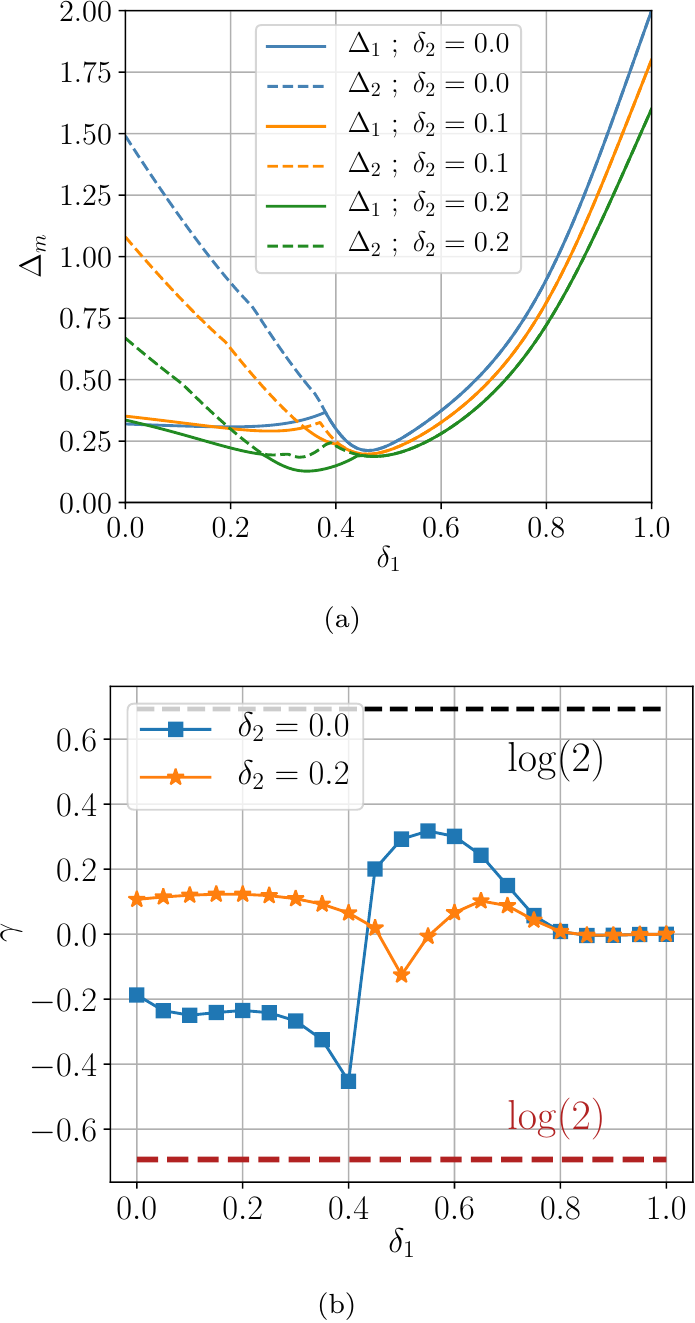}
\caption{(a) The first ($\Delta_1$) and the second ($\Delta_2$) gap to the GS energy following the Eq.~\ref{eq_gamma_zz_x} where we keep $\delta_2$ constant and vary the $\delta_1$ to tune the $\Gamma$-phase ($\delta_1=0$) to a $\tau^x$-paramagnet ($\delta_1=1$). The calculation is done for a system size $L_x\times\L_y=2\times4$. (b) The behavior of topological correction to the entanglement entropy ($\gamma$) is shown for the Hamiltonian in Eq.~\ref{eq_gamma_zz_x} as a function of $\delta_1$ for constant $\delta_2=0.0,~0.2$. The calculation is done in a $(L_x,L_y)=(4\times 3)$ lattice.}
\label{fig_gap_gamma_zz_x}
\end{figure}

To show that the large $\Gamma$ phase is indeed smoothly connected to the 
paramagnet, we tune it to the $x-$ paramagnet (via parameter $\delta_1$) in presence of Ising perturbation ($\propto \delta_2$) (see Eq.~\ref{eq_gamma_zz_x} in the main text). While the suscpetibility comes down with increasing strength of Ising perturbation(see Fig.~\ref{fig_sus_gamma_zz_x}), one finds that the minima of energy gaps ($min(\Delta_m) = m^{th}$ excitation gap) remains finite as a function of $\delta_1$ for different values of $\delta_2$ (see Fig.~\ref{fig_gap_gamma_zz_x}). One also finds that the topological entanglement entropy($\gamma$) remains close to zero across the complete interpolation showing that the large $\Gamma$ phase is not a gapped topologically ordered state.

\section{Additional numerical results for the $KJ\Gamma$ Hamiltonian}
\label{sec_JKGnumerical}

\begin{figure}
\centering
\includegraphics[]{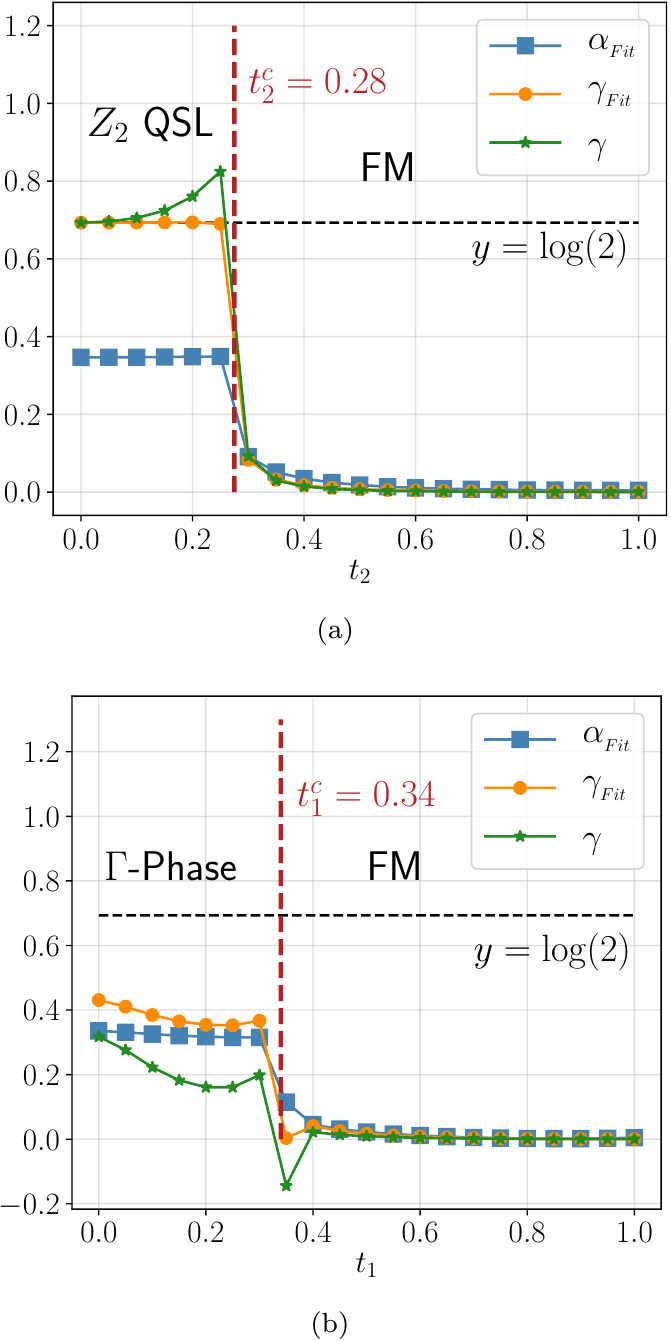}
\caption{Following Eq.~\ref{eq_ent_entropy} the values of $\alpha_{_{Fit}},~\gamma_{Fit},~\gamma$ are shown as we vary $t_2$ (see Fig.~\ref{fig_full_pd_43}) for a constant $t_1=0.0$ in (a) and $t_1=0.6$ in (b), the calculations are done in an $(L_x,L_y)=(4,3)$ system. Here a perturbing $\tau^z$ magnetic field has been applied to break the degeneracy between the two fold symmetry braking GS manifold.}
\label{fig_ent_ent}
\end{figure}

Here we present additional results for the behavior of bipartite entanglement for different cuts in the $KJ\Gamma$ phase diagram (see Fig.~\ref{fig_full_pd_43}). 
Following Eq.~\ref{eq_ent_entropy}, we calculate the different scaling coefficients of entanglement entropy (dubbed as $X_{_{Fit}}~;~X=\alpha,\gamma$) along with the topological entanglement entropy ($\gamma$) calculated using the Kitaev-Preskil method \cite{kitaev2006topological, kitaev2006topological}. The behavior of these quantities in the $t_2$ direction for  $t_1=0.0,~0.6$ is shown in Fig.~ \ref{fig_ent_ent}. Clearly both in FM and $\Gamma$ phase, $\gamma \sim 0$ while in $Z_2$ QSL, $\gamma \sim \log(2)$. 

\section{Field theory for the transition from TC to Heisenberg limit}\label{sec_jk_no_pm_ft}

Here we present additional details for the Gauge Mean Field Theory that describes the transition between the TC and the Heisenberg limit (discussed in section.~\ref{sec_phase_tran_tc_heisenberg}).

\subsection{Gauge Mean Field Theory}
\label{subsec_gmft}

Following Ref. \cite{nanda2020phases} we start our analysis by decoupling the first term in Eq.~\ref{eq_leading} within gauge mean field theory where the gauge fluctuations has been neglected. The first term in Eq.~\ref{eq_leading} is written using this decoupling as:  $\left[\mu^x_a\rho^z_{ab}\mu^x_b\right]\left[\rho^x_{bc}\right]\rightarrow \langle\mu^x_a\rho^z_{ab}\mu^x_b\rangle\rho^x_{bc} +\mu^x_a\rho^z_{ab}\mu^x_b\langle\rho^x_{bc}\rangle$. Thus the Eq.~\ref{eq_leading} becomes:
\begin{align}
\tilde{\mathcal{H}}^{AF}_{\Gamma=0}\rightarrow\tilde{\mathcal{H}}_{\Gamma=0}^{\rm GMFT}=\tilde{\mathcal{H}}_{\Gamma=0}^{\rm GMFT}(e)+\tilde{\mathcal{H}}_{\Gamma=0}^{\rm GMFT}(m)
\end{align}
where
\begin{align}
\tilde{\mathcal{H}}_{\Gamma=0}^{\rm GMFT}(e)=-\sum_{\langle ab\rangle\in H}J_{ab} \mu^x_a\rho^z_{ab}\mu^x_b-J_{TC}\sum_a\mu_a^z
\label{eq_egmft}
\end{align}
describes the $e$ sector with 
\begin{align}
J_{ab}= J\left[\langle\rho^x_{b,b-\hat y}\rangle+\langle\rho^x_{b,b+\hat y}\rangle+\langle\rho^x_{a,a-\hat y}\rangle+\langle\rho^x_{a,a+\hat y}\rangle\right]
\label{eq_jeff}
\end{align}
being the effective coupling and 
\begin{align}
\tilde{\mathcal{H}}_{\Gamma=0}^{\rm GMFT}(m)=-\sum_{\langle \bar{a}\bar{b}\rangle\in H}J_{\bar{a}\bar{b}} \tilde\mu^x_{\bar{a}}\tilde\rho^z_{\bar{a}\bar{b}}\tilde\mu^x_{\bar{b}}-J_{TC}\sum_{\bar{a}}\tilde\mu_{\bar{a}}^z
\label{eq_mgmft}
\end{align}
describes the $m$ sector with
\begin{align}
J_{\bar{a}\bar{b}}=J\left[\langle\tilde\rho^x_{\bar{b},\bar{b}-\hat y}\rangle+\langle\tilde\rho^x_{\bar{b},\bar{b}+\hat y}\rangle+\langle\tilde\rho^x_{\bar{a},\bar{a}-\hat y}\rangle+\langle\tilde\rho^x_{\bar{a},\bar{a}+\hat y}\rangle\right]
\label{eq_jmff}
\end{align}

Upto the first order this becomes a series of transverse field Ising chains in the horizontal direction, we choose the following gauge:
\begin{align}
\rho^z_{a,a+\hat x}=\tilde\rho^z_{\bar{a},\bar{a}+\hat x}=+1
\label{eq_gaugechoice}
\end{align}
Clearly in the presence of the Heisenberg term, the single excitation sector of {\it e} \& {\it m} acquires a dispersion, the condensation of these soft modes give rise to $\langle\mu^x\rangle\neq 0$  and $\langle\tilde\mu^x\rangle\neq 0$  
for the respective chains.

\begin{figure}
\includegraphics[]{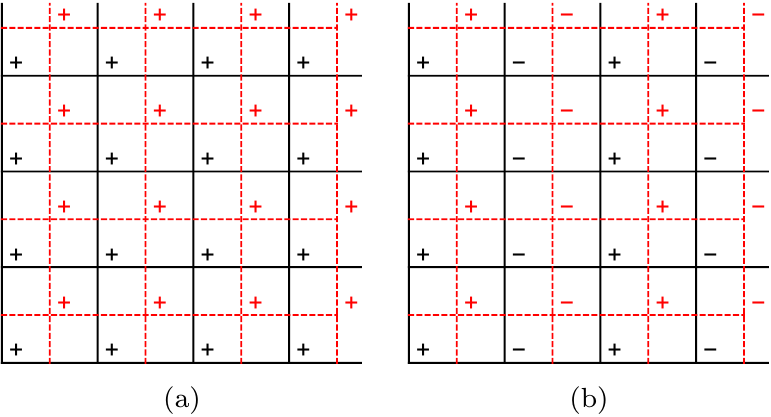}
\caption{The electric (black) and the magnetic (red) soft modes on the direct and dual lattice respectively. The $\boldsymbol{\pm}$ denotes $\mu^x=\pm 1$ and $\tilde\mu^x=\pm 1$ respectively. (a) and (b) shows the two time reversal partners respectively, $(\hat{\nu}_e^{(1)},\hat{\nu}_e^{(2)})$ for the electric and $(\hat{\nu}_m^{(1)},\hat{\nu}_m^{(2)})$ the magnetic sectors.}
\label{fig_sm}
\end{figure}

For the above gauge the soft mode develops at zero momentum as shown in Fig.~\ref{fig_sm} for both the $e$ and $m$  sectors. This can be denoted by 
\begin{align}\label{eq_soft_nu_1}
\hat\nu_e^{(1)}=1;~~~~~~\hat\nu_m^{(1)}=1
\end{align}
for the {\it e (m)} sector on the direct (dual) lattice.

Time reversal symmetry (see Eq.~\ref{eq_gauge_trans_tr}) gives the partner soft mode for both the $e$ and $m$ sectors as shown in Fig.~\ref{fig_sm} which are given by
\begin{align}\label{eq_soft_nu_2}
\hat{\nu}_e^{(2)}=e^{i\pi x};~~~~\hat{\nu}_m^{(2)}=e^{i\pi X}
\end{align}
for the $e$ sector and $m$ sectors. The cartesian coordinates of the direct and dual lattices are given by $(x,y)$ and ($X,Y$) with $X=x+1/2$ and $Y=y+1/2$ (red dashed line in Fig.~\ref{fig_sm}(a) and \ref{fig_sm}(b)). Since no further soft modes are generated by the remaining symmetry, the transition out of the $Z_2$-QSL is described using these soft modes.

\subsection{Symmetry transformations of the soft modes}
\label{sec:smAFM}

Now, using the symmetry transformations of the gauge degrees of freedoms in Eqs. \ref{eq_gauge_trans_transl}-\ref{eq_gauge_trans_rpi}, the transformations of the complex soft modes in Eqs. \ref{eq_soft_e} and \ref{eq_soft_m} are obtained:
\begin{align}
&{\bf T_{d_1}} :\left\{\begin{array}{l}
\Phi_e\rightarrow\Phi_m\\
\Phi_m\rightarrow\Phi_e^*\\
\end{array}\right.~~~
&{\bf T_{d_2}} :\left\{\begin{array}{l}
\Phi_e\rightarrow\Phi_m^*\\
\Phi_m\rightarrow\Phi_e\\
\end{array}\right.\nonumber\\
&{\bf T_{x}} :~\left\{\begin{array}{l}
\Phi_e\rightarrow\Phi_e^*\\
\Phi_m\rightarrow\Phi_m^*\\
\end{array}\right.~~~
&~{\bf T_{y}} :~\left\{\begin{array}{l}
\Phi_e\rightarrow\Phi_e\\
\Phi_m\rightarrow\Phi_m\\
\end{array}\right.\nonumber\\
&{\mathcal{T}} :~~\left\{\begin{array}{l}
\Phi_e\rightarrow-i\Phi_e\\
\Phi_m\rightarrow-i\Phi_m\\
\end{array}\right.~~~
&~{\sigma_v} :~
\left\{\begin{array}{l}
\Phi_e\rightarrow \Phi_e\\
\Phi_m\rightarrow \Phi_m\\
\end{array}\right.\nonumber\\
&C_{2z} :\left\{\begin{array}{l}
\Phi_e\rightarrow i\Phi_e^*\\
\Phi_m\rightarrow i\Phi_m^*\\
\end{array}\right.~~~
&R_{\pi} :\left\{\begin{array}{l}
\Phi_e\rightarrow i\Phi_e^*\\
\Phi_m\rightarrow i\Phi_m^*\\
\end{array}\right. 
\label{eq_emme}
\end{align}

Here we note that the $\sigma^v$ and $R_{\pi}$ symmetries acts differently on the soft modes compared to the Ref. \cite{nanda2020phases}.The gauge invariant spin order parameter in terms of the above soft modes are \cite{lannert2001quantum,xu2009global,PhysRevB.84.104430}:
\begin{align}
&\ttau_i^z\sim |\Phi_e|^2\cos(2\theta^e)~~~~~~\forall i\in {\rm Horizontal~bonds}\nonumber\\
&\ttau_i^x\sim |\Phi_m|^2\cos(2\theta^m)~~~~~~\forall i\in {\rm Vertical~bonds}
\label{eq_magord}
\end{align} 
Crucially, the two spin order parameters are odd under $\mathcal{T},~C_{2z},~R_{\pi}$ symmetry transformations.

\subsection{Symmetry transformation of the gauge fields}
\label{sec:mCS}

Following the $U(1)\times U(1)$ mutual CS formalism, we introduce two internal gauge fields $A_{\mu}$ and $B_\mu$ in Eq. \ref{eq_u1cs} that minimally couples to the electric ($\Phi_e$) and magnetic ($\Phi_m$) soft modes respectively. The transformation rules for the gauge fields follows from Eqs. \ref{eq_gauge_trans_transl}-\ref{eq_gauge_trans_rpi}.

\begin{align}
&{\bf T_{d_1}} :\left\{\begin{array}{l}
A_\mu\rightarrow B_\mu\\
B_\mu\rightarrow-A_\mu\\
\end{array}\right.~~~
{\bf T_{d_2}} :\left\{\begin{array}{l}
A_\mu\rightarrow-B_\mu\\
B_\mu\rightarrow A_\mu\\
\end{array}\right.\nonumber\\
&{\bf T_{x}} :~\left\{\begin{array}{l}
A_\mu\rightarrow -A_\mu\\
B_\mu\rightarrow-B_\mu\\
\end{array}\right.~~~
{\bf T_y} :~\left\{\begin{array}{l}
A_\mu\rightarrow A_\mu\\
B_\mu\rightarrow B_\mu\\
\end{array}\right.\nonumber\\
&{\mathcal{T}} :~~\left\{\begin{array}{l}
A_{\mu}\rightarrow -A_{\mu}\\
B_{\mu}\rightarrow -B_{\mu}\\
\end{array}\right.\nonumber\\
&{\sigma_v} :~\left\{\begin{array}{l}
 A_{x}\rightarrow A_{x},~~A_{y}\rightarrow -A_{y},~~A_\tau\rightarrow A_\tau\\
 B_{x}\rightarrow B_{x},~~B_{y}\rightarrow -B_{y},~~B_\tau\rightarrow B_\tau\\
\end{array}\right.\nonumber\\
&{C_{2z}} :\left\{\begin{array}{l}
 A_{x}\rightarrow A_{x},~~A_{y}\rightarrow -A_{y},~~A_\tau\rightarrow -A_\tau\\
 B_{x}\rightarrow B_{x},~~B_{y}\rightarrow -B_{y},~~B_\tau\rightarrow -B_\tau\\
\end{array}\right.\nonumber\\
&R_\pi :~\left\{\begin{array}{l}
 A_{x}\rightarrow A_{x},~~A_{y}\rightarrow A_{y},~~A_\tau\rightarrow -A_\tau\\
 B_{x}\rightarrow B_{x},~~B_{y}\rightarrow B_{y},~~B_\tau\rightarrow -B_\tau\\
\end{array}\right. \nonumber \\
\label{eq_abba}
\end{align}

\subsection{The phases}
\label{sec:phases}

To capture the phases at the mean field level, for $u>0$, we have
\begin{align}
\langle\Phi_e\rangle=\langle\Phi_m\rangle=0
\end{align}
Thus the complex soft modes can be integrated out so that the effective theory is described by $\mathcal{S}_{CS}$, which is the $Z_2$ QSL phase.

For $u<0$ both the electric and magnetic modes condense, {\it i.e.},
\begin{align}
\langle\Phi_e\rangle,\langle\Phi_m\rangle\neq 0
\end{align}
In this case due to the Anderson-Higgs mechanism the gauge fields acquire a mass and their dynamics is dropped. The four fold terms in Eqs. \ref{eq_e4} and \ref{eq_m4} becomes
\begin{align}
\sim -\lambda\left(|\Phi_e|^4\cos(4\theta^e)+|\Phi_m|^4\cos(4\theta^m)\right]
\label{eq_theta4}
\end{align}

Therefore, for $\lambda>0$ the free energy minima occurs for 
\begin{align}
\theta^e,\theta^m=0,\pm \pi/2, \pi
\end{align}
which gives the two possible the symmetry broken partner spin ordered states as:
\begin{align}
&\langle\ttau_i^z\rangle\sim \langle|\Phi_e|^2\cos(2\theta^e)\rangle\sim \pm 1~~~~~~\forall i\in {\rm Horizontal~bonds}\nonumber\\
&\langle\ttau_i^x\rangle\sim \langle|\Phi_m|^2\cos(2\theta^m)\rangle\sim \pm 1~~~~~~\forall i\in {\rm Vertical~bonds}
\label{eq_magord2}
\end{align} 
Further the state breaks $\mathcal{T}$, $C_{2z}$ and $R_{\pi}$.  In this phase, the interaction between the electric and the magnetic modes (Eq.~\ref{eq_emint}) can be written as 
\begin{align}
\mathcal{L}_{em}\sim w|\Phi_e|^2|\Phi_m|^2\cos(2\theta^e)~\cos(2\theta^m)
\label{eq_emint2}
\end{align}
For $w<0(>0)$, this results in ferromagnetic (antiferromagnetic) spin ordering in terms of $\ttau^x$ (on horizontal bonds) and $\ttau^z$ (on the vertical bonds). The antiferromagnetic order also breaks translation symmetry under ${\bf T_{d_1}}$ and ${\bf T_{d_2}}$ which interchanges a vertical and horizontal bond. The above phenomenology suggest $w\sim \text{sgn}(J)$. Therefore the above critical theory indeed reproduces the right phases.

\subsection{The details of the mutual \texorpdfstring{$Z_2$}{} gauge theory formulation}
\label{appen_z2}

The partition function corresponding to the mutual $Z_2$ action (Eq.~\ref{eq_mutualz2action}) is given by
\begin{align}
    \mathcal{Z}=\sum_{\{\rho\}}\sum_{\{\tilde{\rho}\}}\int \left[\mathcal{D}\theta^e\right]\left[\mathcal{D}\theta^m\right]~\exp\left[-\mathcal{S}\right]
\end{align}
where $\mathcal{S}$ is given by Eq.~\ref{eq_mutualz2action}. 

For further manipulation, we re-write the above partition function as
\begin{align}
    \mathcal{Z}=\sum_{\{\tilde{\rho}\}}\int \left[\mathcal{D}\theta^m\right]~\exp\left[-\mathcal{S}_m\right]~\mathcal{Z}_e
\end{align}
where
\begin{align}
    \mathcal{Z}_e=\sum_{\{\rho\}}\int \left[\mathcal{D}\theta^e\right]~\exp\left[-\mathcal{S}_e-\mathcal{S}_{CS}\right]
\end{align}
We now write the electric action, $\mathcal{S}_e$, as in Eq.~\ref{eq_electricz2} and perform standard steps of XY duality in presence of a $Z_2$ gauge field~\cite{Senthil_PRB_2006,Senthil_2001,Bhattacharjee_PRB_2011}

Starting with writing it down within a Villain approximation as
\begin{align}
    \mathcal{Z}_e=\sum_{\{\rho\}}\sum_{\{m_{ab}\}}\int \left[\mathcal{D}\theta^e\right]~\exp\left[-\mathcal{S}_{CS}\right]~\exp\left[-\mathcal{S}_e^{(1)}\right]
\end{align}
$m_{ab}$ is an integer field living on the links of the direct lattice and 
\begin{align}
    \mathcal{S}_e^{(1)}&=-t\sum_{ab} \left(\theta_a^e-\theta_b^e+\frac{\pi}{2}(1-\rho_{ab})+2\pi m_{ab}\right)^2
\end{align}
which we can decouple via an auxiliary link field $L_{ab}$ to get
\begin{align}
    \mathcal{Z}_e=\sum_{\{\rho\}}\sum_{\{m_{ab}\}}\int \left[\mathcal{D}\theta^e\right]~\left[\mathcal{D}L\right]~\exp\left[-\mathcal{S}_{CS}\right]~\exp\left[-\mathcal{S}_e^{(2)}\right]
\end{align}
where 
\begin{align}
    \mathcal{S}_e^{(2)}&=\frac{1}{2t}\sum_{ab} L_{ab}^2+iL_{ab}\left(\Delta_j\theta_a^e+\frac{\pi}{2}(1-\rho_{ab})+2\pi m_{ab}\right)
\end{align}

The integer field $m_{ab}$ can be integrated out and restricts $L_{ab}$ to an integer leading to Eq.~\ref{eq_intermediatese} in the main text.

 \begin{figure}
 \centering
 \includegraphics[]{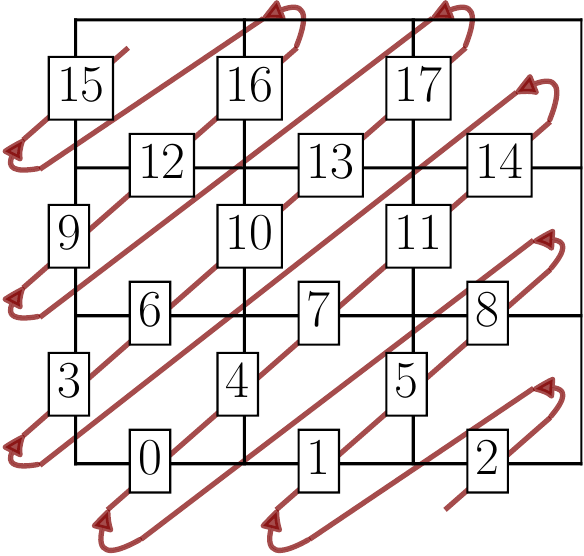}
 \caption{A single Jordan Wigner string running through the $H_1$ direction in an open system.}
 \label{JWstring}
 \end{figure} 

\section{A possible transition between the \texorpdfstring{$Z_2$}{} QSL and a cluster SPT}
\label{appen_majoranamft}

It is interesting to consider the phase transition between the $Z_2$ QSL in the Toric code limit and each of the cluster SPTs given by Eq.~\ref{eq_four_spt}. Such transitions provide examples of yet a new class of novel phase transitions even though presently we do not know a mechanism to stabilise either the cluster SPT phase or this transition in the spin system that we have considered. As we discuss below, this transition is naturally described in terms of Majorana fermions rather than the spins, which makes the transition different from the above class of {\it bosonic} transitions discussed above-- hence we describe them.

In this Appendix, we develop the mean field theory, in particular, for the transition between the $Z_2$ QSL with Hamiltonian given by Eq.~\ref{eq_tc_rot_unrot} and the stacked cluster SPT $H_1$, given by Eq.~\ref{eq_four_spt} such that the Hamiltonian is given by :
\begin{align}
    H'=&\sum_{i}\left[\alpha~ \tau^z_{i+d_1}\tau^x_{i}\tau^z_{i-d_1}+(1-\alpha)\tau^z_{i+d_1}\tau^z_{i-d_2}\tau^y_i\tau^y_{i+d_1-d_2}\right]
\end{align}
where $\alpha$ is the parameter which can be tuned to drive the phase transition. For an open system using the Jordan-Wigner transformations of Eq.~\ref{eq_JWmajorana} as defined in Fig.~\ref{JWstring} the above Hamiltonian becomes
\begin{align}
    H'=\sum_i\left[\alpha~i \tilde\gamma_{i-d_1}\gamma_{i+d_1}+(1-\alpha)\tilde\gamma_{i-d_2}\tilde\gamma_{i-d_2+d_1}\gamma_{i}\gamma_{i+d_1}\right]
    \label{eq_majoranaham}
\end{align}
where $(\tilde\gamma_i,\gamma_i)$ are the two Majorana fermions at site $i$ such that under time-reversal symmetry $\mathcal{T}~:~(\tilde\gamma_i,\gamma_i)\rightarrow (\tilde\gamma_i,-\gamma_i)$. 

In the transformed language, each chain in the stacked cluster SPT at $\alpha=0$ is a pair of spin-less topological superconductor whereas the $Z_2$ QSL is a cluster Mott insulator. 

A mean field decomposition of the four Majorana term along the time reversal invariant channels leads to
\begin{align}
    \tilde\gamma_{i-d_2}\tilde\gamma_{i-d_2+d_1}\gamma_{i}\gamma_{i+d_1}\rightarrow &\langle i~\tilde\gamma_{i-d_2}\gamma_{i}\rangle~i \tilde\gamma_{i-d_2+d_1}\gamma_{i+d_1}\nonumber\\
    &+ i~\tilde\gamma_{i-d_2}\gamma_{i}~\langle i \tilde\gamma_{i-d_2+d_1}\gamma_{i+d_1}\rangle\nonumber\\
    &-\langle i\tilde\gamma_{i-d_2}\gamma_{i+d_1}\rangle~i\tilde\gamma_{i-d_2+d_1}\gamma_{i}\nonumber\\
    &-i\tilde\gamma_{i-d_2}\gamma_{i+d_1}~\langle i\tilde\gamma_{i-d_2+d_1}\gamma_{i}\rangle
\end{align}

Let us define the following mean-field ansatz:
\begin{align}
    \zeta_1\equiv\langle i~\tilde\gamma_{i-d_2}\gamma_{i}\rangle ~~;~~\zeta_2\equiv \langle i \tilde\gamma_{i-d_2+d_1}\gamma_{i+d_1}\rangle \\ \nonumber
    \zeta_3\equiv \langle i\tilde\gamma_{i-d_2}\gamma_{i+d_1}\rangle~~;~~\zeta_4\equiv \langle i\tilde\gamma_{i-d_2+d_1}\gamma_{i}\rangle
\end{align}
which we consider as variational parameters and study the spectrum of the quadratic  Hamiltonian as a function of $\alpha$. Symmetry dictates that $\zeta_1=\zeta_2 = \zeta$;  fourier transforming and defining $\Psi^T = (\tilde{\gamma}_k ~,~ \gamma_k )$, where $k=(k_1,k_2)$ are the reciprocal lattice vectors in $d_1, d_2$ direction respectively, the Hamiltonian one obtains is
\begin{equation}
    H = \sum_k \Psi^\dagger  
    \begin{pmatrix}
  0 & f(k)\\ 
  f^*(k) & 0
\end{pmatrix}
    \Psi 
\end{equation}
 where
\begin{equation}
 f(k) = i e^{-2 i k_1} \left(\alpha +(\alpha -1) e^{i (k_1-k_2)} \left(\zeta_4 -2 \zeta  e^{i k_1}+\zeta_3  e^{2 i k_1}\right)\right)
\end{equation}
 If $\zeta_3=\zeta_4=0$ for a fixed $\zeta_1=\frac{1}{2}=\zeta$, $f(k) = i \alpha  e^{-2 i k_1}-i (\alpha -1) e^{-i k_2}$ which implies a direct transition with a gap closing along the complete $k_2=2k_1+\pi$ line when $\alpha=0.5$. With a finite value of $\zeta_3,\zeta_4$ the nodal line semimetal opens up into a phase with nodal points hosting anisotropic Dirac dispersion. Generically one therefore expects an intermediate gapless phase in the finite region of $\alpha$ when interpolating between a weak SPT and a toric code $Z_2$ QSL. 
 

\bibliography{biblio.bib}

\end{document}